\documentclass[11pt,a4paper]{article}
\pdfoutput=1
\usepackage{jcappub}
\usepackage[english]{babel}
\usepackage{multirow}
\usepackage{ulem}
\usepackage{url}
\usepackage[utf8]{inputenc}
\usepackage{bm}        
\usepackage{amssymb}   
\usepackage{amsmath}
\usepackage{subfig}
\usepackage{caption,booktabs}
\usepackage{hyperref}
\usepackage{xcolor}
\usepackage{lineno}
\usepackage[symbol]{footmisc}

\newcommand{\de}{\mathrm d}

\newcommand{\g}{$\gamma$}
\newcommand{\bi}{\begin{itemize}}
\newcommand{\ei}{\end{itemize}}
\newcommand{\be}{\begin{equation}}
\newcommand{\ee}{\end{equation}}
\newcommand{\nn}{\mathcal{N}}
\newcommand{\Fermi}{\textit{Fermi}-LAT}

\newcommand{\mdm}{m_{\rm DM}}

\newcommand{\phys}{physical}
\newcommand{\pheno}{phenomenological}

\graphicspath{{./}{figures/}}

\title{High-Significance Detection of Correlation Between the Unresolved Gamma-Ray Background and the Large-Scale Cosmic Structure}
\author[1,2,3]{B.~Thakore\footnote[2]{bhashinashish.thakore@unito.it},}
\author[4]{M.~Negro,}
\author[1,2]{M.~Regis,}
\author[1,2,5,6]{S.~Camera,}
\author[3,7]{D.~Gruen,}
\author[1,2]{N.~Fornengo,}
\author[8,9]{A.~Roodman,}
\author[10,11]{A.~Porredon,}
\author[8,9,12]{T.~Schutt,}
\author[1,2]{A.~Cuoco,}
\author[13]{A.~Alarcon,}
\author[14]{A.~Amon,}
\author[15]{K.~Bechtol,}
\author[16]{M.~R.~Becker,}
\author[17]{G.~M.~Bernstein,}
\author[18,19]{A.~Campos,}
\author[20,21,22]{A.~Carnero~Rosell,}
\author[23,24]{M.~Carrasco~Kind,}
\author[25]{R.~Cawthon,}
\author[26,27]{C.~Chang,}
\author[28]{R.~Chen,}
\author[29]{A.~Choi,}
\author[30]{J.~Cordero,}
\author[9]{C.~Davis,}
\author[31]{J.~DeRose,}
\author[32]{H.~T.~Diehl,}
\author[26,27,32]{S.~Dodelson,}
\author[17,33]{C.~Doux,}
\author[26,27,32]{A.~Drlica-Wagner,}
\author[17]{K.~Eckert,}
\author[34]{J.~Elvin-Poole,}
\author[35]{S.~Everett,}
\author[8]{A.~Fert\'e,}
\author[17]{M.~Gatti,}
\author[27,36]{G.~Giannini,}
\author[23,24]{R.~A.~Gruendl,}
\author[37]{I.~Harrison,}
\author[38]{W.~G.~Hartley,}
\author[39]{E.~M.~Huff,}
\author[17]{M.~Jarvis,}
\author[32]{N.~Kuropatkin,}
\author[9]{P.-F.~Leget,}
\author[40]{N.~MacCrann,}
\author[3,8,9,14]{J.~McCullough,}
\author[14]{J.~Myles,}
\author[41]{A. Navarro-Alsina,}
\author[17]{S.~Pandey,}
\author[26,42]{J.~Prat,}
\author[43]{M.~Raveri,}
\author[30]{R.~P.~Rollins,}
\author[44]{A.~J.~Ross,}
\author[8,9]{E.~S.~Rykoff,}
\author[17]{C.~S{\'a}nchez,}
\author[27]{L.~F.~Secco,}
\author[11]{I.~Sevilla-Noarbe,}
\author[45]{E.~Sheldon,}
\author[46]{T.~Shin,}
\author[28]{M.~A.~Troxel,}
\author[47]{I.~Tutusaus,}
\author[32]{B.~Yanny,}
\author[18]{B.~Yin,}
\author[48]{Y.~Zhang,}
\author[21]{M.~Aguena,}
\author[49]{D.~Brooks,}
\author[36]{J.~Carretero,}
\author[21]{L.~N.~da Costa,}
\author[50]{T.~M.~Davis,}
\author[11]{J.~De~Vicente,}
\author[51]{S.~Desai,}
\author[49]{P.~Doel,}
\author[32]{B.~Flaugher,}
\author[27,32]{J.~Frieman,}
\author[52]{J.~Garc\'ia-Bellido,}
\author[13,53,54]{E.~Gaztanaga,}
\author[32]{G.~Gutierrez,}
\author[50]{S.~R.~Hinton,}
\author[55]{D.~L.~Hollowood,}
\author[44,56]{K.~Honscheid,}
\author[57]{D.~J.~James,}
\author[58,59]{K.~Kuehn,}
\author[49]{O.~Lahav,}
\author[39]{S.~Lee,}
\author[21,60]{M.~Lima,}
\author[61]{J.~L.~Marshall,}
\author[62]{J. Mena-Fern{\'a}ndez,}
\author[36,63]{R.~Miquel,}
\author[64]{R.~L.~C.~Ogando,}
\author[18]{A.~Palmese,}
\author[21,64]{A.~Pieres,}
\author[8,9]{A.~A.~Plazas~Malag\'on,}
\author[36,65]{S.~Samuroff,}
\author[11]{E.~Sanchez,}
\author[11]{D.~Sanchez Cid,}
\author[66]{M.~Smith,}
\author[67]{E.~Suchyta,}
\author[68]{G.~Tarle,}
\author[69]{V.~Vikram,}
\author[48]{A.~R.~Walker,}
\author[31,70]{and N.~Weaverdyck}
\affiliation[1]{Dipartimento di Fisica, Universit\`a degli Studi di Torino, via P. Giuria 1, 10125 Torino, Italy}
\affiliation[2]{INFN -- Istituto Nazionale di Fisica Nucleare, Sezione di Torino, via P. Giuria 1, 10125 Torino, Italy}
\affiliation[3]{University Observatory, Faculty of Physics, Ludwig-Maximilians-Universit\"at, Scheinerstr. 1, 81679 Munich, Germany}
\affiliation[4]{Department of Physics \& Astronomy, Louisiana State University, Baton Rouge, LA 70803, USA}
\affiliation[5]{Department of Physics \& Astronomy, University of the Western Cape, Cape Town 7535, South Africa}
\affiliation[6]{INAF -- Istituto Nazionale di Astrofisica, Osservatorio Astrofisico di Torino, strada Osservatorio 20, 10025 Pino Torinese, Italy}
\affiliation[7]{Excellence Cluster ORIGINS, Boltzmannstr. 2, 85748 Garching, Germany}
\affiliation[8]{SLAC National Accelerator Laboratory, Menlo Park, CA 94025, USA}
\affiliation[9]{Kavli Institute for Particle Astrophysics \& Cosmology, P. O. Box 2450, Stanford University, Stanford, CA 94305, USA}
\affiliation[10]{Ruhr University Bochum, Faculty of Physics and Astronomy, Astronomical Institute, German Centre for Cosmological Lensing, 44780 Bochum, Germany}
\affiliation[11]{Centro de Investigaciones Energ\'eticas, Medioambientales y Tecnol\'ogicas (CIEMAT), Madrid, Spain}
\affiliation[12]{Department of Physics, Stanford University, 382 Via Pueblo Mall, Stanford, CA 94305, USA}
\affiliation[13]{Institute of Space Sciences (ICE, CSIC),  Campus UAB, Carrer de Can Magrans, s/n,  08193 Barcelona, Spain}
\affiliation[14]{Department of Astrophysical Sciences, Princeton University, Peyton Hall, Princeton, NJ 08544, USA}
\affiliation[15]{Physics Department, 2320 Chamberlin Hall, University of Wisconsin-Madison, 1150 University Avenue Madison, WI  53706-1390}
\affiliation[16]{Argonne National Laboratory, 9700 South Cass Avenue, Lemont, IL 60439, USA}
\affiliation[17]{Department of Physics and Astronomy, University of Pennsylvania, Philadelphia, PA 19104, USA}
\affiliation[18]{Department of Physics, Carnegie Mellon University, Pittsburgh, Pennsylvania 15312, USA}
\affiliation[19]{NSF AI Planning Institute for Physics of the Future, Carnegie Mellon University, Pittsburgh, PA 15213, USA}
\affiliation[20]{Instituto de Astrofisica de Canarias, E-38205 La Laguna, Tenerife, Spain}
\affiliation[21]{Laborat\'orio Interinstitucional de e-Astronomia - LIneA, Rua Gal. Jos\'e Cristino 77, Rio de Janeiro, RJ - 20921-400, Brazil}
\affiliation[22]{Universidad de La Laguna, Dpto. Astrofísica, E-38206 La Laguna, Tenerife, Spain}
\affiliation[23]{Center for Astrophysical Surveys, National Center for Supercomputing Applications, 1205 West Clark St., Urbana, IL 61801, USA}
\affiliation[24]{Department of Astronomy, University of Illinois at Urbana-Champaign, 1002 W. Green Street, Urbana, IL 61801, USA}
\affiliation[25]{Physics Department, William Jewell College, Liberty, MO, 64068}
\affiliation[26]{Department of Astronomy and Astrophysics, University of Chicago, Chicago, IL 60637, USA}
\affiliation[27]{Kavli Institute for Cosmological Physics, University of Chicago, Chicago, IL 60637, USA}
\affiliation[28]{Department of Physics, Duke University Durham, NC 27708, USA}
\affiliation[29]{NASA Goddard Space Flight Center, 8800 Greenbelt Rd, Greenbelt, MD 20771, USA}
\affiliation[30]{Jodrell Bank Center for Astrophysics, School of Physics and Astronomy, University of Manchester, Oxford Road, Manchester, M13 9PL, UK}
\affiliation[31]{Lawrence Berkeley National Laboratory, 1 Cyclotron Road, Berkeley, CA 94720, USA}
\affiliation[32]{Fermi National Accelerator Laboratory, P. O. Box 500, Batavia, IL 60510, USA}
\affiliation[33]{Universit\'e Grenoble Alpes, CNRS, LPSC-IN2P3, 38000 Grenoble, France}
\affiliation[34]{Department of Physics and Astronomy, University of Waterloo, 200 University Ave W, Waterloo, ON N2L 3G1, Canada}
\affiliation[35]{California Institute of Technology, 1200 East California Blvd, MC 249-17, Pasadena, CA 91125, USA}
\affiliation[36]{Institut de F\'{\i}sica d'Altes Energies (IFAE), The Barcelona Institute of Science and Technology, Campus UAB, 08193 Bellaterra (Barcelona) Spain}
\affiliation[37]{School of Physics and Astronomy, Cardiff University, CF24 3AA, UK}
\affiliation[38]{Department of Astronomy, University of Geneva, ch. d'\'Ecogia 16, CH-1290 Versoix, Switzerland}
\affiliation[39]{Jet Propulsion Laboratory, California Institute of Technology, 4800 Oak Grove Dr., Pasadena, CA 91109, USA}
\affiliation[40]{Department of Applied Mathematics and Theoretical Physics, University of Cambridge, Cambridge CB3 0WA, UK}
\affiliation[41]{Instituto de F\'isica Gleb Wataghin, Universidade Estadual de Campinas, 13083-859, Campinas, SP, Brazil}
\affiliation[42]{Nordita, KTH Royal Institute of Technology and Stockholm University, Hannes Alfv\'ens v\"ag 12, SE-10691 Stockholm, Sweden}
\affiliation[43]{Department of Physics, University of Genova and INFN, Via Dodecaneso 33, 16146, Genova, Italy}
\affiliation[44]{Center for Cosmology and Astro-Particle Physics, The Ohio State University, Columbus, OH 43210, USA}
\affiliation[45]{Brookhaven National Laboratory, Bldg 510, Upton, NY 11973, USA}
\affiliation[46]{Department of Physics and Astronomy, Stony Brook University, Stony Brook, NY 11794, USA}
\affiliation[47]{Institut de Recherche en Astrophysique et Plan\'etologie (IRAP), Universit\'e de Toulouse, CNRS, UPS, CNES, 14 Av. Edouard Belin, 31400 Toulouse, France}
\affiliation[48]{Cerro Tololo Inter-American Observatory, NSF's National Optical-Infrared Astronomy Research Laboratory, Casilla 603, La Serena, Chile}
\affiliation[49]{Department of Physics \& Astronomy, University College London, Gower Street, London, WC1E 6BT, UK}
\affiliation[50]{School of Mathematics and Physics, University of Queensland,  Brisbane, QLD 4072, Australia}
\affiliation[51]{Department of Physics, IIT Hyderabad, Kandi, Telangana 502285, India}
\affiliation[52]{Instituto de Fisica Teorica UAM/CSIC, Universidad Autonoma de Madrid, 28049 Madrid, Spain}
\affiliation[53]{Institut d'Estudis Espacials de Catalunya (IEEC), 08034 Barcelona, Spain}
\affiliation[54]{Institute of Cosmology and Gravitation, University of Portsmouth, Portsmouth, PO1 3FX, UK}
\affiliation[55]{Santa Cruz Institute for Particle Physics, Santa Cruz, CA 95064, USA}
\affiliation[56]{Department of Physics, The Ohio State University, Columbus, OH 43210, USA}
\affiliation[57]{Center for Astrophysics $\vert$ Harvard \& Smithsonian, 60 Garden Street, Cambridge, MA 02138, USA}
\affiliation[58]{Australian Astronomical Optics, Macquarie University, North Ryde, NSW 2113, Australia}
\affiliation[59]{Lowell Observatory, 1400 Mars Hill Rd, Flagstaff, AZ 86001, USA}
\affiliation[60]{Departamento de F\'isica Matem\'atica, Instituto de F\'isica, Universidade de S\~ao Paulo, CP 66318, S\~ao Paulo, SP, 05314-970, Brazil}
\affiliation[61]{George P. and Cynthia Woods Mitchell Institute for Fundamental Physics and Astronomy, and Department of Physics and Astronomy, Texas A\&M University, College Station, TX 77843,  USA}
\affiliation[62]{LPSC Grenoble - 53, Avenue des Martyrs 38026 Grenoble, France}
\affiliation[63]{Instituci\'o Catalana de Recerca i Estudis Avan\c{c}ats, E-08010 Barcelona, Spain}
\affiliation[64]{Observat\'orio Nacional, Rua Gal. Jos\'e Cristino 77, Rio de Janeiro, RJ - 20921-400, Brazil}
\affiliation[65]{Department of Physics, Northeastern University, Boston, MA 02115, USA}
\affiliation[66]{Physics Department, Lancaster University, Lancaster, LA1 4YB, UK}
\affiliation[67]{Computer Science and Mathematics Division, Oak Ridge National Laboratory, Oak Ridge, TN 37831}
\affiliation[68]{Department of Physics, University of Michigan, Ann Arbor, MI 48109, USA}
\affiliation[69]{Central University of Kerala, Kasargod, Kerala, India - 671316}
\affiliation[70]{Department of Astronomy, University of California, Berkeley,  501 Campbell Hall, Berkeley, CA 94720, USA}      
\abstract{
Our understanding of the \g-ray sky has improved dramatically in the past decade, however, the unresolved \g-ray background (UGRB) still has a potential wealth of information about the faintest \g-ray sources pervading the Universe. Statistical cross-correlations with tracers of cosmic structure can indirectly identify the populations that most characterize the \g-ray background.  In this study, we analyze the angular correlation between the \g-ray background and the matter distribution in the Universe as traced by gravitational lensing, leveraging more than a decade of observations from the \textit{Fermi}-Large Area Telescope (LAT) and 3 years of data from the Dark Energy Survey (DES). 
We detect a correlation at signal-to-noise ratio of 8.9. Most of the statistical significance comes from large scales, demonstrating, for the first time, that a substantial portion of the UGRB aligns with the mass clustering of the Universe as traced by weak lensing. Blazars provide a plausible explanation for this signal, especially if those contributing to the correlation reside in halos of large mass ($\sim 10^{14} M_{\odot}$) and account for approximately 30-40\% of the UGRB above 10 GeV. Additionally, we observe a preference for a curved \g-ray energy spectrum, with a log-parabolic shape being favored over a power-law. We also discuss the possibility of modifications to the blazar model and the inclusion of additional \g-ray sources, such as star-forming galaxies, misalinged active galactic nuclei, or particle dark matter. 
}

\begin{document}
\begin{nolinenumbers}
\vspace*{-\headsep}\vspace*{\headheight}
\footnotesize \hfill FERMILAB-PUB-25-0018-PPD\\
\vspace*{-\headsep}\vspace*{\headheight}
\footnotesize \hfill DES-2024-0866
\end{nolinenumbers}

\maketitle

\renewcommand{\thefootnote}{\arabic{footnote}}
\section{Introduction}
\label{sec:intro}
High-energy astronomy is an endeavour that can potentially provide insights into the disciplines of astrophysics, cosmology and particle physics. Stemming from extremely violent events in the Universe, \g-rays act as messengers, providing information about the mechanisms of rare events as seen in supernovae, as well as those that occur due to matter under extreme stress, such as in the vicinity of pulsars or Active Galactic Nuclei (AGNs). In addition to their use in astrophysics, \g-ray frequencies can also shed light on an important cosmological component, dark matter (DM), representing approximately 25\% of the Universe's energy budget \cite{aghanim2020planck}. DM has been theorized to consist of an exotic fundamental particle which may annihilate or decay into standard-model particles, thereby producing cosmic messengers.
In the case of Weakly Interacting Massive Particles (WIMPs) or any other potential DM particle with a mass in the GeV range or higher, their annihilation or decay is likely to produce photons in the \g-ray spectrum~\cite{ullio2002cosmological}. 
However, due to the small cross-section of DM annihilation, the potential number of detectable events is limited. Furthermore, the \g-ray sky is filled with emissions from various astrophysical sources, such as pulsars, supernova remnants, AGN, and cosmic-ray interaction with Galactic interstellar medium and radiation fields. These emissions create background noise, masking potential signals from DM annihilation. 

A method for distinguishing between the non-thermal \g-ray emissions originating from astrophysical sources and those that might be caused by DM annihilation or decay within the Unresolved Gamma-Ray Background (UGRB) hinges on the concept of cross-correlating UGRB maps with various other maps that trace the underlying large-scale structure of the Universe. Such tracers include cosmic phenomena like the weak gravitational lensing effect~\cite{camera2013novel,camera2015tomographic,shirasaki2014cross,shirasaki2016cosmological,troster2017cross,DES:2019ucp}, the clustering of galaxies~\cite{xia2015tomography,regis2015particle,cuoco2015dark,shirasaki2015cross,cuoco2017tomographic,Ammazzalorso:2018evf,paopiamsap2024constraints} and galaxy clusters~\cite{branchini2017cross,shirasaki2018correlation,hashimoto2019measurement,colavincenzo2020searching,tan2020bounds}, and the lensing effect of the Cosmic Microwave Background (CMB)~\cite{fornengo2015evidence}, which reflect the large-scale distribution of matter across cosmological distances (see also Refs. \cite{ando2014mapping,fornengo2014particle,fornasa2016angular,feng2017planck}).

These techniques are especially promising because they serve as direct gravitational probes of the mass distribution in the Universe, the vast majority of which is expected to be DM \cite{aghanim2020planck}. By examining the angular, energy, and redshift behaviour of the cross-correlations, it becomes possible to disentangle signals that may arise from ordinary astrophysical sources from those associated with DM interactions. Astrophysical sources are hosted within DM halos, and thus exhibit correlations with other direct tracers of the underlying large-scale cosmic structure such as galaxy clustering or weak gravitational lensing. On the other hand, they are typically much smaller than DM halos (thus showing different angular spectra), and mainly follow the redshift dependence of star formation as opposed to the WIMP signals, which peak at very low redshifts. Astrophysical sources also have smooth, power-law like spectra in energy while WIMPs exhibit a cutoff and a more curved spectrum. These characteristics  are handles to tell apart \g-ray emission from astrophysical and dark matter origin.

Beyond its utility in the search for DM, this approach also has broader implications for understanding the population of unresolved \g-ray sources, providing valuable insights into their redshift distribution and clustering properties. This can help to refine our understanding of a variety of \g-ray emitting objects, such as AGN, distant star-forming galaxies (SFGs), and other yet-to-be-resolved populations of astrophysical objects. A key contributor to the UGRB is the blazar population. Blazars are a type of AGN that is estimated to contribute significantly to the UGRB \cite{ajello2015origin}, dominating in particular at fluxes just below the \Fermi\ source detection threshold~\cite{korsmeier2022flat}. Their influence must be therefore carefully accounted for when attempting to identify the composition of the signals in the aforementioned cross-correlations. Blazars exhibit distinct clustering patterns and spectra, which, when accurately modeled, can help distinguish their signals from other \g-ray sources, improving our understanding of their contribution to the UGRB.


In this work, we present a study involving cross-correlations using weak lensing. The weak lensing observable that we consider here is the tangential shear, which is a result of the distortion of background (source) galaxies as a result of the foreground lensing galaxies. This is known as galaxy-galaxy lensing. \\
Over the years, several observational attempts at disentangling cross-correlation signals for both their dark and visible nature have been carried out involving weak lensing \cite{shirasaki2014cross,shirasaki2016cosmological,troster2017cross}. In particular, a cross-correlation study in 2020 by Amazzalorso et al. \cite{DES:2019ucp} obtained the first identification of a cross-correlation signal between the UGRB, as seen by the \Fermi\ 9 year data, and the distribution of mass in the Universe probed by weak gravitational lensing, measured by the DES Y1 datasets \cite{diehl2014dark}. The cross-correlation signal was detected with a significance of 5.3$\sigma$, and found a preference for a DM-inclusive model over a purely astrophysical model at 2.8$\sigma$. Here, we build on the aforementioned detection using 12 years of \g-ray data from \Fermi\ \cite{Fermi-LAT:2019yla,abdollahi2022incremental}, and three-year (Y3) weak lensing shear measurements from DES. 

The paper is organized as follows.
Section \ref{sec:models} describes the phenomenological and physical models used to interpret the cross-correlation signal.
Section \ref{sec:data} describes the \Fermi\ and DES data used for the cross-correlation analysis. Details of the analysis are reported in Section \ref{sec:res}, which also offers an interpretation of the results in terms of phenomenological models, outlining the main properties of the signal. Inferences in terms of astrophysical sources follow in Section \ref{sec:phys}, followed by a conclusion in Section \ref{sec:conc}. Appendix \ref{sec:cov} provides details on the construction of the covariance used in the data analysis, while additional discussions on SFGs, misaligned AGNs (mAGNs), and DM annihilation can be found in Appendices \ref{sec:SFG}, \ref{sec:mAGN}, and \ref{sec:DM} respectively. Finally in Appendix \ref{sec:robustness} we elaborate upon the robustness checks that were performed to validate the results of the analysis.

\section{Theoretical Framework} \label{sec:models}
In this work, we study the 2-point angular cross-correlation function between the UGRB and gravitational shear.
The UGRB is obtained from photon counts measured by the \Fermi\ in different energy bins, after removing the contribution from resolved sources and Galactic foreground. The gravitational shear is given by the tangential ellipticity of galaxies measured by DES, and trace the mass distribution in the Universe. Here we want to test whether the \g-ray background fluctuations are sourced by the matter distribution.

For a map of $\gamma$-rays in the $a$th energy bin and a shear catalog in the $r$th redshift bin, the real-space quantity depicting their cross-correlation $\hat\Xi^{ar}$ can be theoretically computed from the harmonic-space cross-power spectrum $C_\ell^{ar}$ by a Legendre transform \cite{friedrich2021dark}:
\be
\hat\Xi^{ar}(\theta) = \sum_\ell \frac{2\ell+1}{4\pi\ell(\ell+1)} C_\ell^{ar}\, W_\ell^a\,P^{(2)}_\ell(\cos\theta),
\label{eq:Cl2xi}
\ee
with $\theta$ being the angular separation on the sky, $P^{(2)}_\ell$ the Legendre polynomial of order two, and $W_\ell^a$ the beam window function. The latter is computed from a Legendre transform of the \Fermi\ Point-Spread Function (PSF), see Appendix II in \cite{2018PhRvL.121x1101A}, and accounts for the finite resolution of the detector (we neglect the DES and pixel smoothing since they act on much smaller scales than the \Fermi\ PSF).

Correlations that occur at physical scales smaller than the \Fermi\ PSF can be approximated with $C_\ell^{ar}={\rm const}$. On the other hand, correlations on very large scales can be well-described by the clustering in the linear regime, with the angular dependence dictated by the linear matter power spectrum~\cite{Cooray2002}, that we computed from the transfer function in \cite{Eisenstein:1997jh}. In this study, we have adopted a halo model approach to describe both the phenomenological and physical characteristics of the cross-correlations \cite{asgari2023halo}. In the halo-model framework, all mass in the Universe's large-scale structure is assumed to reside within virialised DM halos. Consequently, the correlation function splits into two distinct contributions: the 1-halo and 2-halo terms (abbreviated as `1h' and `2h' in the succeeding equations). The 1-halo term captures correlations between two points within the same halo, dominating at small angular scales. On the other hand, the 2-halo term accounts for correlations between points in different halos, aligning with the broader matter distribution and prevailing at large scales. 

Our first theoretical model is a \pheno\ model constructed as a Power-Law (PL) in energy as follows:
\begin{equation}
\begin{split}
\Xi_{\rm PL}^{ar}(\theta) \  = \Big[A_1  (E_a/E_0)^{-\alpha_1} \Big(\frac{1+z_r}{1+z_0}\Big)^{\beta_1}\ \hat{\Xi}_\textrm{PSF-like}^{a}(\theta) \\
+A_2 (E_a/E_0)^{-\alpha_2} \Big(\frac{1+z_r}{1+z_0}\Big)^{\beta_2}\ \hat\Xi_\textrm{2h-like}^{ar}(\theta)\Big]\frac{\Delta E_a}{\langle I_a \rangle},
\label{eq:phenopl}
\end{split}
\end{equation}
where $E_a$ and $z_r$ are the central values of the energy (measured in  GeV) and redshift bins, $E_0$ is the pivot energy, chosen as the geometric mean of the energy bin centres at $E_0 = 13.7\, \mathrm{GeV}$, $\Delta E_a$ is the width of the energy bin and $\langle I_a \rangle$ is the measured photon flux, provided in Table~\ref{tab:enbins}. Similarly to $E_0$, we have also defined a pivot redshift as a geometric mean of the centres of the four redshift bins at $z_0 = 0.64$.\footnote{The pivot energy and redshift values, while determined arbitrarily, are primarily chosen in order to decorrelate the uncertainties in the slope and the offset. By introducing a pivot energy and redshift, we can remove the degeneracy that arises from the interplay between the normalisation and the spectral and (in the case of a log-parabolic model) curvature indices, which makes the fit less stable. A pivot value removes this degeneracy  and makes the fit more stable and easier to interpret by giving us cleaner uncertainties and their concomitant best fits.}
Note that the term in square bracket is differential in energy. $\hat{\Xi}_\textrm{PSF-like}^{a}(\theta)$ is the Legendre transform of the \Fermi\ PSF (i.e., with $C_\ell^{ar}=1$ in Eq.~\ref{eq:Cl2xi})  and $\hat\Xi_\textrm{2h-like}^{ar}(\theta)$ is a generic large-scale contribution, obtained from linear theory. Correlation functions with a hat have flux units, while those without a hat are normalised to the \g-ray flux, and therefore dimensionless. We make this distinction because the comparison with the measurement is performed with dimensionless Cross-Correlation Function (CCF) as described below, while the comparison with physical models is easier for dimensional quantities. The two normalisations $A_1$ and $A_2$, the spectral indices $\alpha_1$ and $\alpha_2$, and the redshift evolution indices $\beta_1$ and $\beta_2$ are free parameters of the model.
For the aforementioned parameters, we have considered a \pheno\ model with a free relative amplitude between the PSF-like and large-scale terms that can capture the angular behaviour of the signal.
As we will show in the next section, the redshift behaviour is not strongly constrained in our analysis and, for simplicity, we assume a power-law scaling.
Gamma-ray sources typically have energy spectra that can be well approximated by a power-law, and so it is assumed in Eq. (\ref{eq:phenopl}). 
\\
On the other hand, to explore the possibility of a ``curved" spectrum, we devise a second \pheno\ model, considering a Log-Parabola (LP) spectrum in energy (which also describes several \g-ray sources of the \Fermi\ 4FGL catalog~\cite{Fermi-LAT:2019yla}):
\begin{equation}
\begin{split}
\Xi_{\rm \mathrm{LP}}^{ar}(\theta) \  = \Big[A_1  (E_a/E_0)^{-\alpha_1+\gamma_1\log_{10} (E/E_0)} \Big(\frac{1+z_r}{1+z_0}\Big)^{\beta_1}\ \hat{\Xi}_\textrm{PSF-like}^{a}(\theta) \\
+A_2 (E_a/E_0)^{-\alpha_2+\gamma_2\log_{10}(E/E_0)} \Big(\frac{1+z_r}{1+z_0}\Big)^{\beta_2}\ \hat\Xi_\textrm{2h-like}^{ar}(\theta)\Big]\frac{\Delta E_a}{\langle I_a \rangle},
\label{eq:phenolp}
\end{split}
\end{equation}
where $\gamma_1$ and $\gamma_2$ are the spectral indices providing the deviation from a power-law (henceforth referred to as curvature indices) . 

A physical model of the angular power spectrum of the cross-correlation between \g-ray sources and the gravitational shear can be derived as~\cite{Fornengo2014}
\be
C_\ell^{ar} = \int\de E\, \de z\,\frac{1}{H(z)}\frac{W_\textrm{\rm gamma}^a(E,z)W_{\rm shear}^r(z)}{\chi(z)^2}
P_{\gamma\delta}\!\!\left[k=\frac{\ell}{\chi(z)},z\right],
\label{eq:clgen}
\ee
where $\chi(z)$ is the comoving distance to redshift $z$, obeying $\de z/\de\chi=H(z)$ with $H(z)$ the Hubble rate, $W_\textrm{\rm gamma}^a(E,z)$ and $W_{\rm shear}^r(z)$ are the so-called window functions, providing the redshift distribution of the signals, and $P_{\gamma\delta}$ is the three-dimensional cross-power spectrum between a given \g-ray population sourcing the UGRB emission and the matter density contrast $\delta$. In the Limber approximation, the physical scale $k$ (the modulus of the physical wavenumber) and the angular multipole $\ell$ are related by $k=\ell/\chi(z)$. 
All the ingredients entering Eq.~(\ref{eq:clgen}) are detailed in Ref.~\cite{DES:2019ucp}, with cosmological parameters from Ref.~\cite{DES:2021wwk}.

In the main analysis, we will consider blazars (BLZ) to dominate the \g-ray source population. We discuss the addition of SFGs and mAGNs in Appendices \ref{sec:SFG} and \ref{sec:mAGN}. The window function is related to the \g-ray luminosity function (GLF) $\phi$, which depends on luminosity, energy and redshift. For BLZ we will follow Ref.~\cite{korsmeier2022flat} and, unless otherwise specified, we set all the parameters describing $\phi$ to the best-fit of the angular auto-correlation found in Ref.~\cite{korsmeier2022flat}, see the BLL 4FGL+$C_{\rm P}$ ﬁt in their Table 2. 
Specifically, we adopt the following decomposition of the GLF $\Phi(L_{\gamma}, z, \Gamma) = dN/dL_\gamma dV d\Gamma$ (defined as the number of sources per unit of luminosity $L_{\gamma}$,  co-moving volume $V$ at resdhift $z$ and photon spectral index $\Gamma$) in terms of its expression at $z=0$ and a redshift-evolution function $e(L_{\gamma}, z)$:
\begin{equation}
\Phi(L_\gamma,z,\Gamma)= \Phi(L_\gamma,0,\Gamma) \times e(L_{\gamma}, z),
\label{eq:GLF}
\end{equation}
where $L_\gamma$ is the rest-frame luminosity in the energy range $(0.1-100)$ GeV. At redshift $z=0$: 
\begin{eqnarray}
\Phi(L_\gamma,0,\Gamma) &=& \frac{A}{\ln(10) L_\gamma} \left[ \left( \frac{L_\gamma}{L_0} \right)^{{ \kappa_1}} + 
\left( \frac{L_\gamma}{L_0} \right)^{{ \kappa_2}} \right]^{-1}  \times
  \exp \left[ - \frac{(\Gamma-\mu_{\rm BLZ})^2}{2\sigma^2} \right] \;,\label{eq:phies} 
\end{eqnarray}
where $A$ is a normalization factor, the indices $\kappa_1$ and $\kappa_2$ govern the evolution of the GLF with the luminosity $L_{\gamma}$ and the Gaussian term takes into account the distribution of the spectral indices $\Gamma$ around their mean $\mu_{\rm BLZ}$, with a dispersion $\sigma$. 

The redshift behaviour is given by
\begin{equation}
e(L_\gamma,z)=\left[\left( \frac{1+z}{1+z_c}\right)^{-p_1}+\left( \frac{1+z}{1+z_c}\right)^{-p_2}\right]^{-1}\;.
\label{eq:phiz}
\end{equation}

Then the \g-ray window function can be then written as~\cite{DES:2019ucp}:
\be
    W_\textrm{{\rm gamma}}^a(E,z) = \chi(z)^2 \int\, \de \Gamma \,\int_{L_{\gamma}^{\rm min}}^{L_{\gamma}^{\rm max}} \de L_{\gamma}
   \, \Phi_{\rm S}(L_{\gamma},z,\Gamma) \, \frac{\de N}{\de E}\times e^{-\tau\left[E(1+z),z\right]}  \ ,
    \label{eq:win_astro}
\ee
where $\de N/\de E$ is the \g-ray spectrum, taken to be a power-law, $\Phi_\mathrm{S}$ is the GLF of the unresolved source population, $\tau$ is the optical depth due to absorption, and $L_{\gamma}^{\rm max}$ ensures we are considering only unresolved sources. 
Let us note here that, in the following, when we mention the \g-ray spectrum, we will refer to the differential (in energy) photon spectrum, i.e., the quantity denoted by $dN/dE$ in Eq.~\ref{eq:win_astro}.

As already mentioned, we can separate the 1-halo and 2-halo contributions of the power spectrum, so we can write the  {\it physical} model as:

\begin{align}
&\Xi_{\rm phys}^{ar} (\theta) \ \langle I_a \rangle =  A_{\rm BLZ}^{\rm 1h}\ \hat\Xi_{\rm BLZ,1h}^{ar} (\theta,\mu_{\rm BLZ},p_1)+  A_{\rm BLZ}^{\rm 2h}\  \hat\Xi_{\rm BLZ,2h}^{ar} (\theta,\mu_{\rm BLZ},p_1) 
\label{eq:physmdl}
\end{align}
The model parameters, that will be constrained through our Markov Chain Monte Carlo (MCMC) scan, are: two free normalizations for the 1-halo and 2-halo terms, $A_{\rm BLZ}^{\rm 1h}$ and $A_{\rm BLZ}^{\rm 2h}$, which effectively describe the normalization $A$ in Eq.~\ref{eq:phies} and the normalization of the halo-matter bias (see eg. Ref. \cite{asgari2023halo} for a complete description of halo biases), the spectral index $\mu_{\rm BLZ}$ in Eq.~\ref{eq:phies}, governing the energy dependence, and the redshift parameter $p_1$ in Eq.~\ref{eq:phiz}\footnote{We also investigated the possibility of using two different energy and redshift parameters for the 1- and 2-halo terms, but we found that this does not add more information.}.
All the other parameters of the GLF are set following the best-fit of BL Lacs\footnote{BL Lacs, short for BL Lacertae, are a type of  AGN specifically categorized as blazars. Calling a blazar a BL Lac separates it from the other blazar category of Flat Spectrum Radio Quasars (FSRQs). For more information about blazars see eg. Mukherjee et al. \cite{mukherjee1997egret}.} from the analysis of \g-ray number counts and angular auto-correlation in ~\cite{korsmeier2022flat}.

As for the \pheno\ model, all terms are computed in different energy and redshift bins, labeled by indices $a$ and $r$, respectively.
\section{Data}
\label{sec:data}
\subsection{Weak lensing data} \label{sec:DESdata}
The Dark Energy Survey is a six-year observing program that was carried out with a 570 megapixel camera, the Dark Energy Camera (DECam for short), mounted on the Blanco 4m telescope at the Cerro-Tololo  Inter-American Observatory in Chile \cite{flaugher2015dark}. Spanning a total area of 4143 ${\rm deg}^2$ after masking, the Year 3 survey provides 2.76 times the weak lensing survey area compared to the 1500 ${\rm deg}^2$ for DES Y1 \cite{gatti2021dark,maccrann2022dark}. The observations are carried out in five broadband filters, namely, the \textit{griz}Y, ranging from $\sim 400$ nm to $\sim 1060$ nm in wavelength. The DES Year 3 (Y3) analysis utilizes the data acquired over the first three years of observations, amounting to a total of 319 nights. 
 \begin{figure}[!h]
     \centering
     \includegraphics[width=0.8\linewidth]{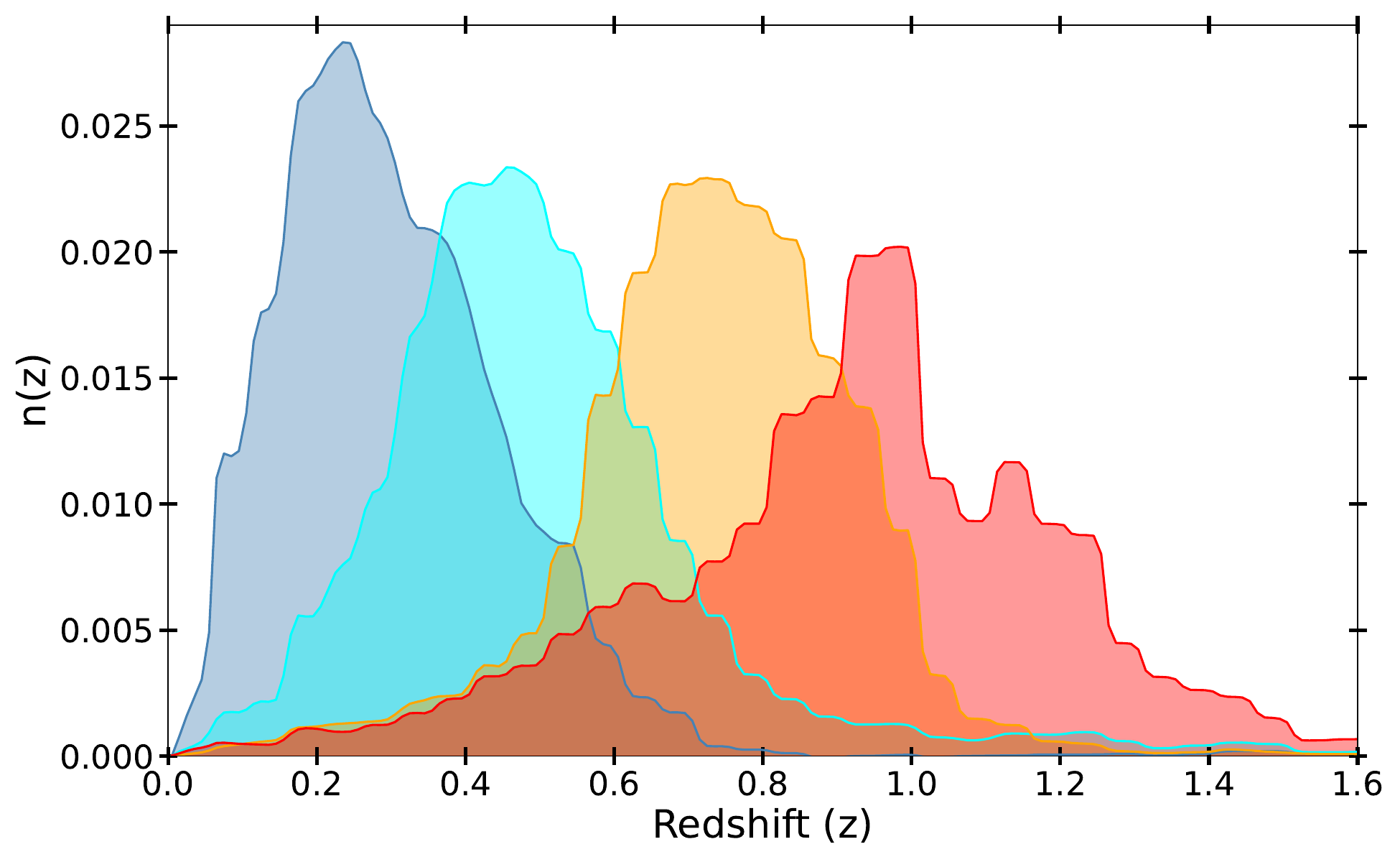}
     \caption{ Normalized redshift distribution $n(z)$ of the DES Y3 source galaxies, with the data taken from  Ref. \cite{DES:2021wwk}.}
     \label{fig:nz_plot}
 \end{figure}
In our analysis we make use of the \textsc{metacalibration} shear catalogue, 
which is fully described in Ref. \cite{sheldon2017practical,huff2017metacalibration}, 
and included in the DES Y3 shape catalogue 
of Ref. \cite{gatti2021dark}. The DES Y3 shape catalog that is used in our analysis is a subset of the objects in the \textsc{Gold} catalog \cite{sevilla2021dark} that pass the \textsc{metacalibration} cuts, and provides calibrated ellipticity measurements that describe the shapes of individual galaxies after correcting for observational biases. It also includes redshift bin assignments for each galaxy, grouping them by estimated distance to enable studies of large-scale structure and galaxy-galaxy lensing.
By introducing artificial shear to images and tracking the response of the estimator to the applied shear, one can address both model and noise biases through the introduction of a mean response factor $R$. The methodology indicates, therefore, that \textsc{Metacalibration} can be applied to calibrate any shear estimator, including the shapes that have been derived from model fitting or weighted moments. \textsc{metacalibration} has been shown to be accurate at the part-per-thousand level in the absence of blending with other galaxies \cite{sheldon2017practical}, and at the part-per-hundred level for the blending present in the DES Y3 data \cite{gatti2021dark}.

After accounting for the additive and multiplicative biases, the wide-field data for the weak lensing shape catalogue consists of approximately 100 million galaxies, having an effective source number density of $n_\text{eff}$ = 5.59 gal/arcmin$^2$, and a corresponding shape noise of $\sigma_e = 0.261$ \cite{gatti2021dark}. The measurements were carried out in the \textit{riz} bands, with the $g$-band excluded due to insufficient PSF modelling.

In order to establish reliable cross-correlation constraints, it is necessary to calibrate the redshift measurements for the source galaxies properly, allowing for a reliable estimate of the tangential shear that becomes one of the two key components of the two-point estimator that generates the cross-correlation values. For DES Y3, the method used to determine and calibrate the photometric redshift distribution of the wide-field galaxies is a combination of two methods: the Self-Organizing Maps $p(z)$ (abbreviated as SOMPZ) \cite{buchs2019phenotypic}, and the clustering redshift technique WZ \cite{gatti2022dark}.  \\
Some uncertainties arise from factors like the limited coverage of the Deep Fields and the finite number of simulated \textsc{Balrog} sources \cite{everett2022dark}, which inject mock galaxies into real survey images to account for systematic biases (see, for example, Ref.~\cite{suchyta2016no}). To mitigate these uncertainties, an ensemble of redshift distribution realisations, $n_i(z)$, is constructed for each redshift bin. By generating multiple realisations, the analysis captures a range of possible outcomes, ensuring more robust estimates of the underlying redshift distribution. Additionally, the clustering redshift technique (WZ) improves redshift constraints by cross-correlating the weak lensing sources with galaxies of known redshift.
\\
The computation of the likelihood functions also relies on shear ratios (the full methodology is described in 
Ref. \cite{sanchez2022dark}).
Shear ratios provide additional constraining and validation power through the measurement of the galaxy-galaxy lensing signal of a lens galaxy redshift bin at small scales. They therefore reflect the ratio of mean lensing efficiencies of objects in those source bins with respect to the lens bin redshift. This, in turn, depends on the redshift distribution of the sources. Such a method is essentially independent from SOMPZ and clustering redshifts, because of its utilization of lensing signals. 

The four DES Y3 redshift bins are depicted in Fig. \ref{fig:nz_plot}. The mean redshift for each bin is chosen from Ref. \cite{myles2021dark}, with $\langle z_1 \rangle = 0.339$, $\langle z_2 \rangle = 0.528$, $\langle z_3 \rangle = 0.752$, and $\langle z_4 \rangle = 0.952$ \footnote{The DES Y3 component of the analysis uses updated tomographic binning compared to older Y3 analyses. This is implemented due to the changes observed in the $\Delta \chi^2_{\rm min}$ and cosmology results, as explained in  footnote 5 of Ref. \cite{mccullough2024dark}.}.

The purpose of this work primarily concerns the detection of the cross-correlation between the UGRB and the tangential shear, as opposed to robust constraints on cosmological parameters. Due to the low Signal-to-Noise Ratio (SNR) of the work, there is an uncertainty in the shear bias due to blending (as discussed in 
Ref. \cite{maccrann2022dark}) 
that is not propagated in this paper. The bias due to blending, however, is negligible here and will therefore not affect the interpretations arising from the analysis.   

\subsection{Gamma-ray data \label{sec:FERMIdata}} 

\begin{figure*}[!htbp]
  \includegraphics[width=0.49\textwidth]{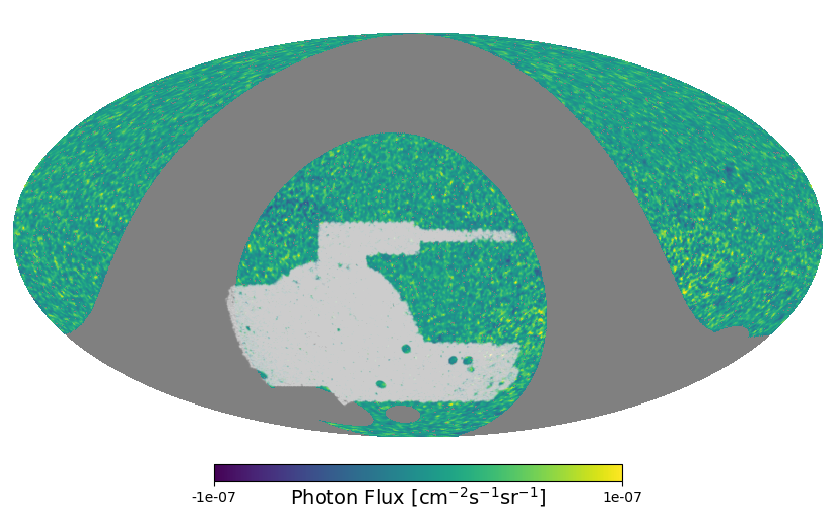}
  \includegraphics[width=0.49\textwidth]{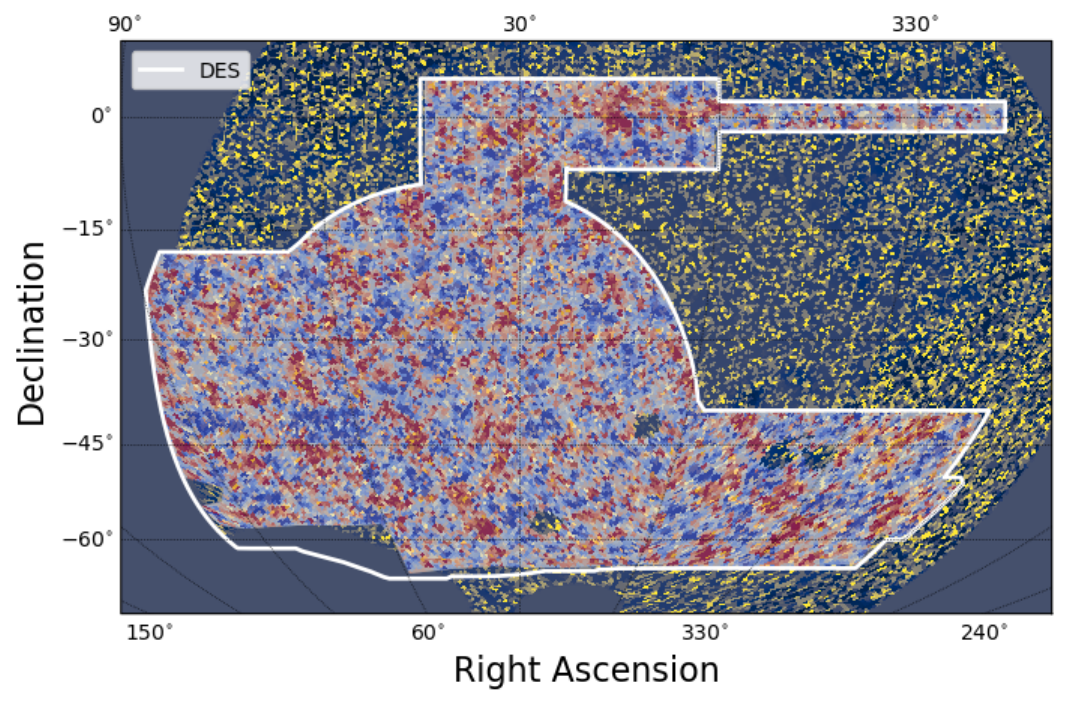}
  \caption{Left: The DES Y3 sky coverage (in light gray) superimposed on the \Fermi\ $\gamma$-ray intensity map (after Galactic foreground masking and subtraction) for photons in the (9.120-17.38) GeV energy range. The plot is in a Mollweide projection with equatorial coordinates, and has been downgraded to $\mathrm{N_{side}}=256$ and smoothed with a 
  Gaussian beam  of size $\sigma = 0.4^\circ$ for visualization purposes. The Galactic foreground has been masked and subtracted with the procedure described in Section \ref{sec:FERMIdata}. Right: The map zoomed in on the surveys' overlapping regions, showing the DES Y3 mass map \cite{jeffrey2021dark} along with the overlapping region in the \Fermi\ 12 year \g-ray background. The map refers to the same energy bin as the image on the left.} 
  \label{fig:maps}
\end{figure*}
\Fermi\ is a \g-ray pair-conversion telescope that has been in operation since 2008. With its broad energy range ($\sim$30 MeV to $>1 \rm TeV$) and effective rejection of charged cosmic-ray background, it is well-suited for studying the UGRB. The telescope scans the entire sky every three hours, achieving an angular resolution for gamma rays of about 0.1 degrees above 10 GeV.
The \g-ray data used in the cross-correlation measurement rely on a statistic of 12 years of observation of the \Fermi\ telescope, from August 4, 2008 to August 8, 2020 (Mean Elapsed Time (MET) = 239557417--618050000.0). 
The all-sky count and exposure maps are obtained with the Fermitools (v2.2.0)\footnote{\href{https://fermi.gsfc.nasa.gov/ssc/data/analysis/software/}{https://fermi.gsfc.nasa.gov/ssc/data/analysis/software/}}. We use Pass8-R3 processed data and select \textsc{sourceveto\_v2} event class\footnote{See 
\href{https://www.slac.stanford.edu/exp/glast/groups/canda/lat_Performance.htm}{http://www.slac.stanford.edu/exp/glast/groups/canda/lat\_ Performance.htm}.} and the combination of PSF1+2+3 event types, i.e., excluding the quartile of events with the worst angular reconstruction (labelled as PSF0) that would impact negatively on the measurement of the angular correlation at small scales. We have chosen \textsc{sourceveto\_v2} as it provides the best compromise in terms of acceptance and background rejection. In fact \textsc{sourceveto\_v2} has an acceptance comparable to \texttt{P8R2\_clean\_v6}, with a residual contamination almost equal to that of \texttt{P8R2\_ultracleanveto\_v6} at all energies\footnote{see \href{https://www.slac.stanford.edu/exp/glast/groups/canda/archive/pass8r3v2/lat\_Performance.htm}{https://www.slac.stanford.edu/exp/glast/groups/canda/archive/pass8r3v2/lat\_Performance.htm}}. This results in a clean photon dataset with sufficient event statistics, which is particularly critical when conducting cross-correlation analyses in order to ensure enough events when suppressing background is essential.
The model of the PSF as a function of the energy is obtained using the {\tt gtpsf} routine of the Fermitools, using the appropriate Instrument Response Function (IRFs) for the selected events\footnote{\href{https://fermi.gsfc.nasa.gov/ssc/data/analysis/documentation/Cicerone/Cicerone_LAT_IRFs/index.html}{https://fermi.gsfc.nasa.gov/ssc/data/analysis/documentation/Cicerone/Cicerone\_LAT\_IRFs/index.html}}. For each energy bin we determine an effective average PSF, by weighting the differential $\text{PSF}(E)$ by the UGRB spectrum \cite{Ackermann_2015}. We generate \g-ray maps in \texttt{HEALPix}\ \cite{gorski2005healpix} format with $N_{\rm side}=1024$. Such pixeling is optimal for this analysis as it is comparable to the smallest angular scales probed by the DES Y3 weak lensing analyses, and is better than the \Fermi\ PSF at any energy. As mentioned in Section~\ref{sec:models}, we account for the \Fermi\ PSF in our modelling.

To get the final flux maps, we follow the same approach as in Refs.~\cite{fornasa2016angular,2018PhRvL.121x1101A, DES:2019ucp}, which we summarize here for convenience. First, we produced counts maps in 100 micro, logarithmically spaced, energy bins between $100\,\mathrm{MeV}$ and $1\,\mathrm{TeV}$; then we divide the count maps of each micro bin by the corresponding average-exposure maps computed in the same micro bin, and corrected by the pixel area, to obtain flux maps in units of photons/cm$^2$/s/sr. The resulting flux maps are then summed up to produce maps for 9 macro energy bins between $631\,\mathrm{MeV}$ to $1\,\mathrm{TeV}$, as reported in Table \ref{tab:enbins}. We discard $E<0.6$ GeV because at low energy the angular resolution is too poor for our purposes (see following paragraphs).

To extract the UGRB component in the \g-ray maps, we exclude the majority of the Galactic foreground emission and the resolved point sources listed in the 4FGL-DR2 Catalog \cite{Fermi-LAT:2019yla}. While the former is not expected to contribute to the cross-correlation signal with the (extragalactic) gravitational lensing,  it nevertheless provides a noise term to our measurement. Therefore, in order to remove the majority of the bright emission from the Galactic plane, we apply a flat mask on latitudes $|b|<30\,\deg$. Furthermore,
to remove Galactic large-scale emission at higher latitudes, we perform a template fitting of the Galactic diffuse emission model and its subsequent  subtraction at the micro-energy-bin level, following the procedure adopted in Ref.~\cite{2018PhRvL.121x1101A} and utilizing the new Galactic emission model {\tt gll\_iem\_v07.fits}\footnote{\href{https://fermi.gsfc.nasa.gov/ssc/data/access/lat/BackgroundModels.html}{https://fermi.gsfc.nasa.gov/ssc/data/access/lat/BackgroundModels.html}}. 
Finally, in order to extract the unresolved emission we are interested in,
resolved point sources are removed from the maps following the masking approach described in Ref.~\cite{2018PhRvL.121x1101A}, which conservatively masks  a region around each catalogue source determined by taking into account both the source's brightness and the PSF in the specific energy bin. Each energy bin has therefore a unique mask, and the fractions of sky available at different energies can be found in Appendix \ref{sec:cov}.



In Fig.~\ref{fig:maps}, we show an example of the \Fermi\ \g-ray intensity map in equatorial coordinates in the (9.120-17.378) GeV energy bin with the application of the mask described above and illustrating the effect of the Galactic foreground subtraction. In the maps we also illustrate the overlapping and the mass distribution of the DES footprint (as detailed in Ref. \cite{jeffrey2021dark}).

\section{Analysis and results} 
\label{sec:res}

To measure the cross-correlation between the UGRB and gravitational shear, we compute the following estimator of the 2-point correlation function (see also Ref. \cite{gruen2018density}):
\begin{align}
&\Xi^{ar} (\theta) = \Xi^{\mathrm{signal}}_{\Delta \theta_h, \Delta E_a, \Delta z_r}-\Xi^{\mathrm{random}}_{\Delta \theta_h, \Delta E_a, \Delta z_r}
=  \frac{\sum_{i,j} \, e^r_{\mathrm{t},ij} \, I^a_j}{R \sum_{i,j} \, I^a_j }-\frac{\sum_{i,j}  \, e^r_{\mathrm{t},ij} \, I^a_{j, \rm random}}{R \sum_{i,j} I^a_{j, \rm random} } ,
\label{eq:crossshear}
\end{align}
where $ \Xi^{\mathrm{signal}}_{\Delta \theta_h, \Delta E_a, \Delta z_r}$ is the correlation function in configuration space of the two observables measured in different angular ($\Delta \theta_h$), \g-ray energy ($\Delta E_a$) and lensing source-galaxy redshift ($\Delta z_r$) bins. The correlation is obtained by summing the products of tangential ellipticity of source galaxies $i$ in redshift bin $r$ relative to a pixel $j$, $e^r_{ij,\mathrm{t}}$, multiplied by the \Fermi\ photon intensity flux in the $a$-th energy bin and in pixel $j$, $I^a_j$. The sum runs over all unmasked pixels $j$ and all sources $i$ in the DES shear catalogue, and it is performed in each of the different photon energy bins and source galaxies redshift bins. The mean response $R$ is determined as described above.

From the correlation function, we remove $ \Xi^{\mathrm{random}}_{\Delta \theta_h, \Delta E_a, \Delta z_r}$, the measurement of tangential shear around random lines of sight. This is done by setting $I^a_{j, \rm random}=1$ anywhere within the sky region used for \g-ray measurements in that energy bin and 0 elsewhere. This reduces additive shear systematic effects, random very-large-scale structures, or chance shear alignments relative to the mask. The random subtraction, while not affecting the expected signal, lowers the variance at large angular separations (see, e.g., Refs. \cite{singh2017galaxy,gruen2018density}). 
\begin{table*}
\centering
{\footnotesize
\begin{tabular}{llllllllll}
 \hline
 & \multicolumn{9}{c}{Bin number} \\
 \cline{2-10}
 & 1 & 2 & 3 & 4 & 5 & 6 & 7 & 8 & 9 \\
 \hline
 \hline
$E_{\rm min}$ [GeV] & $0.631$ & $1.202$ & $2.290$ & $4.786$ & $9.120$ & $17.38$ & $36.31$ & $69.18$ & $131.8$ \\
$E_{\rm max}$ [GeV] & $1.202$ & $2.290$ & $4.786$ & $9.120$ & $17.38$ & $36.31$ & $69.18$ & $131.8$ & $1000$ \\
$\theta_{\rm cont}$ 68\% [deg] & $1.00$ & $0.58$ & $0.36$ & $0.22$ & $0.15$ & $0.12$ & $0.11$ & $0.10$ & $0.10$ \\
Photon counts & $360865
$ & $780820$ & $551998
$ & $221181$ & $89897$ & $38277$ & $11990$ & $3757$ & $1619$ 
\\
$\langle I_a \rangle \mathrm{[10^{-7}cm^{-2}s^{-1}sr^{-1}
]}$ & $5.69$ & $2.18$ & $0.991$ & $0.375$ & $0.157$ & $0.0670$ & $0.0197$ & $0.0062$ & $0.0032$ 
\\
\hline
\end{tabular}
}
\caption{Gamma-ray energy bins over which the analysis is performed, 68\% containment angles $\theta_{\rm cont}$ of the {\it Fermi}-LAT PSF, and photon counts in the unmasked Fermi area in each energy bin along with the average measured intensity.}
\label{tab:enbins}
\end{table*}

Using the estimator shown in Eq.~\ref{eq:crossshear}, we perform the cross-correlation measurement in 12 logarithmically-spaced angular bins, with radii between $5$ and $600\,\mathrm{arcmin}$, 9 photon energy bins, detailed in Table~\ref{tab:enbins}, and the 4 redshift bins introduced in Sect. \ref{sec:DESdata}, for a total of 432 bins. The cross-correlations have been computed using the \texttt{TreeCorr} \cite{jarvis2004skewness,jarvis2015treecorr} package, with the MCMC scans performed using the affine-invariant ensemble sampler as configured in the Python package \texttt{emcee} \cite{foreman2013emcee}, and the contour plots obtained using   \texttt{ChainConsumer}\ \cite{hinton2016chainconsumer}. For the MCMC analysis, the total number of walkers considered are twice the number of the free parameters of the model. We consider a uniform prior for each parameter and a Gaussian likelihood.
For the spectral index $\mu_{\rm BLZ}$, the prior range is $[1.5,3.0]$, based on findings in Ref.~\cite{korsmeier2022flat}. We assume broad, uninformed priors for the remaining physical and phenomenological parameters. The convergence of the chains is assessed by computing the autocorrelation and the concomitant autocorrelation time as described in Ref. \cite{goodman2010ensemble}. 
\begin{figure*}[t]
    \centering
    \includegraphics[width=0.49\textwidth]{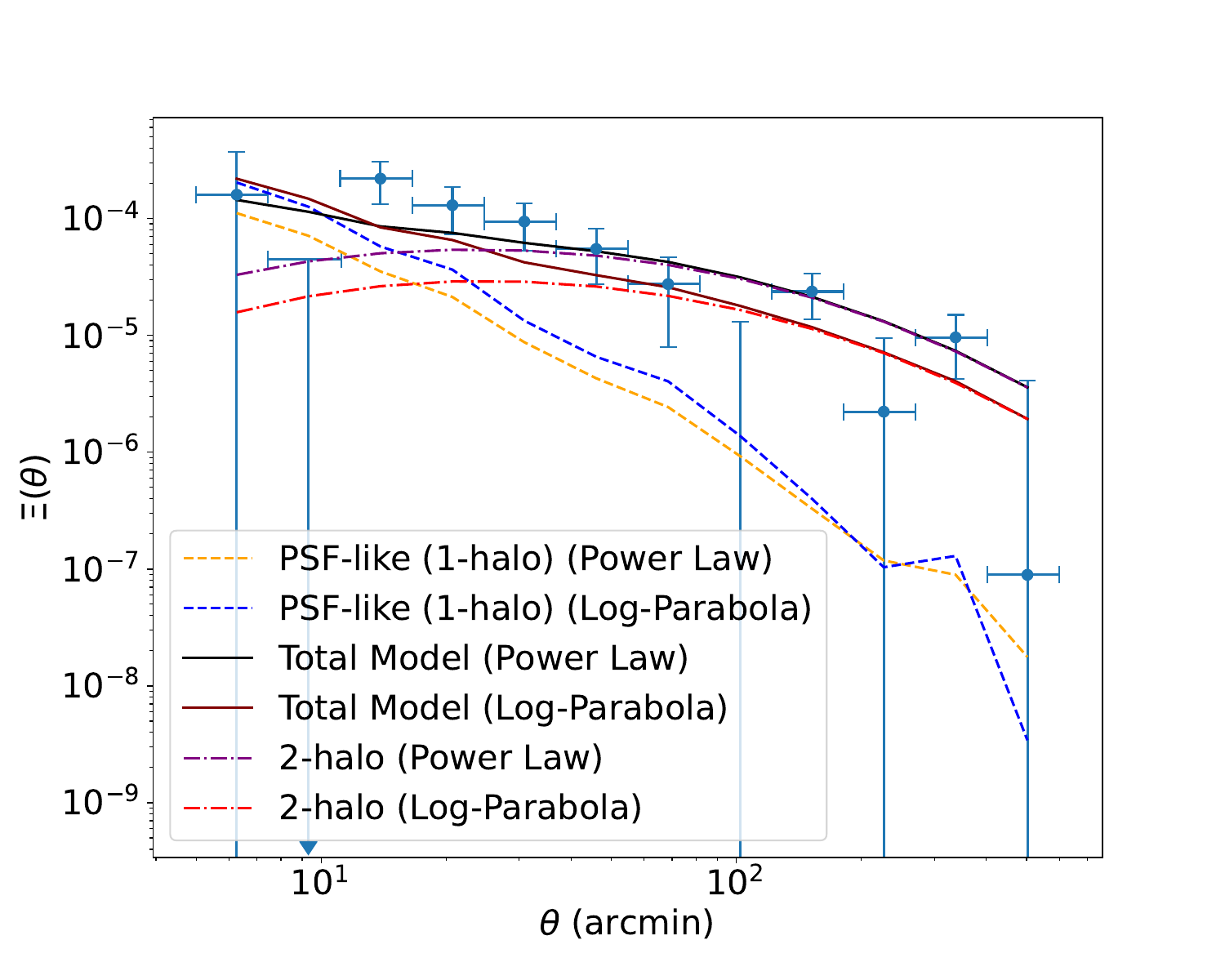}
    \includegraphics[width=0.49\textwidth]{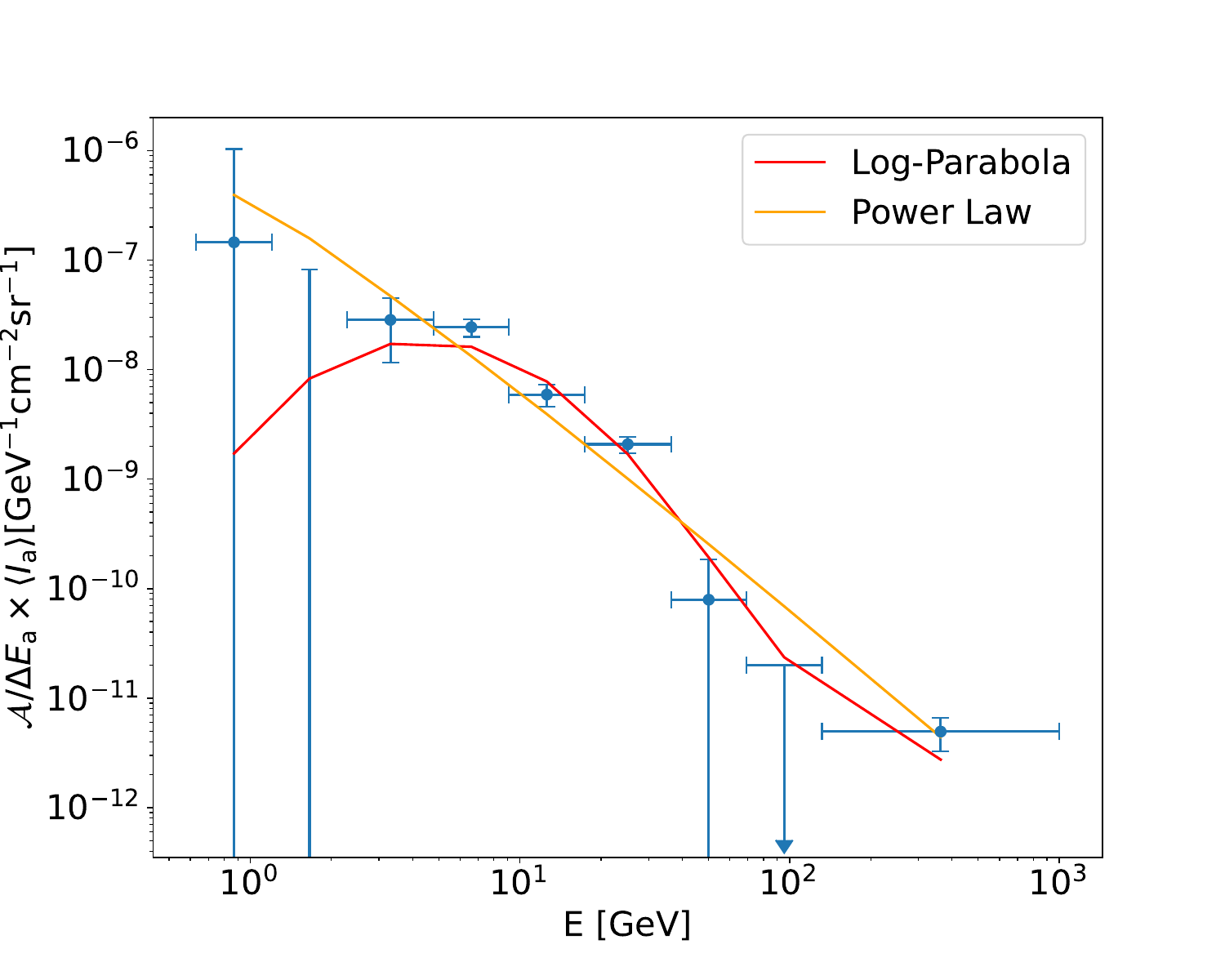}
    \includegraphics[width=0.49\textwidth]{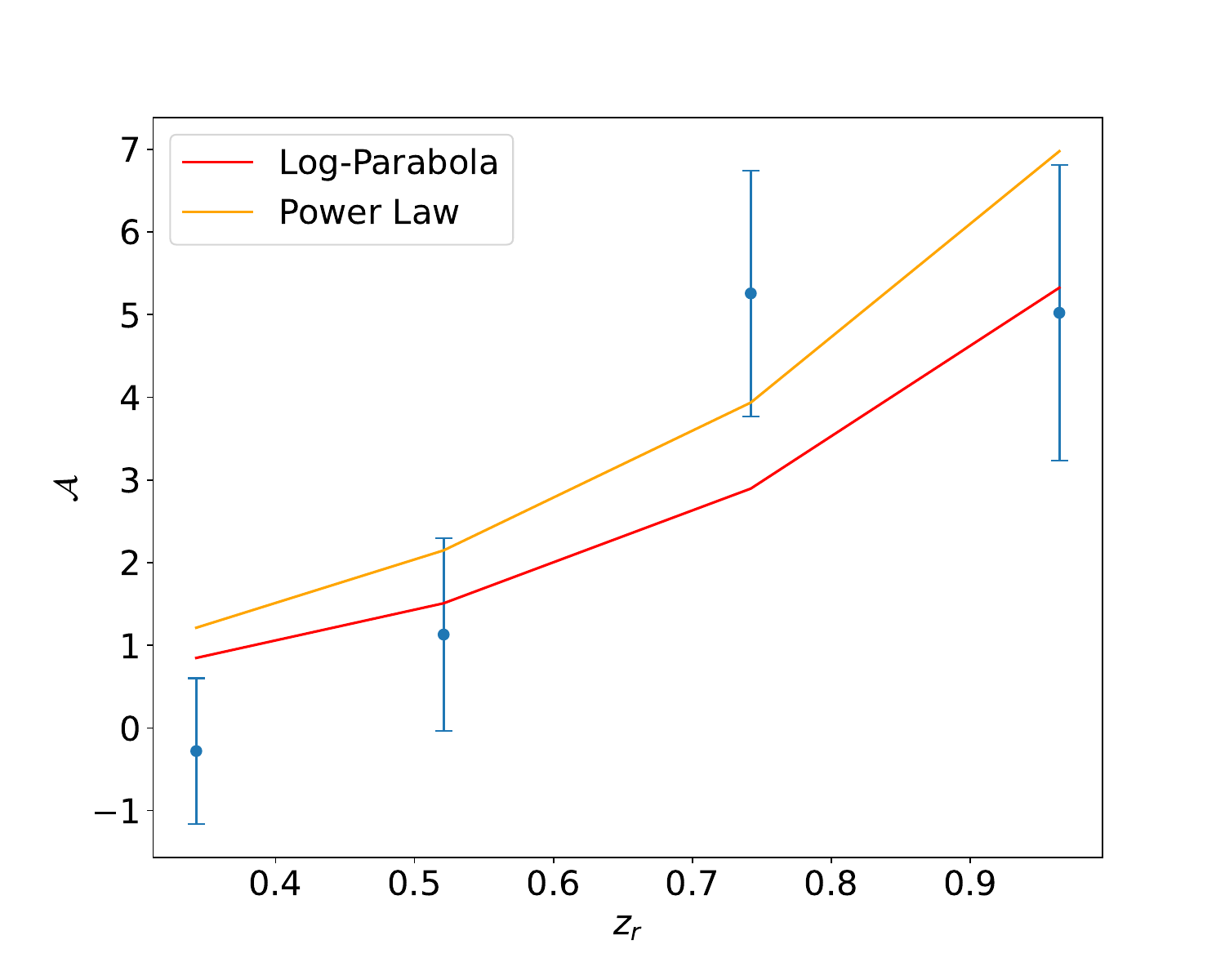}
    \caption{Measurement and models of the cross-correlation between \g-ray photons and gravitational tangential shear, for the log-parabola and power-law models, showing the angular behaviour (top left), along with the energy  (top right) and redshift (bottom) dependences. The second angular bin and the second highest energy bins consisted of negative data points when averaged across energy and redshift/angle and redshift. They have thus been displayed with their respective 2-sigma upper limits, followed by a downward arrow. It is important to note here that these are one-dimensional projections of a multi-dimensional fit, shown to provide a visual representation of the models used. See text for details on the derivation of the reported quantities.}
    \label{fig:integratedsignal}
\end{figure*}

\begin{figure*}[t]
    \centering
    \includegraphics[width=0.50\textwidth]{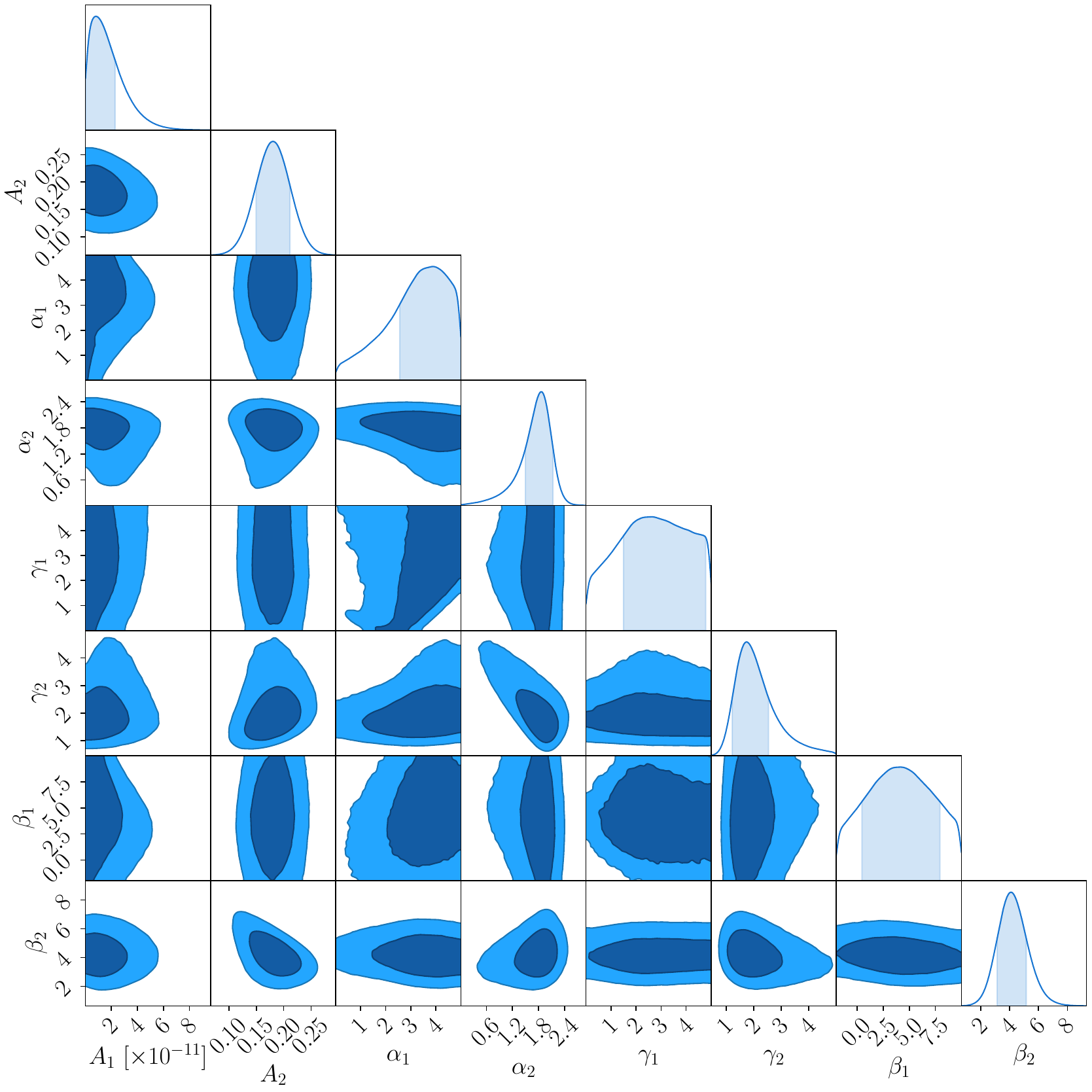}
    \includegraphics[width=0.49\textwidth]{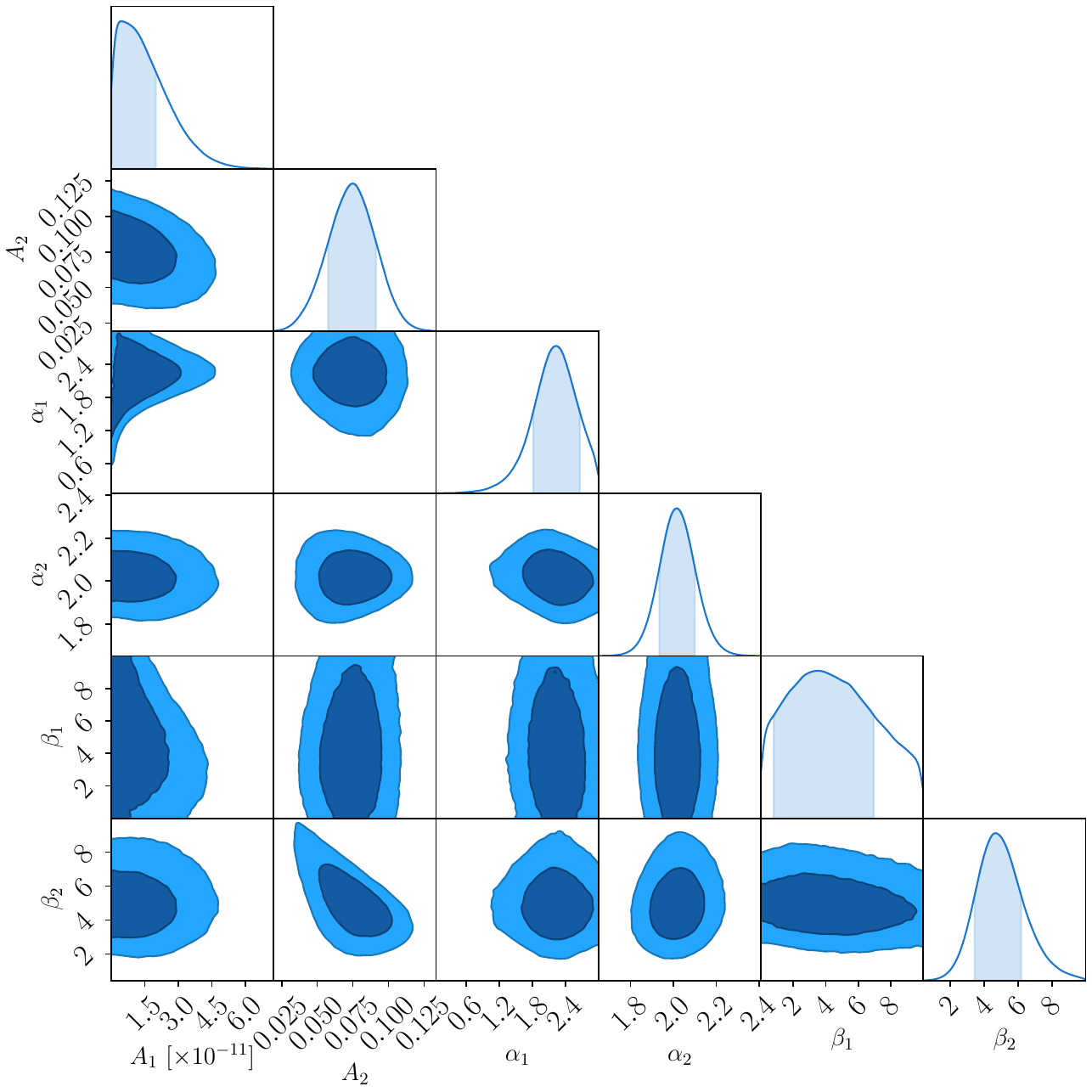}
    \caption{Left: Constraints on the parameters of the log-parabola \pheno\ model. Right: Constraints on the parameters of the power-law \pheno\ model. In both panels, the 2D contours refer to the 68\% and 95\% credible regions, with the shaded areas in the 1-D subplots denoting the 68\% credible interval for the associated posterior distributions.}
    \label{fig:contourplotspheno}
\end{figure*}

For a qualitative assessment of the measured signal, we plot the angular, energy and redshift dependence of the cross-correlation measurements in Fig. \ref{fig:integratedsignal}, along with their 68\% and 95\% confidence intervals in Fig. \ref{fig:contourplotspheno}. For the angular dependence, we simply average the cross-correlations and their concomitant phenomenological and physical model counterparts along energy and redshift. The energy and redshift dependencies, on the other hand, have been calculated using a matched filter amplitude ${\cal A}={\bm \Xi}^{\sf T}\,{\bm \Gamma}^{-1}\,\bar {\bm \Xi}_{\rm M}/(\bar {\bm \Xi}_{\rm M}^{\sf T}\,{\bm \Gamma}^{-1}\,\bar {\bm \Xi}_{\rm M})$\footnote{More information on the derivation of the matched filter SNR can be found on \url{https://github.com/BhashinT/Matched_Filter_FermixDES.git}
}, where $\Gamma$ is the covariance matrix (more information in Appendix \ref{sec:cov}), and $\bar {\bm \Xi}_{\rm M}$ is given by a simple model that we choose to be flat in energy and redshift, while scaling as $1/\theta$ in angle, to approximately reproduce the expected signal, and ${\bm \Xi}$ is the estimator defined in Eq.~(\ref{eq:crossshear}), or the best fit phenomenological and physical models. Note that with boldface notation we indicate the full vector of the correlation function, while so far we have indicated the individual components of the vector. The simple model is then given as $\Xi(\theta) = A_{\mathrm{s}}/\theta$, where $A_{\mathrm{s}}$ is the normalization amplitude of the simple model.  The error on ${\cal A}$ is given by $\sigma^2_{\cal A}=({\bm \Gamma}^{-1}\,\bar {\bm \Xi}_{\rm M})^{\sf T}\,{\bm \Gamma}\,({\bm \Gamma}^{-1}\,\bar {\bm \Xi}_{\rm M})/(\bar {\bm \Xi}_{\rm M}^{\sf T}\,{\bm \Gamma}^{-1}\,\bar {\bm \Xi}_{\rm M})^2$.
In order to calculate the evolution of the energy (redshift) bins, we take all the data vectors (i.e., as cross-correlation signals or as their best fit \pheno\ or \phys\ models) and their covariances in each redshift (energy) bin, and calculate the amplitudes and standard deviations for the selected data vectors and models. We can clearly see a positive detection with a differential signal scaling in energy with a spectral index around two and some curvature, and a mildly increasing redshift behaviour.

To be more quantitative, and in order to determine the statistical significance of the signal, we test the deviation of the measurement from a null signal (pure noise) by means of the \pheno\ model introduced above, and using two statistical methods.  

First, we perform a $\Delta\chi^2$ test statistics, with the chi-squared defined as: 
\begin{equation}
\left.\begin{aligned}
\chi^2({\bm P}_{\rm mod}) = & 
\left[{\bm \Xi}_{\rm data} -{\bm \Xi}_{\rm th}({\bm P}_{\rm mod})\right]^{\sf T}{\bm \Gamma}^{-1}\left[{\bm \Xi}_{\rm data} -{\bm \Xi}_{\rm th}({\bm P}_{\rm mod})\right], 
\end{aligned}\right.
\end{equation}
where ${\bm \Xi}_{\rm data}$ is the data vector, and ${\bm \Xi}_{\rm th}$ is the theoretical cross-correlation for the models outlined above, described by the parameter set ${\bm P}_{\rm mod}$.
The $\Delta\chi^2$ is defined as $\Delta\chi^2_{\rm mod}=\chi^2_{\rm null}- \chi^2({\bm P}_{\rm mod}^\star)$, with $\chi^2({\bm P}_{\rm mod}^\star)$ computed at the model parameter values ${\bm P}_{\rm mod}^\star$ that best fit the data, and $\chi^2_{\rm null}$ referring to no signal, i.e.\ ${\bm \Xi}_{\rm th}=0$. 
The best fits and confidence intervals of the parameters are found in an MCMC likelihood analysis.
The second estimator of the significance of the signal is the matched filter signal-to-noise ratio (see e.g. Ref. \cite{becker2016cosmic}),
\be
{\rm SNR}({\bm P}_{\rm mod})=\frac{{\bm \Xi}_{\rm data}^{\sf T}{\bm \Gamma}^{-1}{\bm \Xi}_{\rm th}({\bm P}_{\rm mod})}{\sqrt{{\bm \Xi}_{\rm th}^{\sf T}({\bm P}_{\rm mod}){\bm \Gamma}^{-1} {\bm \Xi}_{\rm th}({\bm P}_{\rm mod})}};
\ee
and we will evaluate ${\rm SNR}_{\rm mod}\equiv{\rm SNR}({\bm P}_{\rm mod}^\star)$.

In Table~\ref{tab:chi2comp} we present the results on detection significance. The \pheno\ model results for the full data show clear evidence for the presence of a cross-correlation signal at the level of $\mathrm{SNR}_{\rm mod}=8.9$ for the log-parabola model, and $\mathrm{SNR}_{\rm mod}=7.2$ for the power-law model.  
For both model types, we display in Table \ref{tab:bestfit_pheno} the best-fit values of all parameters, obtained from the maximum of the joint posterior distribution, and the 68\% credible intervals, obtained from the 1D marginalized posterior distribution. 
In order to investigate the features of the signal in more detail, we repeat the tests by subdividing the data set according to redshift, energy, and angular separation. Specifically, Low/High-$z$ refers to the first two and last two redshift bins; Low/High-$E$ bins are defined by being below/above $9\,\mathrm{GeV}$, i.e., first four/last five energy bins; and Small/Large-$\theta$ separates angular scales below/above 3 times the 68\% containment angle of the \Fermi\ PSF, reported in Table~\ref{tab:enbins}.
\begin{table}
\centering
\begin{tabular}{lc|ll|ll|lll}
\hline
 & \multicolumn{7}{c}{Data set} \\
 \cline{2-8}
 & Full & Low-$z$ & High-$z$ & Low-$E$ & High-$E$ & Small-$\theta$ & Large-$\theta$ & Model\\
\hline
\hline
$\Delta \chi^2_{\rm lp}$ & $78.9$ & $3.40$ & $75.3$ & $23.3$ & $55.4$ & $8.09$ & $71.26$ & Log-parabola\\
${\rm SNR}_{\rm lp}$ & $8.89$ & $2.45$ & $8.70$ & $4.89$ & $7.49$ & $2.86$ & $8.45$ & Log-parabola \\
\hline
$\Delta \chi^2_{\rm pl}$ & $51.4$ & $0.92$ & $53.7$ & $15.8$ & $38.7$ & $11.58$ & $44.10$ & Power-law \\
${\rm SNR}_{\rm pl}$ & $7.17$ & $1.66$ & $7.34$ & $4.03$ & $6.24$ & $3.40$ & $6.64$ & Power-law  \\
 \hline
$\Delta \chi^2_{\rm phys-BLZ}$ & $52.4$ & $3.19$ & $53.7$ & $15.6$ & $41.3$ & $10.69$ & $47.39$ & Physical\\
${\rm SNR}_{\rm phys-BLZ}$ & $7.23$ & $1.83$ & $7.32$ & $4.04$ & $6.47$ & $3.27$ & $6.88$ & Physical\\
 \hline
\end{tabular}
\caption{$\Delta \chi^2_{\rm mod}$ and ${\rm SNR}_{\rm mod}$ computed for the \pheno\ and \phys\ models, using either the full data set or the various subsamples discussed in the text. For the Low-$z$ case we selected the two first redshift bins ($z\lesssim 0.6$), while for the High-$z$ case the last two bins ($z\gtrsim 0.6$); the Low-$E$ subset refers to the first four energy bins, i.e., energies below 9 GeV, while the High-$E$ to the bins at higher energies; finally, the {Small-$\theta$/Large-$\theta$} cases correspond to data points below/above 3 times $68\%$ containment angle of the \Fermi\ PSF.}  
\label{tab:chi2comp}
\end{table}
From Table~\ref{tab:chi2comp} we infer that the signal is mostly concentrated at high energies, large angles, and high redshifts. 
Higher significance from higher redshift bins is somewhat expected (see also the physical model below), because those bins have a higher lensing signal by integrating over longer line-of-sight distances. The evidence for correlation at large angles suggests that the measurement is not dominated by few very massive and very bright objects, but rather it comes from a clustered population of extragalactic sources.
The low energy bins suffer from the poor \Fermi\ angular resolution which prevents to obtain a signal at large significance. The evidence at large energies points towards an interpretation in terms of sources with a hard spectrum, namely with a relatively large amount of photons at high energy.
While the obtained redshift and angular features are common to any \g-ray source population following large scale structures, the \g-ray spectrum might suggest a preference for blazars, rather than softer sources, such as SFGs and mAGNs. The best-fit for the spectral index of the 2-halo component for the log-parabolic model, $\alpha_2=1.94^{+0.19}_{-0.45}$ as well as the spectral index component for the power-law model $\alpha_2=2.01^{+0.08}_{-0.08}$ is quite hard with respect to the spectral index of the average intensity of the UGRB, which is $\alpha \approx 2.3$ \cite{Ackermann:2014usa}, but compatible with BL Lac emission, which is the source population expected to be the most relevant in the range of fluxes probed by this analysis, just below the \Fermi\ flux sensitivity threshold. 

The spectral scaling has a noticeable curvature, with the log-parabola model having a significantly higher SNR compared to its power-law counterpart. The log-parabola model is strongly favored, at a $\Delta \chi^2 \sim 27$. This could point towards multiple physical phenomena, that will be discussed in the next section.

\begin{table}[]
\centering
\begin{tabular}{|l|l|l|l|l|}
\hline
\textbf{Parameter} & \textbf{$\mathbf{68\%}$ C.I. (PL)} & \textbf{Best fit (PL)} & \textbf{$\mathbf{68\%}$ C.I. (LP)} & \textbf{Best fit (LP)} 
        \\ \hline
        \hline
$\rm A_1$ ($\times 10^{-12}$)   & $[0.2,19.6]$                      & $17.3$                         & $[0.35,22.84]$                  & $1.48$                            \\ \hline
$\rm A_2$                 & $[0.058,0.091]$               & $0.077$                    & $[0.15,0.21]$                  & $0.20$                       \\ \hline
$\alpha_1$             & $[1.82,2.67]$                  & $2.13$                      & $[2.53,4.98]$                     & $0.871$                         \\ \hline
$\alpha_2$             & $[1.93,2.09]$                  & $2.01$                      & $[1.49,2.13]$                     & $1.94$                      \\ \hline
$\gamma_1$             & N.A.                             & N.A.                                 & $[1.55,4.80]$                     & $0.073$                                   \\ \hline
$\gamma_2$             & N.A.                             & N.A.                                 & $[1.22,2.50]$                     & $1.61$                         \\ \hline
$\beta_1$              & $[0.76,6.91]$                  & $4.63$                      & $[0.40,7.91]$                     & $5.17$                         \\ \hline
$\beta_2$              & $[3.45,6.20]$                  & $4.83$                      & $[3.46,6.20]$                     & $4.83$                         \\ \hline
\end{tabular}
\caption{The $68\%$ credible interval (C.I.) and global best fit values of the parameters for the power-law and log-parabola \pheno\ models, denoted as PL and LP respectively. }
\label{tab:bestfit_pheno}
\end{table}
\section{Physical interpretation} 
\label{sec:phys}

Having clearly assessed the presence of the cross-correlation signal, in this section, we attempt to disentangle the  astrophysical \g-ray components of the cross-correlation signal. Different emitters can lead to different shapes of the cross-correlation signal as a function of the angular separation, energy, and/or redshift. In this work, we consider three astrophysical components, namely, BLZs, SFGs and mAGNs, and discuss also the possibility of a particle DM contribution. 

We anticipate that when confronting data with expectations from astrophysical populations, 
we find that BLZ are preferred by the fit, with negligible contributions from SFG and mAGNs, as found also in \cite{DES:2019ucp}. This is essentially due to the harder \g-ray spectrum of the signal, compatible with BLZ but not with SFG and mAGNs, as already discussed above.
We therefore start with the most economical choice by including only BLZ in our \phys\ model, and come back 
to SFGs, mAGNs, and DM towards the end of this section and in the Appendix.

\begin{figure*}[t]
    \centering
    \includegraphics[width=0.45\textwidth]{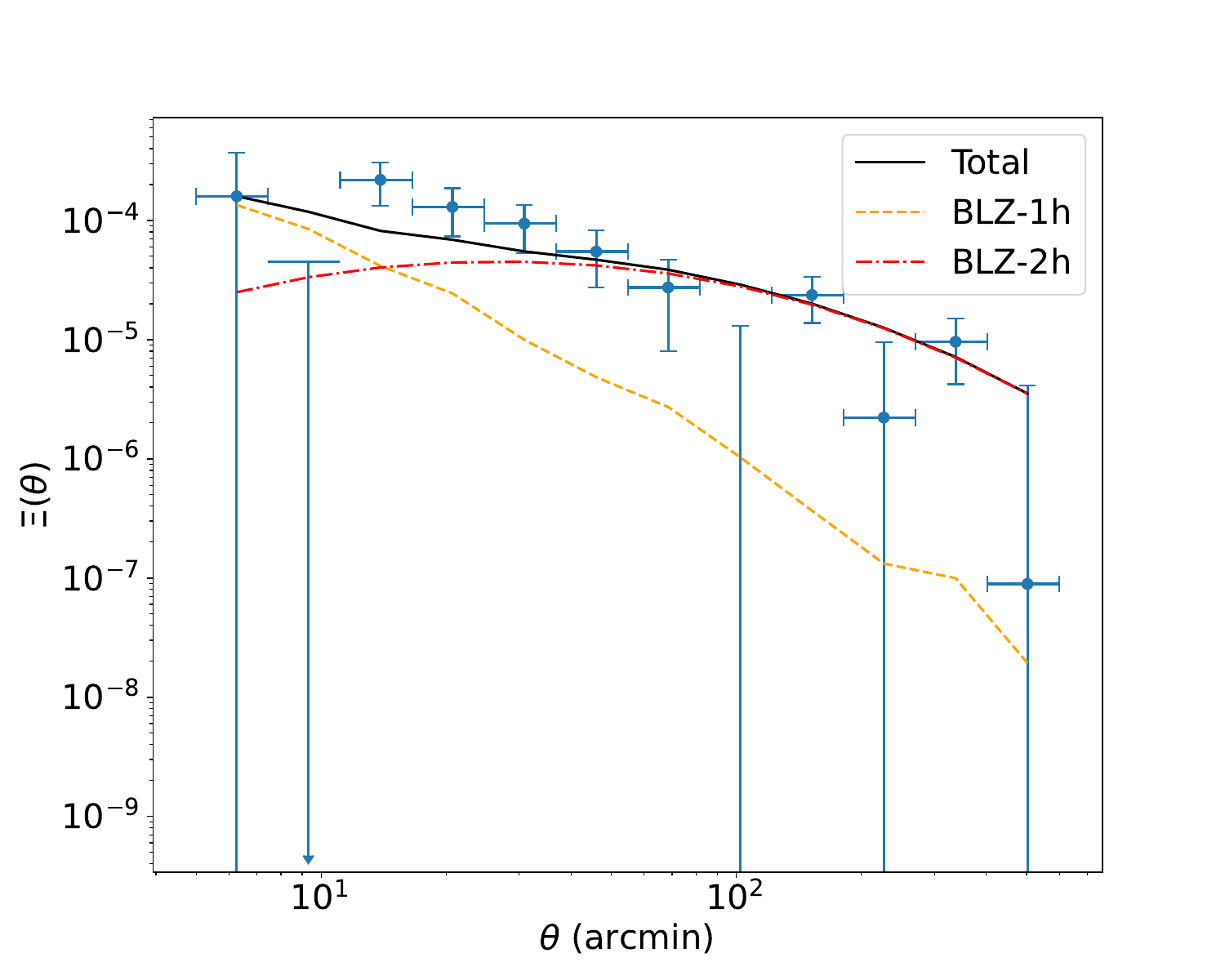}
    \includegraphics[width=0.45\textwidth]{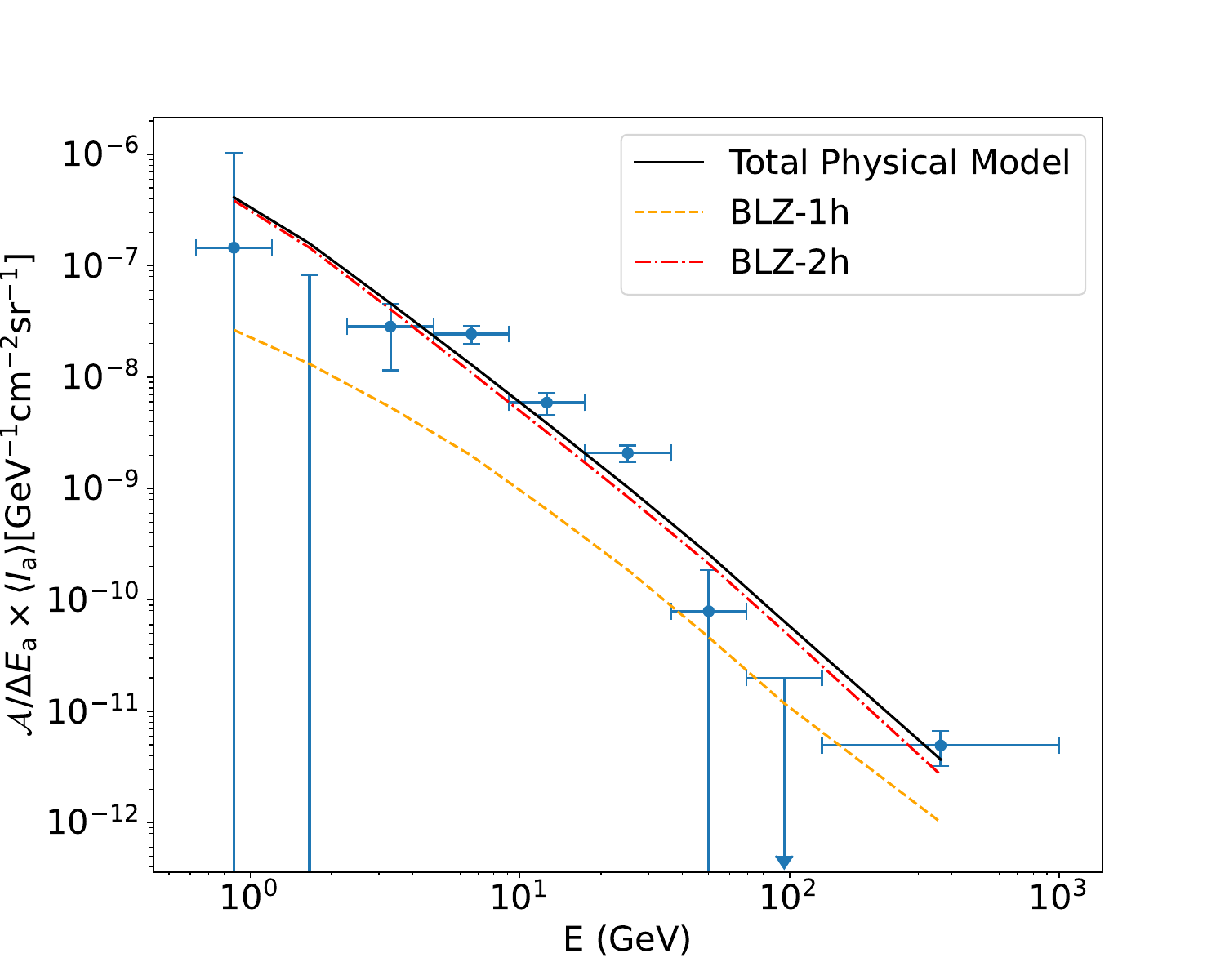}
    \includegraphics[width=0.45\textwidth]{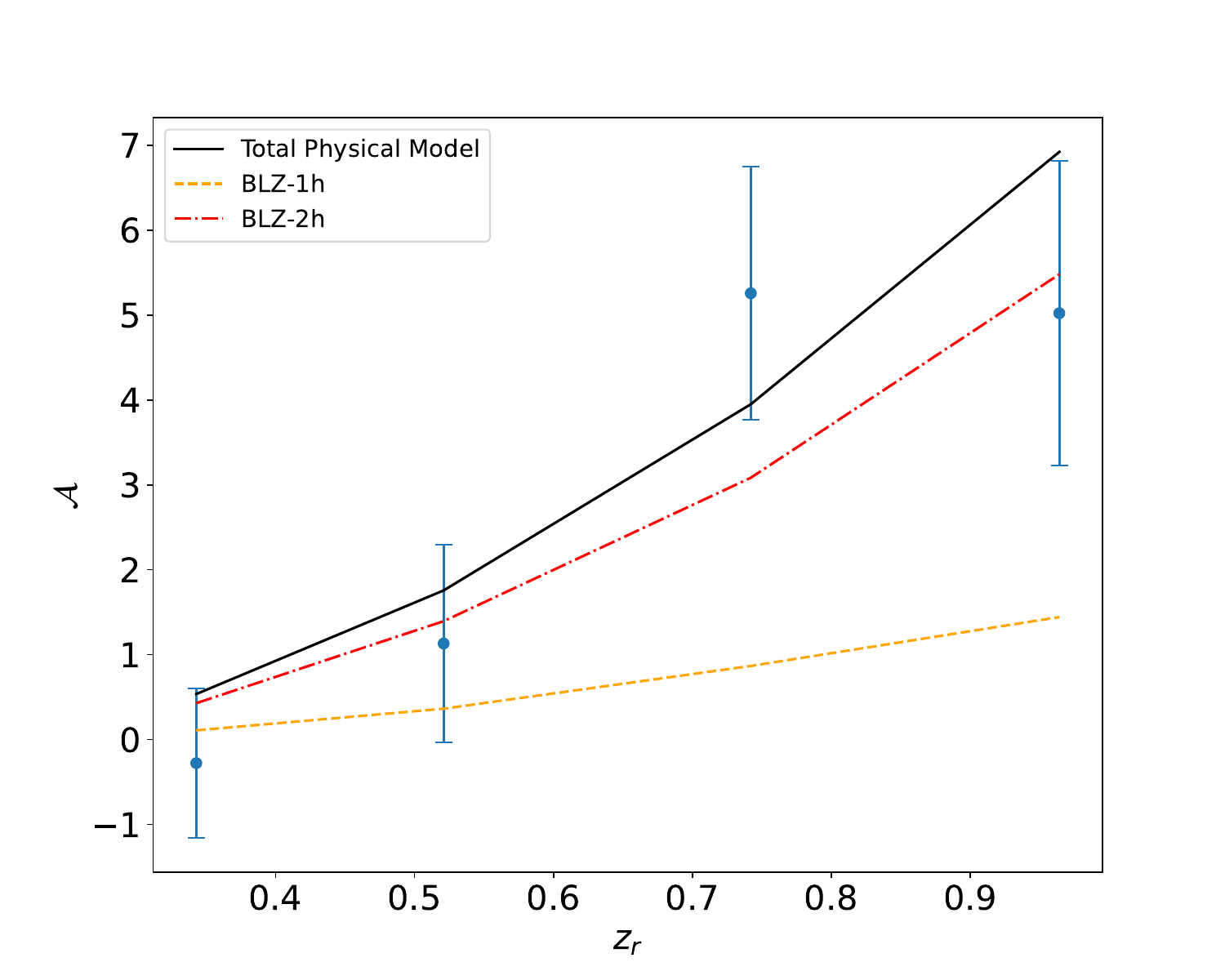}
    \caption{Similar to Fig.~\ref{fig:integratedsignal} but for the \phys\ model with BLZ-only contribution. We display the angular (top left), energy (top right), and redshift dependence (bottom) of the cross-correlation.}
    \label{fig:physmodelblz}
\end{figure*}

For the purpose of this work, blazars can be considered as point-like sources---i.e.\ their size is much smaller than the \Fermi\ PSF. Additionally, the size of the halo hosting blazars rarely exceeds the \Fermi\ PSF. As a consequence, the angular correlation function for the 1-halo term essentially follows from the detector PSF. However, as seen from Table \ref{tab:chi2comp} and Fig. \ref{fig:physmodelblz}, a cross-correlation on larger angular scales is required by the fit, and the statistical significance of the signal is actually driven by the 2-halo component. 
This means that we are able to probe the clustering of blazars, i.e., their large scale distribution.
In order to investigate the BLZ properties needed to account for the measured cross-correlation signal, we perform the statistical tests discussed in the previous section, but now with a \phys\ model, based on a  characterisation of the 1- and 2-halo cross-correlations of weak lensing with \g-rays from blazars, as described in Eq.~\ref{eq:physmdl}. 
 In Ref.~\cite{korsmeier2022flat} it was found that two BLZ populations are needed to explain the angular power spectrum (APS) data, with FSRQs dominating at GeV energies and BL Lacs taking over from a few GeV. As discussed in the previous section, we have inconclusive results and little constraining power in the first two energy bins. Therefore we focus our modeling on BL Lacs\footnote{The UGRB also contains unresolved Gamma-Ray Burst (GRB) components; however, it is highly unlikely that they are relevant in the energy bins below 100 GeV, i.e, the contributions from GRBs can only notably come from the final energy bin, and occupy a very small fraction of the UGRB (see e.g.  Refs. \cite{guetta2023tev,min2024contribution,yao2020contribution}). Keeping in mind that we do not expect significant UGRB contribution from GRB in the energy range we are looking at with integrated time of observation of 16 years, we can reliably conclude that the current GRB models cannot explain the spectral curvature that we observe in our work, and therefore have not been considered.}. 

Note that we allow the 1-halo and the 2-halo terms to be separately adjusted in the fit against the data. 
The two normalization parameters are a simple effective way to account for a different luminosity dependence of the GLF (leading to a different amplitude of the signal) and for a different function for the host-halo mass versus \g-ray luminosity (which impacts differently the linear and non-linear bias, thus altering the ratio between 1- and 2-halo terms).

The results are shown in Table~\ref{tab:chi2comp} (along with the best fit parameter values shown in Table~\ref{tab:bestfit_phys} in a manner similar to that of Table~\ref{tab:bestfit_pheno}), where the overall significance of the presence of a signal, and its scalings as a function of energy, angular scales and redshift are all confirmed. Actually, the BLZ case provides results that are very similar to the \pheno\ power-law model. 
\begin{figure*}[t]
    \centering
    \includegraphics[width=0.45\textwidth]{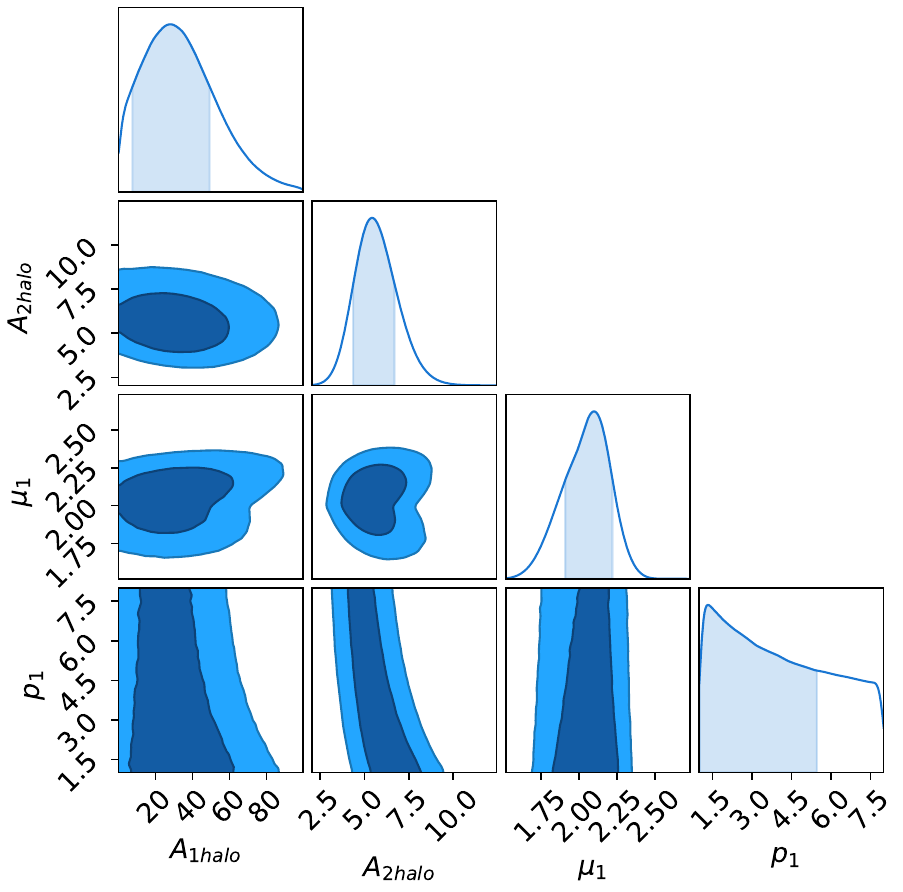}
   \caption{Constraints on the parameters describing the physical model BLZ, based on the BL Lacs model in \cite{korsmeier2022flat}, and assuming a single population accounting for the measured cross-correlation. As before, the 2D contours refer to the 68\% and 95\% credible regions, with the shaded areas in the 1D subplots denoting the 68\% credible interval for the 1D posterior distribution.}
    \label{fig:contourplotsphysmodelBLZ}
\end{figure*}

More details of the parameter constraints are shown in Fig.~\ref{fig:contourplotsphysmodelBLZ}, where the triangular plot of the posterior distributions of the model parameters are reported. The posterior exhibits a preference for a large 2-halo term of blazars with normalisation $A_{\rm BLZ}^{\rm 2h}=6.59^{+0.11}_{-2.23}$, while the normalisation of the blazar 1-halo term $A_{\rm BLZ}^{\rm 1h}=34.17^{+26.66}_{-15.05}$ shows somewhat weaker constraints. The evidence for a 2-halo term is much greater than what was observed 
in Ref. \cite{DES:2019ucp}. 
This nicely meets expectations since the improvement of the DES Y3 dataset with respect to DES Y1 is the larger portion of the sky covered.
Note that $A_{\rm BLZ}^{\rm 2h}$ can be seen as the amplification of the signal due to a larger linear bias and/or GLF with respect to our reference choice. With both being $\mathcal{O}(1)$, one can account for our findings. Given the expected size of systematic uncertainties (see discussion below on $M(\mathcal{L})$ and the bias), we consider it as a good indication that our reference model is able to describe the physical picture.

The spectral index $\mu_{\rm BLZ} = 2.07_{-0.16}^{+0.15}$ is consistent with BLZ emissions, as already mentioned when discussing the power-law \pheno\ model from Sec. \ref{sec:res}. 
The redshift behaviour is also compatible with the BLZ model, see Fig.~\ref{fig:physmodelblz}. The larger significance of high-$z$ bins with respect to Ref. \cite{DES:2019ucp} might be again attributed to the enhanced sensitivity to large scale structures of the current dataset, which means that the signal is no longer dominated by a few bright (and so typically closer) sources. 
We tested this conclusion by measuring the correlation of \Fermi\ maps with the 4FGL sources included, i.e., without masking the extragalactic sources. In this case, due to the presence of bright, low-redshift sources, we observe a much more dramatic increase of the signal in the low-$z$ bin as compared to the higher-$z$ bins, which confirms the aforementioned explanation of the redshift behaviour.
\begin{table}[]
\centering
\begin{tabular}{|l|l|l|}
\hline
\textbf{Parameter} & \textbf{$\mathbf{68\%}$ C.I.} & \textbf{Best-fit} \\ \hline
\hline
$A_{1\rm halo}$                & $[7.51,49.22]$   & $34.17$       \\ \hline
$A_{2\rm halo}$                  & $[4.36,6.70]$      & $6.59$          \\ \hline
$\mu_{\rm BLZ}$                & $[1.91,2.22]$       & $2.07$          \\ \hline
$p_1$                 & $[1.02,5.46]$         & $1.02$                    \\ \hline
\end{tabular}
\caption{The $68\%$ credible interval and global best fit values of the \phys\ model parameters of Eq. (\ref{eq:physmdl}). }
\label{tab:bestfit_phys}
\end{table}
There are only two properties that do not exactly align with our expectations of blazar behaviour.
The first has been already emphasized in the previous section, namely, that a curved \g-ray spectrum is preferred over a power-law (note that the absorbed blazar spectrum has a curvature, which however occurs at high energies, see discussion below). Second, the amplitude of the normalization parameters is surprisingly large, especially for the 2-halo case. To be compatible with the findings in Ref. \cite{korsmeier2022flat}, concerning, e.g., the total contribution of BLZ to the UGRB intensity, we would need $A_{\rm BLZ}^{\rm 2h}\lesssim 2$, which is about $3\sigma$ away from our results. 
Let us now try to address these two points.

A different \g-ray population might have a curved spectrum.
The emission from SFGs has a significant contribution from $\pi_0$ decay, leading to a log-parabolic shape for the \g-ray spectrum, see, e.g., Ref. \cite{roth2021diffuse}. 
On the other hand, the peak occurs below a few GeV, whilst, to fit our data, one needs a peak around 10 GeV, see Fig.~\ref{fig:integratedsignal}.
Therefore, the attempt to provide an explanation using SFGs as the log-parabolic component remaines unsuccessful. More details are provided in Appendix \ref{sec:SFG}. 

A more appropriate \g-ray spectrum can be provided by particle DM in terms of WIMPs. Indeed, by choosing the appropriate DM mass and annihilation channel, a significant preference over the BLZ-only model can be obtained.
We discuss this case in Appendix \ref{sec:DM}.

A less exotic explanation is to fine-tune the BLZ model. One could obtain a log-parabolic spectrum as a result of the combination of different BLZ sub-populations or by directly implementing a curved behavior. On the other hand, it is then not trivial to satisfy bounds from other probes, like the 1-point statistics or the auto-correlation, see, e.g., \cite{korsmeier2022flat}. A comprehensive test of this hypothesis is beyond the scope of this work. 

As a side note, we also emphasize that we detect the presence of a signal at energies above 100 GeV\footnote{To consider the energy dependence in the plots as the physical \g-ray spectrum is a bit misleading. Since at high energies \Fermi\ is less sensitive to source detection, the masking becomes less effective (and in order to alleviate for it we added the 3FHL catalog \cite{TheFermi-LAT:2017pvy}, which is specifically devised to the detection of high-energy sources).}.
This might suggest the presence of a very hard source population on top of the main BLZ contribution (although, currently, there is no statistically significant preference).
For example, the magnetohydrodynamical simulations performed in \cite{hussain2023diffuse} suggest that the majority of the UGRB at these energies originate from cosmic-rays in galaxy clusters, having a very hard spectral index. 

Finally, instead of a curvature in the original \g-ray spectrum, a stronger absorption in the energy range between $\sim 10$ and 50 GeV can generate the required behaviour. Gamma-ray absorption is due to pair production from the interaction of \g-rays with optical and infrared light. To describe absorption, one has thus to estimate the extragalactic background light (EBL). This is not a trivial task, and requires to account for important systematic effects, e.g., the contamination from large zodiacal light associated with interplanetary dust in the Solar system (see e.g. Ref. \cite{cooray2016extragalactic} for more information on EBL measurements).
The curvature seen in the \g-ray spectrum of our correlation could be associated to a higher level of EBL at ultra-violet frequencies (that are the most relevant ones to generate pair production when interacting with 10-50 GeV photons) with respect to the reference case we are using \cite{Finke2010,Finke:2022uvv}. To this end, we implement the optical depth obtained from the EBL evaluation in Ref.~\cite{Stecker:2016fsg}. The latter differs by a factor of a few from the EBL in Ref.~\cite{Finke2010,Finke:2022uvv} in the far-UV. 

\begin{figure*}[t]
    \centering
    \includegraphics[width=0.45\textwidth]{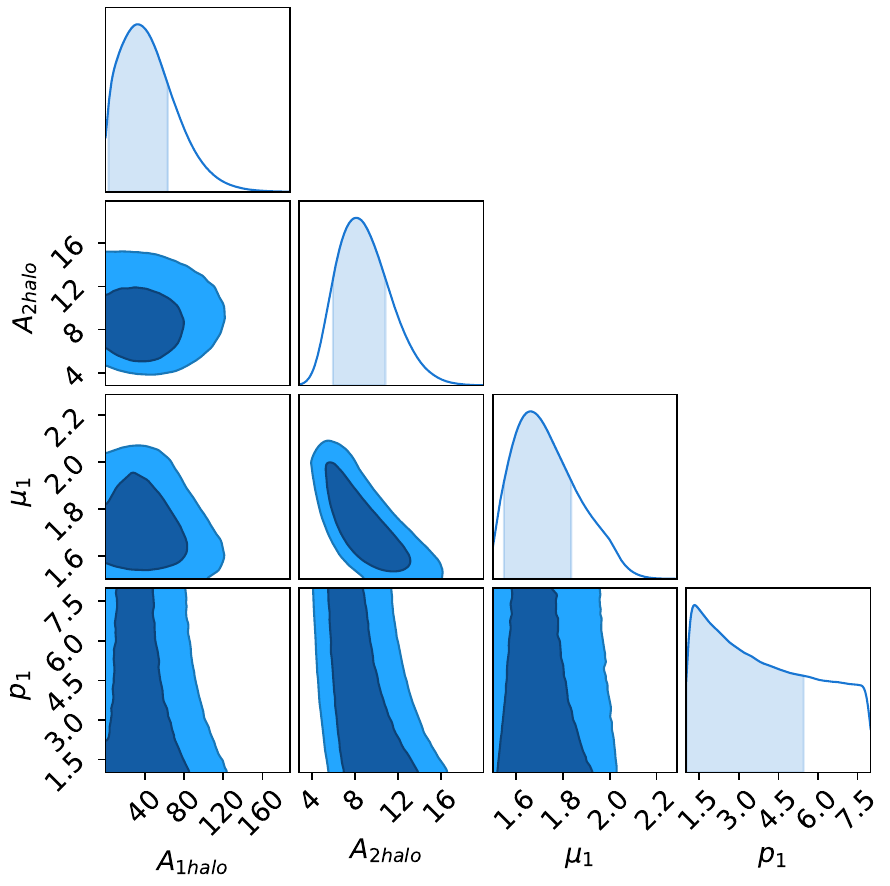}
    \includegraphics[width=0.50\textwidth]{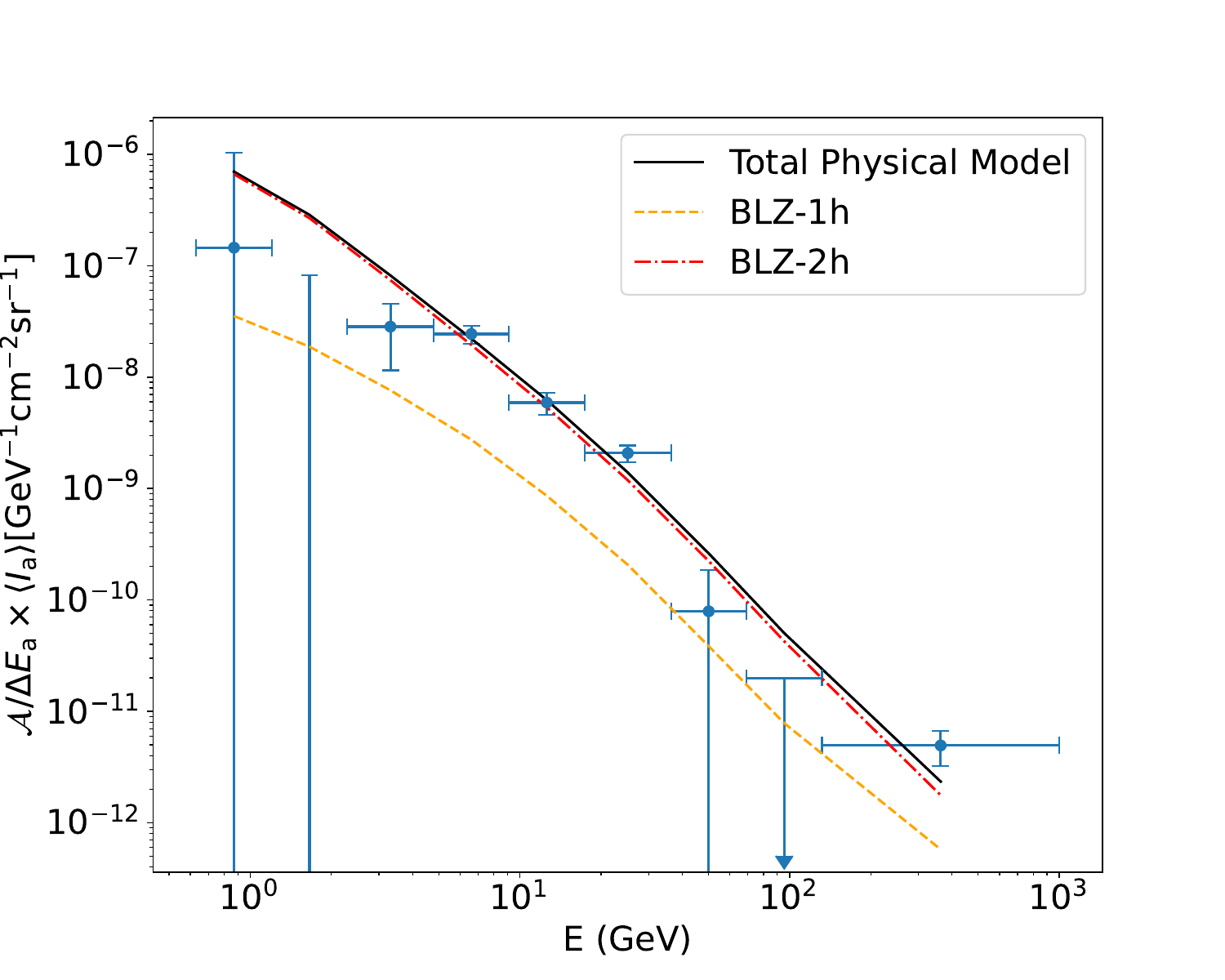}
    \includegraphics[width=0.50\textwidth]{Y3x12y-Phys_Model_Figures/BLZ_only_contours.pdf}
    \caption{Left: Constraints on the parameters of the BLZ model, as in Fig.~\ref{fig:contourplotsphysmodelBLZ}, but replacing the absorption optical depth of \cite{Finke2010} with the one of \cite{Stecker:2016fsg}. The contour plot shown in Fig. \ref{fig:contourplotsphysmodelBLZ} is also displayed below for comparison. Right: The energy behaviour of the 1- and 2-halo components, to show  that the alternative absorption model can provide a slightly more curved spectrum in the 10-50 GeV range, compared to the case of Fig.~\ref{fig:physmodelblz}.}
    \label{fig:contourplotsblzstecker}
\end{figure*}

Results are shown in Fig.~\ref{fig:contourplotsblzstecker}. The spectrum is indeed slightly more curved than the case in Fig.~\ref{fig:physmodelblz}. The SNR grows to 7.8 with a $\Delta\chi^2\simeq 8$ with respect to the reference absorption model (compared to the previous SNR of 7.23). Therefore, a larger amplitude of the EBL in the UV can help in recovering the curvature we measured, as expected, even though additional modifications seem to be needed.


We now discuss the possible origin for the large value of $A_{\rm BLZ}^{\rm 2h}$.
The blazar-shear cross-correlation depends on the relation between the blazar \g-ray luminosity and the host-halo mass, a quantity which is rather uncertain. We followed Ref. \cite{camera2015tomographic}, where it was derived by associating the \g-ray luminosity of blazars to the mass of the supermassive black hole powering the AGN and then relating the black hole mass to the mass of the DM halo: $M(\mathcal{L})=2\times 10^{13}M_\odot\left[\mathcal{L}/(10^{47}\,\mathrm{erg\,s^{-1}})\right]^{0.23}(1+z)^{-0.9}$, where $\mathcal{L}$ is the rest-frame luminosity of blazars in the energy range $0.1$ to $100\,\mathrm{GeV}$. 
By considering larger values for the DM halo mass, one can obtain both larger 1-halo term and larger linear bias (i.e., larger 2-halo term), therefore reducing the amplitude needed for $A_{\rm BLZ}^{\rm 1h}$ and $A_{\rm BLZ}^{\rm 2h}$.

\begin{figure*}[t]
    \centering
    \includegraphics[width=0.45\textwidth]{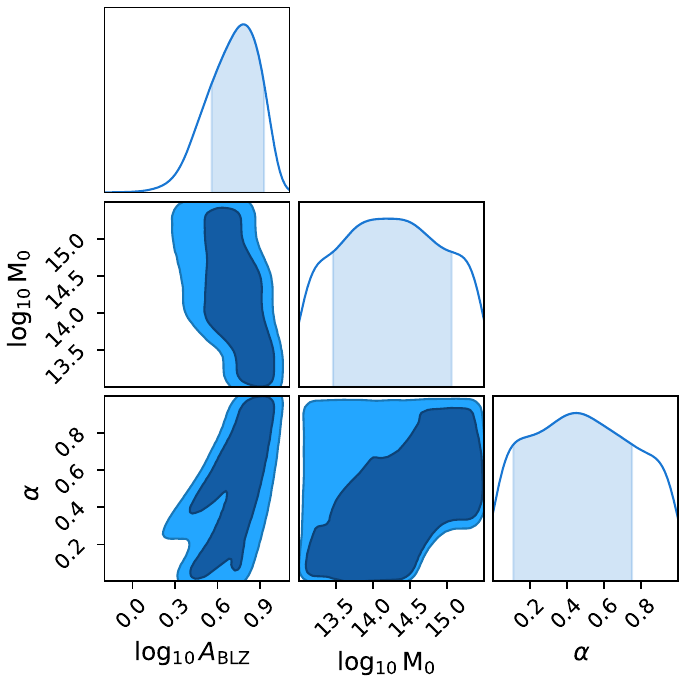}
   \caption{Constraints on the parameters describing the relation between the host halo mass and blazar luminosity, $M(\mathcal{L}) = M_0(L_{\mathrm{BLZ}}/10^{47} \mathrm{erg/s})^{\alpha}(1+z)^{0.9}$.
    The amplitude $A_{\rm BLZ}$ is in log-scale (in contrast to the linear scale in the previous plots) to illustrate better its relation with $M_0$.}
    \label{fig:Mhaloalpha}
\end{figure*}

In order to test this scenario we allow for a more generic $M(\mathcal{L})$ relation: $M(\mathcal{L}) = M_0(L_{\mathrm{BLZ}}/10^{47} \mathrm{erg/s})^{\alpha}(1+z)^{-0.9}$, with $M_0$ and $\alpha$ being free parameters.
For this test, we set the energy and redshift parameters of the BLZ model in Eq.~(\ref{eq:physmdl}) to the best-fit values found from the above analysis, and set a common normalization to both the 1- and 2-halo terms, $A_{\rm BLZ}^{\rm 1h}=A_{\rm BLZ}^{\rm 2h}=A_{\rm BLZ}$. 
Along with the best fit value of $\alpha = 0.44^{+0.30}_{-0.34}$, we find the best-fit halo mass parameter value to be $M_0 = 10^{14.1^{+0.98}_{-0.61}} M_\odot$, suggesting that the average mass of the halo hosting the unresolved blazars we are probing with this analysis is larger than the average one derived in Ref. \cite{camera2015tomographic}, tailored to galaxy-size halos. Our result agrees with results in Ref.~\cite{DES:2019ucp} where $M_0 > 10^{14} M_\odot$ was found.

We can note from Fig.~\ref{fig:Mhaloalpha} that if
the average halo mass is above $10^{14}M_\odot$, then a value of $A_{\rm BLZ}\simeq 2$ is compatible with data. We have also found this to be true for the alternative BLZ model that we considered, with the model having a best-fit halo mass $M_0 = 10^{14.08^{+0.34}_{-0.81}} M_\odot$. 
Following Ref.~\cite{korsmeier2022flat}, one can see that $A_{\rm BLZ}\lesssim 2$ is required in order to be consistent with number counts and auto-correlation analyses and makes the BLZ model providing about 30-40\% of the UGRB above 10 GeV.
Therefore, if the BLZ-only model is the correct interpretation of our measurement, the weak lensing signal has to be provided by cluster-size halos (i.e., with mass $\gtrsim 10^{14}\,M_\odot$) where the \g-ray blazars should reside. 

The angular correlation between clusters of galaxies and the UGRB has been detected in Refs.~\cite{Branchini:2016glc,Colavincenzo:2019jtj} and the origin of the signal is compatible with \g-ray emission from blazars hosted by the clusters. Relatively massive hosts are also found for the blazars above detection threshold, i.e., the ones in the \Fermi\ catalogue \cite{Allevato:2014qga}.

\section{Conclusion} 
\label{sec:conc}
In the present work, we measured and interpreted the angular correlation between the \g-ray sky and the matter distribution in the Universe. We employed 12 years of observation of the \Fermi\ telescope and 3 years of data from DES.
Building upon the results obtained with \Fermi\- 9yr and DES Y1 in Ref.~\cite{DES:2019ucp}, where a signal of cross-correlation between \g-ray sources and gravitational lensing shear was found with an SNR of 5.3, we improved the statistical significance and found an SNR of 8.9.

Thanks to the larger portion of the sky available from DES Y3, we detect for the first time a large-scale correlation, i.e., a proof that a significant fraction of the UGRB follows the mass clustering of the Universe traced by weak lensing.

Blazars are a sufficiently good explanation of the signal, provided that the ones responsible for the measured correlation reside in halos of large mass ($\sim 10^{14}M_\odot$),
and that BLZ account for about 30-40\% of the UGRB above 10 GeV.

We found a significant preference for a curved \g-ray spectrum, with the log-parabolic shape preferred over the power-law at $\Delta \chi^2 \sim 27$.
This can have different physical interpretations, including a modification of the BLZ model with respect to the current state of the art, additional \g-ray sources, like SFG and DM, discussed in Appendices~\ref{sec:SFG} and \ref{sec:DM}, or considering different UV EBL models to the default model implemented in this text.

In order to ascertain the properties of the \g-ray source population providing the reported signal, we plan to perform an analysis adding the cross-correlation of the \g-ray sky with the DES Y3 galaxy clustering. 
For this, we expect an even larger statistical significance with respect to the one reported here for the case of the lensing shear. Thus we should be able to work with sub-samples of galaxies that can emphasize the correlation with a given population, e.g., luminous red galaxies are more suited to explore the DM interpretation since they are not supposed to be SFGs or to host blazars, and to better characterize the redshift dependence of the signal to assess the impact of absorption. 
\\

\newpage

\section{Acknowledgments}

BT, MR, and NF acknowledge support from the  Research grant TAsP (Theoretical Astroparticle Physics) funded by \textsc{infn}. The work of MR and NF is supported by the Italian Ministry of University and Research (MUR) via the PRIN 2022 Project No. 20228WHTYC – CUP: D53C24003550006.
SC acknowledges support from the Italian Ministry of University and Research (\textsc{mur}), PRIN 2022 `EXSKALIBUR – Euclid-Cross-SKA: Likelihood Inference Building for Universe's Research', Grant No.\ 20222BBYB9, CUP D53D2300252 0006, from the Italian Ministry of Foreign Affairs and International
Cooperation (\textsc{maeci}), Grant No.\ ZA23GR03, and from the European Union -- Next Generation EU. This research was also funded by the Deutsche Forschungsgemeinschaft (DFG, German Research Foundation) under Germany's Excellence Strategy – EXC-2094 – 390783311. SA was supported by JSPS KAKENHI Grant Numbers JP17H04836, JP18H04578, and JP18H04340. This work was supported in part by the U.S. Department of Energy under contract number DE-AC02-76SF00515.

Funding for the DES Projects has been provided by the U.S. Department of Energy, the U.S. National Science Foundation, the Ministry of Science and Education of Spain, 
the Science and Technology Facilities Council of the United Kingdom, the Higher Education Funding Council for England, the National Center for Supercomputing 
Applications at the University of Illinois at Urbana-Champaign, the Kavli Institute of Cosmological Physics at the University of Chicago, 
the Center for Cosmology and Astro-Particle Physics at the Ohio State University,
the Mitchell Institute for Fundamental Physics and Astronomy at Texas A\&M University, Financiadora de Estudos e Projetos, 
Funda{\c c}{\~a}o Carlos Chagas Filho de Amparo {\`a} Pesquisa do Estado do Rio de Janeiro, Conselho Nacional de Desenvolvimento Cient{\'i}fico e Tecnol{\'o}gico and 
the Minist{\'e}rio da Ci{\^e}ncia, Tecnologia e Inova{\c c}{\~a}o, the Deutsche Forschungsgemeinschaft and the Collaborating Institutions in the Dark Energy Survey. 

The Collaborating Institutions are Argonne National Laboratory, the University of California at Santa Cruz, the University of Cambridge, Centro de Investigaciones Energ{\'e}ticas, 
Medioambientales y Tecnol{\'o}gicas-Madrid, the University of Chicago, University College London, the DES-Brazil Consortium, the University of Edinburgh, 
the Eidgen{\"o}ssische Technische Hochschule (ETH) Z{\"u}rich, 
Fermi National Accelerator Laboratory, the University of Illinois at Urbana-Champaign, the Institut de Ci{\`e}ncies de l'Espai (IEEC/CSIC), 
the Institut de F{\'i}sica d'Altes Energies, Lawrence Berkeley National Laboratory, the Ludwig-Maximilians Universit{\"a}t M{\"u}nchen and the associated Excellence Cluster Universe, 
the University of Michigan, NSF NOIRLab, the University of Nottingham, The Ohio State University, the University of Pennsylvania, the University of Portsmouth, 
SLAC National Accelerator Laboratory, Stanford University, the University of Sussex, Texas A\&M University, and the OzDES Membership Consortium.

Based in part on observations at NSF Cerro Tololo Inter-American Observatory at NSF NOIRLab (NOIRLab Prop. ID 2012B-0001; PI: J. Frieman), which is managed by the Association of Universities for Research in Astronomy (AURA) under a cooperative agreement with the National Science Foundation.

The DES data management system is supported by the National Science Foundation under Grant Numbers AST-1138766 and AST-1536171.
The DES participants from Spanish institutions are partially supported by MICINN under grants PID2021-123012, PID2021-128989 PID2022-141079, SEV-2016-0588, CEX2020-001058-M and CEX2020-001007-S, some of which include ERDF funds from the European Union. IFAE is partially funded by the CERCA program of the Generalitat de Catalunya.

We  acknowledge support from the Brazilian Instituto Nacional de Ci\^encia
e Tecnologia (INCT) do e-Universo (CNPq grant 465376/2014-2).

This document was prepared by the DES Collaboration using the resources of the Fermi National Accelerator Laboratory (Fermilab), a U.S. Department of Energy, Office of Science, Office of High Energy Physics HEP User Facility. Fermilab is managed by Fermi Forward Discovery Group, LLC, acting under Contract No. 89243024CSC000002.


The \textit{Fermi} LAT Collaboration acknowledges generous ongoing support
from a number of agencies and institutes that have supported both the
development and the operation of the LAT as well as scientific data analysis.
These include the National Aeronautics and Space Administration and the
Department of Energy in the United States, the Commissariat \`a l'Energie Atomique
and the Centre National de la Recherche Scientifique / Institut National de Physique
Nucl\'eaire et de Physique des Particules in France, the Agenzia Spaziale Italiana
and the Istituto Nazionale di Fisica Nucleare in Italy, the Ministry of Education,
Culture, Sports, Science and Technology (MEXT), High Energy Accelerator Research
Organization (KEK) and Japan Aerospace Exploration Agency (JAXA) in Japan, and
the K.~A.~Wallenberg Foundation, the Swedish Research Council and the
Swedish National Space Board in Sweden.
Additional support for science analysis during the operations phase is gratefully
acknowledged from the Istituto Nazionale di Astrofisica in Italy and the Centre
National d'\'Etudes Spatiales in France. This work performed in part under DOE
Contract DE-AC02-76SF00515.

\textbf{Author Contributions}: B. Thakore performed the measurement with D. Gruen, and analyzed the comparison to models with M. Regis. M. Negro provided the Fermi-LAT gamma-ray data. They wrote the manuscript for this paper with constructive feedback throughout the process from the other core authors, namely, S. Camera, N. Fornengo and A. Roodman.
The document has been through internal reviews within the DES collaboration, with A. Porredon and T. Schutt acting as internal reviewers, and within the Fermi-LAT collaboration with A. Cuoco acting as the internal reviewer. All of the reviewers have  provided valuable expertise and notable feedback that contributed to an improvement in the quality of the paper. The remaining authors have made contributions to the DES Y3 Key Project analysis pipeline, including but not limited to the DES instruments, data collection, processing and calibration, and various analysis pipelines.

\newpage

\bibliographystyle{unsrt}
\bibliography{bibliography}

\begin{thebibliography}{10}

\bibitem{aghanim2020planck}
Nabila Aghanim, Yashar Akrami, Mark Ashdown, Jonathan Aumont, Carlo Baccigalupi, Mario Ballardini, Anthony~J Banday, RB~Barreiro, Nicola Bartolo, S~Basak, et~al.
\newblock Planck 2018 results-vi. cosmological parameters.
\newblock {\em Astronomy \& Astrophysics}, 641:A6, 2020.

\bibitem{ullio2002cosmological}
Piero Ullio, Lars Bergstr{\"o}m, Joakim Edsj{\"o}, and Cedric Lacey.
\newblock Cosmological dark matter annihilations into $\gamma$ rays: A closer look.
\newblock {\em Physical Review D}, 66(12):123502, 2002.

\bibitem{camera2013novel}
Stefano Camera, Mattia Fornasa, Nicolao Fornengo, and Marco Regis.
\newblock A novel approach in the weakly interacting massive particle quest: Cross-correlation of gamma-ray anisotropies and cosmic shear.
\newblock {\em The Astrophysical Journal Letters}, 771(1):L5, 2013.

\bibitem{camera2015tomographic}
Stefano Camera, Mattia Fornasa, Nicolao Fornengo, and Marco Regis.
\newblock Tomographic-spectral approach for dark matter detection in the cross-correlation between cosmic shear and diffuse $\gamma$-ray emission.
\newblock {\em Journal of Cosmology and Astroparticle Physics}, 2015(06):029, 2015.

\bibitem{shirasaki2014cross}
Masato Shirasaki, Shunsaku Horiuchi, and Naoki Yoshida.
\newblock Cross correlation of cosmic shear and extragalactic gamma-ray background: Constraints on the dark matter annihilation cross section.
\newblock {\em Physical Review D}, 90(6):063502, 2014.

\bibitem{shirasaki2016cosmological}
Masato Shirasaki, Oscar Macias, Shunsaku Horiuchi, Satoshi Shirai, and Naoki Yoshida.
\newblock Cosmological constraints on dark matter annihilation and decay: Cross-correlation analysis of the extragalactic $\gamma$-ray background and cosmic shear.
\newblock {\em Physical Review D}, 94(6):063522, 2016.

\bibitem{troster2017cross}
Tilman Tr{\"o}ster, Stefano Camera, Mattia Fornasa, Marco Regis, Ludovic Van~Waerbeke, Joachim Harnois-D{\'e}raps, Shin'ichiro Ando, Maciej Bilicki, Thomas Erben, Nicolao Fornengo, et~al.
\newblock Cross-correlation of weak lensing and gamma rays: implications for the nature of dark matter.
\newblock {\em Monthly Notices of the Royal Astronomical Society}, 467(3):2706--2722, 2017.

\bibitem{DES:2019ucp}
S.~Ammazzalorso et~al.
\newblock {Detection of Cross-Correlation between Gravitational Lensing and $\gamma$ Rays}.
\newblock {\em Phys. Rev. Lett.}, 124(10):101102, 2020.

\bibitem{xia2015tomography}
Jun-Qing Xia, Alessandro Cuoco, Enzo Branchini, and Matteo Viel.
\newblock Tomography of the fermi-lat $\gamma$-ray diffuse extragalactic signal via cross correlations with galaxy catalogs.
\newblock {\em The Astrophysical Journal Supplement Series}, 217(1):15, 2015.

\bibitem{regis2015particle}
Marco Regis, Jun-Qing Xia, Alessandro Cuoco, Enzo Branchini, Nicolao Fornengo, and Matteo Viel.
\newblock Particle dark matter searches outside the local group.
\newblock {\em Physical Review Letters}, 114(24):241301, 2015.

\bibitem{cuoco2015dark}
Alessandro Cuoco, Jun-Qing Xia, Marco Regis, Enzo Branchini, Nicolao Fornengo, and Matteo Viel.
\newblock Dark matter searches in the gamma-ray extragalactic background via cross-correlations with galaxy catalogs.
\newblock {\em The Astrophysical Journal Supplement Series}, 221(2):29, 2015.

\bibitem{shirasaki2015cross}
Masato Shirasaki, Shunsaku Horiuchi, and Naoki Yoshida.
\newblock Cross-correlation of the extragalactic gamma-ray background with luminous red galaxies.
\newblock {\em Physical Review D}, 92(12):123540, 2015.

\bibitem{cuoco2017tomographic}
Alessandro Cuoco, Maciej Bilicki, Jun-Qing Xia, and Enzo Branchini.
\newblock Tomographic imaging of the fermi-lat $\gamma$-ray sky through cross-correlations: A wider and deeper look.
\newblock {\em The Astrophysical Journal Supplement Series}, 232(1):10, 2017.

\bibitem{Ammazzalorso:2018evf}
Simone Ammazzalorso, Nicolao Fornengo, Shunsaku Horiuchi, and Marco Regis.
\newblock {Characterizing the local gamma-ray Universe via angular cross-correlations}.
\newblock {\em Phys. Rev. D}, 98(10):103007, 2018.

\bibitem{paopiamsap2024constraints}
Anya Paopiamsap, David Alonso, Deaglan~J Bartlett, and Maciej Bilicki.
\newblock Constraints on dark matter and astrophysics from tomographic $\gamma$-ray cross-correlations.
\newblock {\em Physical Review D}, 109(10):103517, 2024.

\bibitem{branchini2017cross}
Enzo Branchini, Stefano Camera, Alessandro Cuoco, Nicolao Fornengo, Marco Regis, Matteo Viel, and Jun-Qing Xia.
\newblock Cross-correlating the $\gamma$-ray sky with catalogs of galaxy clusters.
\newblock {\em The Astrophysical Journal Supplement Series}, 228(1):8, 2017.

\bibitem{shirasaki2018correlation}
Masato Shirasaki, Oscar Macias, Shunsaku Horiuchi, Naoki Yoshida, Chien-Hsiu Lee, and Atsushi~J. Nishizawa.
\newblock Correlation of extragalactic $\gamma$ rays with cosmic matter density distributions from weak gravitational lensing.
\newblock {\em Physical Review D}, 97(12):123015, 2018.

\bibitem{hashimoto2019measurement}
Daiki Hashimoto, Atsushi~J Nishizawa, Masato Shirasaki, Oscar Macias, Shunsaku Horiuchi, Hiroyuki Tashiro, and Masamune Oguri.
\newblock Measurement of redshift-dependent cross-correlation of hsc clusters and fermi $\gamma$-rays.
\newblock {\em Monthly Notices of the Royal Astronomical Society}, 484(4):5256--5266, 2019.

\bibitem{colavincenzo2020searching}
Manuel Colavincenzo, Xiuhui Tan, Simone Ammazzalorso, Stefano Camera, Marco Regis, Jun-Qing Xia, and Nicolao Fornengo.
\newblock Searching for gamma-ray emission from galaxy clusters at low redshift.
\newblock {\em Monthly Notices of the Royal Astronomical Society}, 491(3):3225--3244, 2020.

\bibitem{tan2020bounds}
Xiuhui Tan, Manuel Colavincenzo, and Simone Ammazzalorso.
\newblock Bounds on wimp dark matter from galaxy clusters at low redshift.
\newblock {\em Monthly Notices of the Royal Astronomical Society}, 495(1):114--122, 2020.

\bibitem{fornengo2015evidence}
Nicolao Fornengo, Laurence Perotto, Marco Regis, and Stefano Camera.
\newblock Evidence of cross-correlation between the cmb lensing and the $\gamma$-ray sky.
\newblock {\em The Astrophysical journal letters}, 802(1):L1, 2015.

\bibitem{ando2014mapping}
Shin’ichiro Ando, Aur{\'e}lien Benoit-L{\'e}vy, and Eiichiro Komatsu.
\newblock Mapping dark matter in the gamma-ray sky with galaxy catalogs.
\newblock {\em Physical Review D}, 90(2):023514, 2014.

\bibitem{fornengo2014particle}
Nicolao Fornengo and Marco Regis.
\newblock Particle dark matter searches in the anisotropic sky.
\newblock {\em Frontiers in Physics}, 2:6, 2014.

\bibitem{fornasa2016angular}
Mattia Fornasa, Alessandro Cuoco, Jes{\'u}s Zavala, Jennifer~M. Gaskins, Miguel~A. S{\'a}nchez-Conde, German Gomez-Vargas, Eiichiro Komatsu, Tim Linden, Francisco Prada, Fabio Zandanel, et~al.
\newblock Angular power spectrum of the diffuse gamma-ray emission as measured by the fermi large area telescope and constraints on its dark matter interpretation.
\newblock {\em Physical Review D}, 94(12):123005, 2016.

\bibitem{feng2017planck}
Chang Feng, Asantha Cooray, and Brian Keating.
\newblock Planck lensing and cosmic infrared background cross-correlation with fermi-lat: Tracing dark matter signals in the gamma-ray background.
\newblock {\em The Astrophysical Journal}, 836(1):127, 2017.

\bibitem{ajello2015origin}
M.~Ajello, D.~Gasparrini, Miguel S{\'a}nchez-Conde, G.~Zaharijas, M.~Gustafsson, J.~Cohen-Tanugi, CD~Dermer, Yoshiyuki Inoue, D.~Hartmann, M.~Ackermann, et~al.
\newblock The origin of the extragalactic gamma-ray background and implications for dark matter annihilation.
\newblock {\em The Astrophysical Journal Letters}, 800(2):L27, 2015.

\bibitem{korsmeier2022flat}
Michael Korsmeier, Elena Pinetti, Michela Negro, Marco Regis, and Nicolao Fornengo.
\newblock Flat-spectrum radio quasars and bl lacs dominate the anisotropy of the unresolved gamma-ray background.
\newblock {\em The Astrophysical Journal}, 933(2):221, 2022.

\bibitem{diehl2014dark}
H.T. Diehl, T.M.C. Abbott, J.~Annis, R.~Armstrong, L.~Baruah, A.~Bermeo, G.~Bernstein, E.~Beynon, C.~Bruderer, E.J. Buckley-Geer, et~al.
\newblock The dark energy survey and operations: Year 1.
\newblock In {\em Observatory Operations: Strategies, Processes, and Systems V}, volume 9149, pages 332--346. SPIE, 2014.

\bibitem{Fermi-LAT:2019yla}
S.~Abdollahi et~al.
\newblock {$Fermi$ Large Area Telescope Fourth Source Catalog}.
\newblock {\em Astrophys. J. Suppl.}, 247(1):33, 2020.

\bibitem{abdollahi2022incremental}
Soheila Abdollahi, Fabio Acero, Luca Baldini, Jean Ballet, Denis Bastieri, Ronaldo Bellazzini, Bijan Berenji, Alessandra Berretta, Elisabetta Bissaldi, Roger~D. Blandford, et~al.
\newblock Incremental fermi large area telescope fourth source catalog.
\newblock {\em The Astrophysical Journal Supplement Series}, 260(2):53, 2022.

\bibitem{friedrich2021dark}
Oliver Friedrich, F.~Andrade-Oliveira, Hugo Camacho, O.~Alves, R.~Rosenfeld, J.~Sanchez, Xiao Fang, Tim~F. Eifler, E.~Krause, C.~Chang, et~al.
\newblock Dark energy survey year 3 results: covariance modelling and its impact on parameter estimation and quality of fit.
\newblock {\em Monthly Notices of the Royal Astronomical Society}, 508(3):3125--3165, 2021.

\bibitem{2018PhRvL.121x1101A}
M.~Ackermann et~al.
\newblock {Unresolved Gamma-Ray Sky through its Angular Power Spectrum}.
\newblock {\em Physical Review Letters}, 121(24):241101, December 2018.

\bibitem{Cooray2002}
A.~{Cooray} and R.~{Sheth}.
\newblock {Halo models of large scale structure}.
\newblock {\em Phys. Rep.}, 372:1--129, December 2002.

\bibitem{Eisenstein:1997jh}
Daniel~J. Eisenstein and Wayne Hu.
\newblock {Power spectra for cold dark matter and its variants}.
\newblock {\em Astrophys. J.}, 511:5, 1997.

\bibitem{asgari2023halo}
Marika Asgari, Alexander~J. Mead, and Catherine Heymans.
\newblock The halo model for cosmology: a pedagogical review.
\newblock {\em arXiv preprint arXiv:2303.08752}, 2023.

\bibitem{Fornengo2014}
N.~{Fornengo} and M.~{Regis}.
\newblock {Particle dark matter searches in the anisotropic sky}.
\newblock {\em Frontiers in Physics}, 2:6, 2014.

\bibitem{DES:2021wwk}
T.~M.~C. Abbott et~al.
\newblock {Dark Energy Survey Year 3 results: Cosmological constraints from galaxy clustering and weak lensing}.
\newblock {\em Phys. Rev. D}, 105(2):023520, 2022.

\bibitem{mukherjee1997egret}
R.~Mukherjee, S.D. Bertsch, D.L .and~Bloom, B.L. Dingus, J.A. Esposito, C.E. Fichtel, R.C. Hartman, S.D. Hunter, G.~Kanbach, D.A. Kniffen, et~al.
\newblock Egret observations of high-energy gamma-ray emission from blazars: an update.
\newblock {\em The Astrophysical Journal}, 490(1):116, 1997.

\bibitem{flaugher2015dark}
Brenna Flaugher, H.T. Diehl, K.~Honscheid, T.M.C. Abbott, O.~Alvarez, R.~Angstadt, J.T. Annis, M.~Antonik, O.~Ballester, L.~Beaufore, et~al.
\newblock The dark energy camera.
\newblock {\em The Astronomical Journal}, 150(5):150, 2015.

\bibitem{gatti2021dark}
Marco Gatti, Erin Sheldon, A.~Amon, M.~Becker, M.~Troxel, A.~Choi, Cyrille Doux, Niall MacCrann, A.~Navarro-Alsina, Ian Harrison, et~al.
\newblock Dark energy survey year 3 results: weak lensing shape catalogue.
\newblock {\em Monthly Notices of the Royal Astronomical Society}, 504(3):4312--4336, 2021.

\bibitem{maccrann2022dark}
Niall MacCrann, Matthew~R. Becker, Jamie McCullough, Alexandra Amon, Daniel Gruen, Michael Jarvis, Ami Choi, Michael~A. Troxel, Erin Sheldon, Brian Yanny, et~al.
\newblock Dark energy survey y3 results: blending shear and redshift biases in image simulations.
\newblock {\em Monthly Notices of the Royal Astronomical Society}, 509(3):3371--3394, 2022.

\bibitem{sheldon2017practical}
Erin~S Sheldon and Eric~M Huff.
\newblock Practical weak-lensing shear measurement with metacalibration.
\newblock {\em The Astrophysical Journal}, 841(1):24, 2017.

\bibitem{huff2017metacalibration}
Eric Huff and Rachel Mandelbaum.
\newblock Metacalibration: direct self-calibration of biases in shear measurement.
\newblock {\em arXiv preprint arXiv:1702.02600}, 2017.

\bibitem{sevilla2021dark}
Ignacio Sevilla-Noarbe, K~Bechtol, M~Carrasco Kind, A~Carnero Rosell, MR~Becker, Alex Drlica-Wagner, RA~Gruendl, ES~Rykoff, E~Sheldon, B~Yanny, et~al.
\newblock Dark energy survey year 3 results: Photometric data set for cosmology.
\newblock {\em The Astrophysical Journal Supplement Series}, 254(2):24, 2021.

\bibitem{buchs2019phenotypic}
Romain Buchs, Chris Davis, D.~Gruen, J.~DeRose, A.~Alarcon, G.M. Bernstein, C.~S{\'a}nchez, J.~Myles, A.~Roodman, S.~Allen, et~al.
\newblock Phenotypic redshifts with self-organizing maps: A novel method to characterize redshift distributions of source galaxies for weak lensing.
\newblock {\em Monthly Notices of the Royal Astronomical Society}, 489(1):820--841, 2019.

\bibitem{gatti2022dark}
Marco Gatti, Giulia Giannini, Gary~M. Bernstein, A.~Alarcon, Justin Myles, A.~Amon, R~Cawthon, M~Troxel, J.~DeRose, S.~Everett, et~al.
\newblock Dark energy survey year 3 results: clustering redshifts--calibration of the weak lensing source redshift distributions with redmagic and boss/eboss.
\newblock {\em Monthly Notices of the Royal Astronomical Society}, 510(1):1223--1247, 2022.

\bibitem{everett2022dark}
S.~Everett, B.~Yanny, N.~Kuropatkin, E.M. Huff, Y.~Zhang, J.~Myles, A.~Masegian, J.~Elvin-Poole, S.~Allam, G.M. Bernstein, et~al.
\newblock Dark energy survey year 3 results: measuring the survey transfer function with balrog.
\newblock {\em The Astrophysical Journal Supplement Series}, 258(1):15, 2022.

\bibitem{suchyta2016no}
E.~Suchyta, E.M. Huff, J.~Aleksi{\'c}, P.~Melchior, S.~Jouvel, N.~MacCrann, A.J. Ross, M.~Crocce, E.~Gaztanaga, K.~Honscheid, et~al.
\newblock No galaxy left behind: accurate measurements with the faintest objects in the dark energy survey.
\newblock {\em Monthly Notices of the Royal Astronomical Society}, 457(1):786--808, 2016.

\bibitem{sanchez2022dark}
Carles S{\'a}nchez, Judit Prat, G~Zacharegkas, S~Pandey, E~Baxter, GM~Bernstein, J~Blazek, R~Cawthon, C~Chang, E~Krause, et~al.
\newblock Dark energy survey year 3 results: Exploiting small-scale information with lensing shear ratios.
\newblock {\em Physical Review D}, 105(8):083529, 2022.

\bibitem{myles2021dark}
Justin Myles, Alex Alarcon, Alexandra Amon, Carles S{\'a}nchez, Spencer Everett, Joseph DeRose, J~McCullough, Daniel Gruen, Gary~M Bernstein, Michael~A Troxel, et~al.
\newblock Dark energy survey year 3 results: redshift calibration of the weak lensing source galaxies.
\newblock {\em Monthly Notices of the Royal Astronomical Society}, 505(3):4249--4277, 2021.

\bibitem{mccullough2024dark}
J.~McCullough, A.~Amon, E.~Legnani, D.~Gruen, A.~Roodman, O.~Friedrich, N.~MacCrann, M.R. Becker, J.~Myles, S.~Dodelson, et~al.
\newblock Dark energy survey year 3: Blue shear.
\newblock {\em arXiv preprint arXiv:2410.22272}, 2024.

\bibitem{jeffrey2021dark}
Niall Jeffrey, Marco Gatti, Chihway Chang, Lorne Whiteway, Umut Demirbozan, Andr{\'a}s Kov{\'a}cs, Giorgia Pollina, David Bacon, Nico Hamaus, Tomasz Kacprzak, et~al.
\newblock Dark energy survey year 3 results: Curved-sky weak lensing mass map reconstruction.
\newblock {\em Monthly Notices of the Royal Astronomical Society}, 505(3):4626--4645, 2021.

\bibitem{Ackermann_2015}
M.~Ackermann et~al.
\newblock {THE} {SPECTRUM} {OF} {ISOTROPIC} {DIFFUSE} {GAMMA}-{RAY} {EMISSION} {BETWEEN} 100 {MeV} {AND} 820 {GeV}.
\newblock {\em The Astrophysical Journal}, 799(1):86, jan 2015.

\bibitem{gorski2005healpix}
Krzysztof~M. Gorski, Eric Hivon, Anthony~J. Banday, Benjamin~D. Wandelt, Frode~K. Hansen, Mstvos Reinecke, and Matthia Bartelmann.
\newblock Healpix: A framework for high-resolution discretization and fast analysis of data distributed on the sphere.
\newblock {\em The Astrophysical Journal}, 622(2):759, 2005.

\bibitem{gruen2018density}
Daniel Gruen, O.~Friedrich, E.~Krause, J.~DeRose, R.~Cawthon, C.~Davis, J.~Elvin-Poole, E.S. Rykoff, R.H. Wechsler, A.~Alarcon, et~al.
\newblock Density split statistics: Cosmological constraints from counts and lensing in cells in des y1 and sdss data.
\newblock {\em Physical Review D}, 98(2):023507, 2018.

\bibitem{singh2017galaxy}
Sukhdeep Singh, Rachel Mandelbaum, Uro{\v{s}} Seljak, An{\v{z}}e Slosar, and Jose Vazquez~Gonzalez.
\newblock Galaxy--galaxy lensing estimators and their covariance properties.
\newblock {\em Monthly Notices of the Royal Astronomical Society}, 471(4):3827--3844, 2017.

\bibitem{jarvis2004skewness}
Mike Jarvis, Gary Bernstein, and Bhuvnesh Jain.
\newblock The skewness of the aperture mass statistic.
\newblock {\em Monthly Notices of the Royal Astronomical Society}, 352(1):338--352, 2004.

\bibitem{jarvis2015treecorr}
Mike Jarvis.
\newblock Treecorr: Two-point correlation functions.
\newblock {\em Astrophysics Source Code Library}, pages ascl--1508, 2015.

\bibitem{foreman2013emcee}
Daniel Foreman-Mackey, David~W. Hogg, Dustin Lang, and Jonathan Goodman.
\newblock emcee: the mcmc hammer.
\newblock {\em Publications of the Astronomical Society of the Pacific}, 125(925):306, 2013.

\bibitem{hinton2016chainconsumer}
Samuel~R. Hinton.
\newblock Chainconsumer.
\newblock {\em The Journal of Open Source Software}, 1(4):00045, 2016.

\bibitem{goodman2010ensemble}
Jonathan Goodman and Jonathan Weare.
\newblock Ensemble samplers with affine invariance.
\newblock {\em Communications in applied mathematics and computational science}, 5(1):65--80, 2010.

\bibitem{becker2016cosmic}
Matthew~R. Becker, M.A. Troxel, N.~MacCrann, E.~Krause, T.F. Eifler, O.~Friedrich, A.~Nicola, A.~Refregier, A.~Amara, D.~Bacon, et~al.
\newblock Cosmic shear measurements with dark energy survey science verification data.
\newblock {\em Physical Review D}, 94(2):022002, 2016.

\bibitem{Ackermann:2014usa}
M.~Ackermann et~al.
\newblock {The spectrum of isotropic diffuse gamma-ray emission between 100 MeV and 820 GeV}.
\newblock {\em Astrophys. J.}, 799:86, 2015.

\bibitem{guetta2023tev}
Dafne Guetta, Silvia Gagliardini, Silvia Celli, Angela Zegarelli, Antonio Capone, Stefano Campion, and Irene DiPalma.
\newblock Tev emission from gamma ray bursts, checking the hadronic model.
\newblock In {\em EPJ Web of Conferences}, volume 280, page 01005. EDP Sciences, 2023.

\bibitem{min2024contribution}
Fang-Sheng Min, Yu-Hua Yao, Ruo-Yu Liu, Shi Chen, Hong Lu, and Yi-Qing Guo.
\newblock Contribution of $\gamma$-ray burst afterglow emissions to the isotropic diffuse $\gamma$-ray background.
\newblock {\em The Astrophysical Journal}, 964(2):195, 2024.

\bibitem{yao2020contribution}
Yu-Hua Yao, Xiao-Chuan Chang, Hong-Bo Hu, Yi-Bin Pan, Hai-Ming Zhang, Hua-Yang Li, Bing-Qiang Qiao, Ming-Ming Kang, Chao-Wen Yang, Wei Liu, et~al.
\newblock Contribution of high-energy grb emissions to the spectrum of the isotropic diffuse $\gamma$-ray background.
\newblock {\em The Astrophysical Journal}, 901(2):106, 2020.

\bibitem{roth2021diffuse}
Matt~A. Roth, Mark~R. Krumholz, Roland~M. Crocker, and Silvia Celli.
\newblock The diffuse $\gamma$-ray background is dominated by star-forming galaxies.
\newblock {\em Nature}, 597(7876):341--344, 2021.

\bibitem{TheFermi-LAT:2017pvy}
M.~Ajello et~al.
\newblock {3FHL: The Third Catalog of Hard Fermi-LAT Sources}.
\newblock {\em Astrophys. J. Suppl.}, 232(2):18, 2017.

\bibitem{hussain2023diffuse}
Saqib Hussain, Rafael Alves~Batista, Elisabete~M. de~Gouveia Dal~Pino, and Klaus Dolag.
\newblock The diffuse gamma-ray flux from clusters of galaxies.
\newblock {\em Nature Communications}, 14(1):2486, 2023.

\bibitem{cooray2016extragalactic}
Asantha Cooray.
\newblock Extragalactic background light measurements and applications.
\newblock {\em Royal Society Open Science}, 3(3):150555, 2016.

\bibitem{Finke2010}
J.~D. {Finke}, S.~{Razzaque}, and C.~D. {Dermer}.
\newblock {Modeling the Extragalactic Background Light from Stars and Dust}.
\newblock {\em \apj}, 712:238--249, March 2010.

\bibitem{Finke:2022uvv}
Justin~D. Finke, Marco Ajello, Alberto Dominguez, Abhishek Desai, Dieter~H. Hartmann, Vaidehi~S. Paliya, and Alberto Saldana-Lopez.
\newblock {Modeling the Extragalactic Background Light and the Cosmic Star Formation History}.
\newblock {\em Astrophys. J.}, 941(1):33, 2022.

\bibitem{Stecker:2016fsg}
Floyd~W. Stecker, Sean~T. Scully, and Matthew~A. Malkan.
\newblock {An Empirical Determination of the Intergalactic Background Light from UV to FIR Wavelengths Using FIR Deep Galaxy Surveys and the Gamma-ray Opacity of the Universe}.
\newblock {\em Astrophys. J.}, 827(1):6, 2016.
\newblock [Erratum: Astrophys.J. 863, 112 (2018)].

\bibitem{Branchini:2016glc}
Enzo Branchini, Stefano Camera, Alessandro Cuoco, Nicolao Fornengo, Marco Regis, Matteo Viel, and Jun-Qing Xia.
\newblock {Cross-correlating the $\gamma$-ray sky with Catalogs of Galaxy Clusters}.
\newblock {\em Astrophys. J. Suppl.}, 228(1):8, 2017.

\bibitem{Colavincenzo:2019jtj}
Manuel Colavincenzo, Xiuhui Tan, Simone Ammazzalorso, Stefano Camera, Marco Regis, Jun-Qing Xia, and Nicolao Fornengo.
\newblock {Searching for gamma-ray emission from galaxy clusters at low redshift}.
\newblock {\em Mon. Not. Roy. Astron. Soc.}, 491(3):3225--3244, 2020.

\bibitem{Allevato:2014qga}
Viola Allevato, Alexis Finoguenov, and Nico Cappelluti.
\newblock {Clustering of $\gamma$-ray selected 2LAC Fermi Blazars}.
\newblock {\em Astrophys. J.}, 797(2):96, 2014.

\bibitem{hikage2019cosmology}
Chiaki Hikage, Masamune Oguri, Takashi Hamana, Surhud More, Rachel Mandelbaum, Masahiro Takada, Fabian K{\"o}hlinger, Hironao Miyatake, Atsushi~J Nishizawa, Hiroaki Aihara, et~al.
\newblock Cosmology from cosmic shear power spectra with subaru hyper suprime-cam first-year data.
\newblock {\em Publications of the Astronomical Society of Japan}, 71(2):43, 2019.

\bibitem{murata2018constraints}
Ryoma Murata, Takahiro Nishimichi, Masahiro Takada, Hironao Miyatake, Masato Shirasaki, Surhud More, Ryuichi Takahashi, and Ken Osato.
\newblock Constraints on the mass--richness relation from the abundance and weak lensing of sdss clusters.
\newblock {\em The Astrophysical Journal}, 854(2):120, 2018.

\bibitem{hartlap2007your}
J.~Hartlap, Patrick Simon, and P.~Schneider.
\newblock Why your model parameter confidences might be too optimistic. unbiased estimation of the inverse covariance matrix.
\newblock {\em Astronomy \& Astrophysics}, 464(1):399--404, 2007.

\bibitem{prat2022dark}
Judit Prat, J.~Blazek, C.~S{\'a}nchez, I.~Tutusaus, S.~Pandey, J.~Elvin-Poole, E.~Krause, M.A. Troxel, L.F. Secco, A.~Amon, et~al.
\newblock Dark energy survey year 3 results: High-precision measurement and modeling of galaxy-galaxy lensing.
\newblock {\em Physical Review D}, 105(8):083528, 2022.

\bibitem{troxel2018survey}
Michael~A. Troxel, Elisabeth Krause, Chihway Chang, Tim~F. Eifler, Oliver Friedrich, Daniel Gruen, Niall MacCrann, A.~Chen, Christopher Davis, Joseph DeRose, et~al.
\newblock Survey geometry and the internal consistency of recent cosmic shear measurements.
\newblock {\em Monthly Notices of the Royal Astronomical Society}, 479(4):4998--5004, 2018.

\bibitem{stecker2019extragalactic}
Floyd~W Stecker, Chris~R Shrader, and Matthew~A Malkan.
\newblock The extragalactic gamma-ray background from core-dominated radio galaxies.
\newblock {\em The Astrophysical Journal}, 879(2):68, 2019.

\bibitem{yuan2018determining}
Zunli Yuan, Jiancheng Wang, DM~Worrall, Bin-Bin Zhang, and Jirong Mao.
\newblock Determining the core radio luminosity function of radio agns via copula.
\newblock {\em The Astrophysical Journal Supplement Series}, 239(2):33, 2018.

\bibitem{2022ApJ...933..221K}
Michael {Korsmeier}, Elena {Pinetti}, Michela {Negro}, Marco {Regis}, and Nicolao {Fornengo}.
\newblock {Flat-spectrum Radio Quasars and BL Lacs Dominate the Anisotropy of the Unresolved Gamma-Ray Background}.
\newblock {\em \apj}, 933(2):221, July 2022.

\bibitem{ackermann2015searching}
M.~Ackermann, A.~Albert, Brandon Anderson, W.B. Atwood, Luca Baldini, G.~Barbiellini, Denis Bastieri, K.~Bechtol, R.~Bellazzini, E.~Bissaldi, et~al.
\newblock Searching for dark matter annihilation from milky way dwarf spheroidal galaxies with six years of fermi large area telescope data.
\newblock {\em Physical Review Letters}, 115(23):231301, 2015.

\bibitem{Regis:2021glv}
Marco Regis et~al.
\newblock {The EMU view of the Large Magellanic Cloud: troubles for sub-TeV WIMPs}.
\newblock {\em JCAP}, 11(11):046, 2021.

\bibitem{fermi2015limits}
Fermi~LAT Collaboration et~al.
\newblock Limits on dark matter annihilation signals from the fermi lat 4-year measurement of the isotropic gamma-ray background.
\newblock {\em Journal of Cosmology and Astroparticle Physics}, 2015(09):008, 2015.

\end{thebibliography}

\appendix

\section{Covariance matrix} \label{sec:cov}
The covariance matrix used for the present analysis is obtained from a combination of the shape-noise term generated from random galaxy rotations, and the theoretically estimated Gaussian large-scale structure covariance.
We expect the former term to dominate the covariance, with the latter expected to have a smaller contribution. As in the main text, the redshift bins for the shear are indicated with indices $r,s$ while energy bins for the UGRB are denoted by indices $a,b$. 

In the Gaussian approximation, an element of the theoretical harmonic-space covariance matrix $\widehat{\bm\Gamma}$ reads:
\be
\widehat\Gamma_{ar\ell,bs\ell^\prime} = \frac{ \delta^{\rm K}_{\ell\ell^\prime}}{(2\ell+1)\Delta \ell f_{\rm sky} }\left[C_\ell^{ar}C_{\ell^\prime}^{bs}+ \big(C_{\ell^\prime}^{rs}+ \nn^{rs}\big) \big(C_\ell^{ab} + \nn^{ab}\big)\right].\label{eq:covCl}
\ee
where the $C_l$'s denote (auto and cross) angular power spectra and $\cal N$ the noises. $\Delta l$ refers to the $\ell$-bin widths and $f_{\rm sky}$ the sky coverage fraction. Eq. \ref{eq:covCl} can therefore be divided into two main  parts: the large-scale structure component, taking the form $C^{ar}_{\ell}C^{bs}_{\ell} + C^{rs}_{\ell}(C^{ab}_{\ell}+\nn^{ab})$, and the shape noise terms $\nn^{rs}(C^{ab}_{\ell}+\nn^{ab})$. All theoretical components involving \g-rays have been corrected for the \Fermi\ PSF beam function prior to their inclusion in the covariance calculation. The noise terms $\nn$ have no angular dependence. $f_{\rm sky}$ accounts for the partial coverage of the sky, with $f_{\rm sky}^{\rm DES} = 0.11$ for DES Y3 (independent of the redshift bin), and $f_{\rm sky}^{\rm Fermi} = (0.073, 0.296,0.433,0.519,0.535,0.546,0.554,0.554,0.554)$ denoting the $f_{\rm sky}$ values for the Fermi sky. In order to be conservative, for the cross-correlations analysis we chose the overlap of the DES and \Fermi\ sky, i.e., $f_{\rm sky}=f_{\rm sky}^{\rm DES}\times f_{\rm sky}^{\rm Fermi}$ considering the smallest $f_{\rm sky}$ value in cases where the energy bins (i.e, $a$ and $b$) were different. 
\\
As mentioned previously, in a Gaussian approximation, the shape noise component can be estimated as $\nn^{rs}(C^{ab}_{\ell}+\nn^{ab})$. Expected to be the dominant noise contribution in the covariance, it is important to model this term reliably. To that end, i.e., to go beyond the Gaussian approximations, we generated 2000 realisations of the noise directly from the data in real space. This was done by rotating the ellipticities using an independent random angle between 0 and $\pi$ and calculating the cross-correlations using the ellipticities and the \Fermi\ maps. The signal generated from this represents a random realization of the shape noise \cite{gruen2018density, hikage2019cosmology,murata2018constraints,gatti2021dark}. 

With a major part of the covariance contribution addressed through the shape noise simulations, we complete the covariance calculations by adding the theoretical estimates of the large-scale structure covariance. The contribution of this term is subdominant in the covariance budget, which is verified both a posteriori, as well as in the previous iteration of this experiment of Ref. \cite{DES:2019ucp}. This part of the covariance depends on both the cross-correlation signals the \g-ray-\g-ray and shear-shear autocorrelations $C^{ab}_{\ell}$ and $C^{rs}_{\ell}$. The details on how to measure the \g-ray autocorrelation have been described in Ref. \cite{2018PhRvL.121x1101A}, with $C^{ab}_\ell$ being fitted with a simple model, given by a power-law added to a constant, written as $C^{ab}_{\ell,\textrm{mod}} = A_{ab}\ell^{-\alpha_{ab}}+C_{\textrm{P}}^{ab}$. The \g-ray noise term, $\nn^{ab}$, on the other hand, is computed using Eq. (5) of Ref. \cite{fornasa2016angular}. The shear autocorrelation is calculated using the galaxy redshift distributions described in Section \ref{sec:DESdata}, assuming a $\Lambda\mathrm{CDM}$ cosmology, with the fiducial values of the model parameters taken from \cite{DES:2021wwk}. 

The large-scale structure part of the covariance is then added to the shape noise term through the creation of a set of 2000 simulated noise realisations from a multivariate Gaussian distribution. These simulations are in physical space and are meant to encode the fluctuations in the large-scale structure. They are added to the shape noise realisations, thereby obtaining 2000 samples, each one with different large-scale and shape noise realisations. These 2000 samples therefore provide a robust estimation of the full covariance matrix.
The inverse of the covariance matrix thus obtained is a biased estimator of the inverse of the true covariance, with the bias depending on the number of realisations $N_s$ and bins $N_b$ of the data. The inverse covariance is corrected for bias by multiplying it with the Anderson-Hartlap factor, given as ($N_s-N_b-2$)$/$($N_s-1$) (see Ref. \cite{hartlap2007your} for more information, and Section VI of Ref. \cite{prat2022dark} for an example of its use in nondiagonal covariances). Using the simulated covariance, the cross-shear measurements  returned a chi-squared value of 430, indicating a very close ($<$1$\sigma$) agreement with the expected value of 432 for the noise component that is represented by the cross-shear.

\begin{figure}[!htbp]
    \centering
    \includegraphics[width=0.45\linewidth]{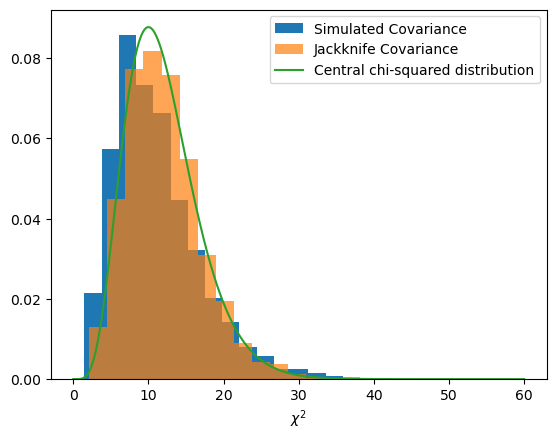}
    \includegraphics[width=0.47\linewidth]{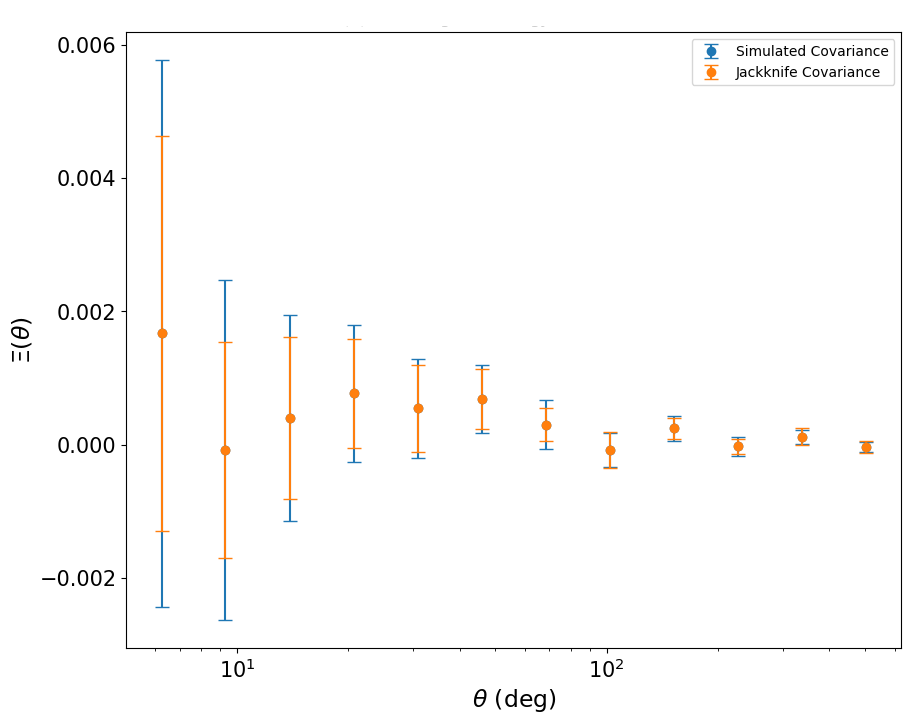}
    \caption{Left: The  chi-squared distribution for 2000 mock datasets in the highest energy and redshift bins as a comparision between the jackknife and simulated covariance matrices. Right: The measured cross-correlation signal for the highest energy and redshift bins, comparing the errors between the jackknife and simulated covariances.}
    \label{chisqjktest}
\end{figure}

\begin{figure*}[t]
    \centering
    \includegraphics[width=0.45\textwidth]{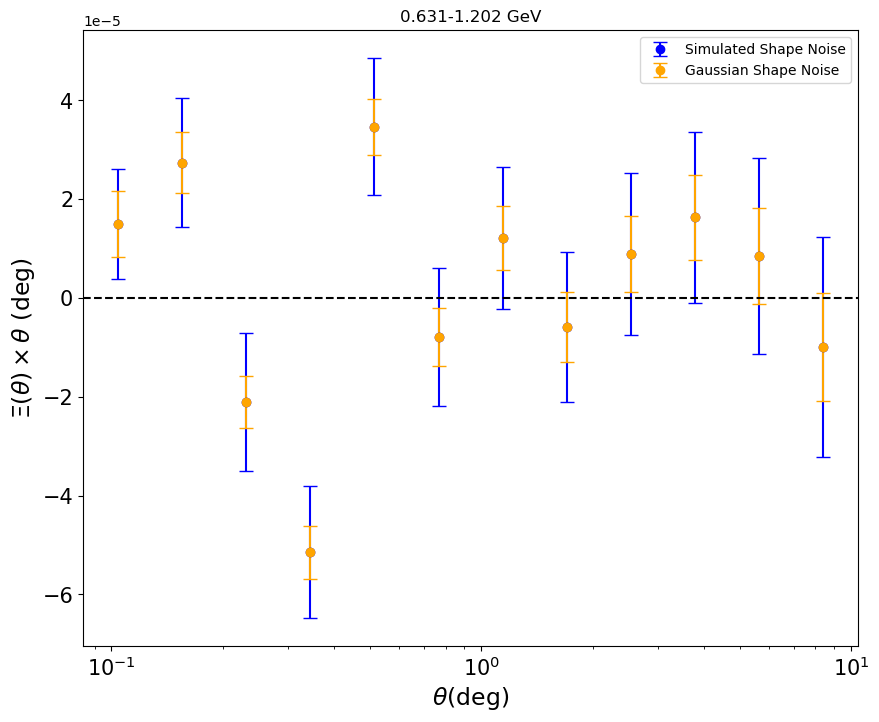}
    \includegraphics[width=0.49\textwidth]{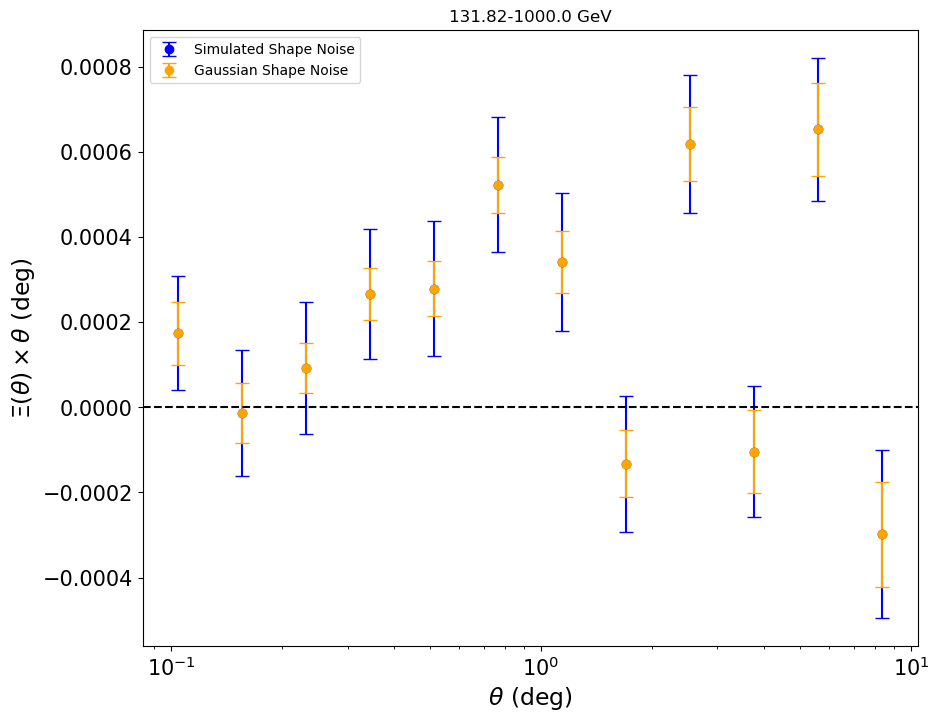}
    \caption{Comparing the shape noise standard deviations generated by simulations against the shape noise standard deviations obtained using the Gaussian approximation, for the lowest (0.631-1.202 GeV, left) and the highest (131.825-1000 GeV, right) energy bins, taking the first redshift bin with a centre of 0.343 for the former and the final redshift bin with a centre of 0.964 for the latter.}
    \label{fig:shapenoisecomparison}
\end{figure*}

In order to establish the reliability of a covariance obtained through simulations, we test it against another covariance obtained through the jackknife method, with each {\it Fermi}-LAT flux map divided into 100 jackknife patches. Since the Anderson-Hartlap factor in this case would become negative in the case of 432 bins, we selected an energy-redshift bin combination with a significant signal, and tested the noise modelling for this bin against that of the simulated covariance. For this purpose, we chose the highest redshift and energy bins, and generated 2000 mock datasets having a mean of zero from a multivariate Gaussian distribution using the jackknife covariance. The left side of Fig. \ref{chisqjktest} shows the results, depicting roughly the same chi-squared distribution, with the standard deviations of the simulated and jackknife covariances being approximately equal. This is also shown in a qualitative manner by the plot on the right, where we have compared the standard deviations for the jackknife and simulated covariances in the measured signal for the aforementioned bin combination. We have also compared the shape noise obtained by the simulated covariance and the Gaussian covariance, as shown in Fig. \ref{fig:shapenoisecomparison} for the lowest energy and redshift as well as the highest energy and redshift bins. This is in contrast to the covariances that were compared in Fig. \ref{chisqjktest}, where the components were not split into their contributions but were tested as a whole. Here, instead, we solely test the robustness of the shape noise obtained using the simulated covariance by using a Gaussian covariance as a reference. In order to depict the behaviour at larger angular scales in a clearer fashion, we have "flattened" the signal by multiplying the cross-correlation $\Xi(\theta)$ with its corresponding angular bin $\theta$. We find that the simulated covariance has a larger shape noise than its Gaussian counterpart, as expected, due to complex masking effects present in the data \cite{troxel2018survey}. By quoting the shape noise term obtained directly from the data, we can ensure that the goodness of fit tests as well as the errors are reliable.

To summarize, we have tested the simulated covariance in two different ways: first, by comparing it to an internally estimated covariance, we ensured that it was able to model the noise accurately using the null chi-squared test. Then we considered the dominant noise component of the simulated matrix, i.e, the shape noise, and tested it against its Gaussian counterpart, finding typically larger error bars for the former owing to masking effects present in the data from which the simulations were generated, which was expected. This allowed us to generate the shape noise errors from the data and ensure that the goodness of fit tests were reliable while giving us a full covariance matrix where the Anderson-Hartlap factor could be applied to de-bias the covariance. This validates the robustness of the simulated covariance, which has been used for the analysis in this work.  
\section{Star forming galaxies} \label{sec:SFG}
\begin{figure*}[t]
    \centering
    \includegraphics[width=0.40\textwidth]{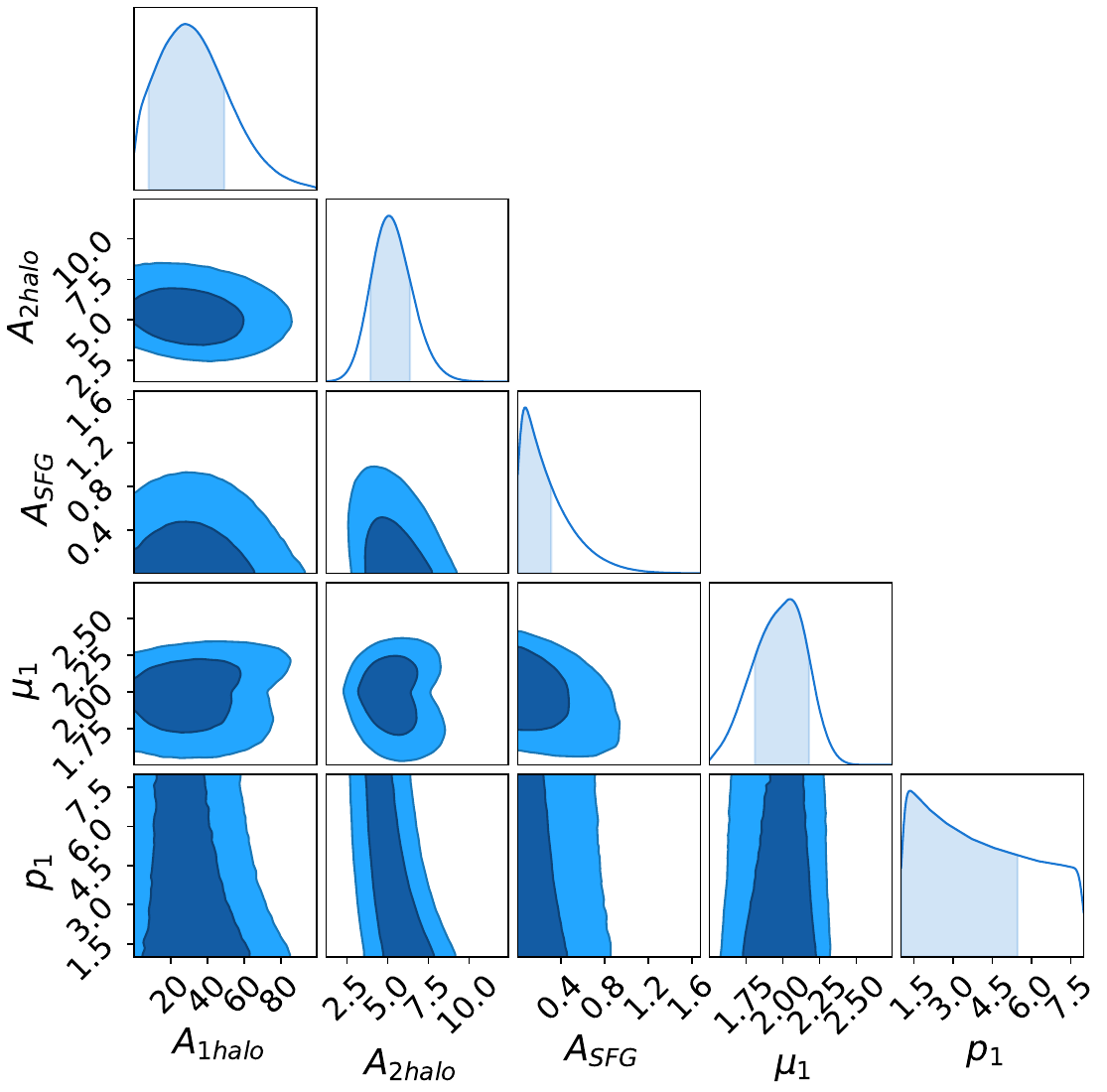}
    \includegraphics[width=0.45\textwidth]{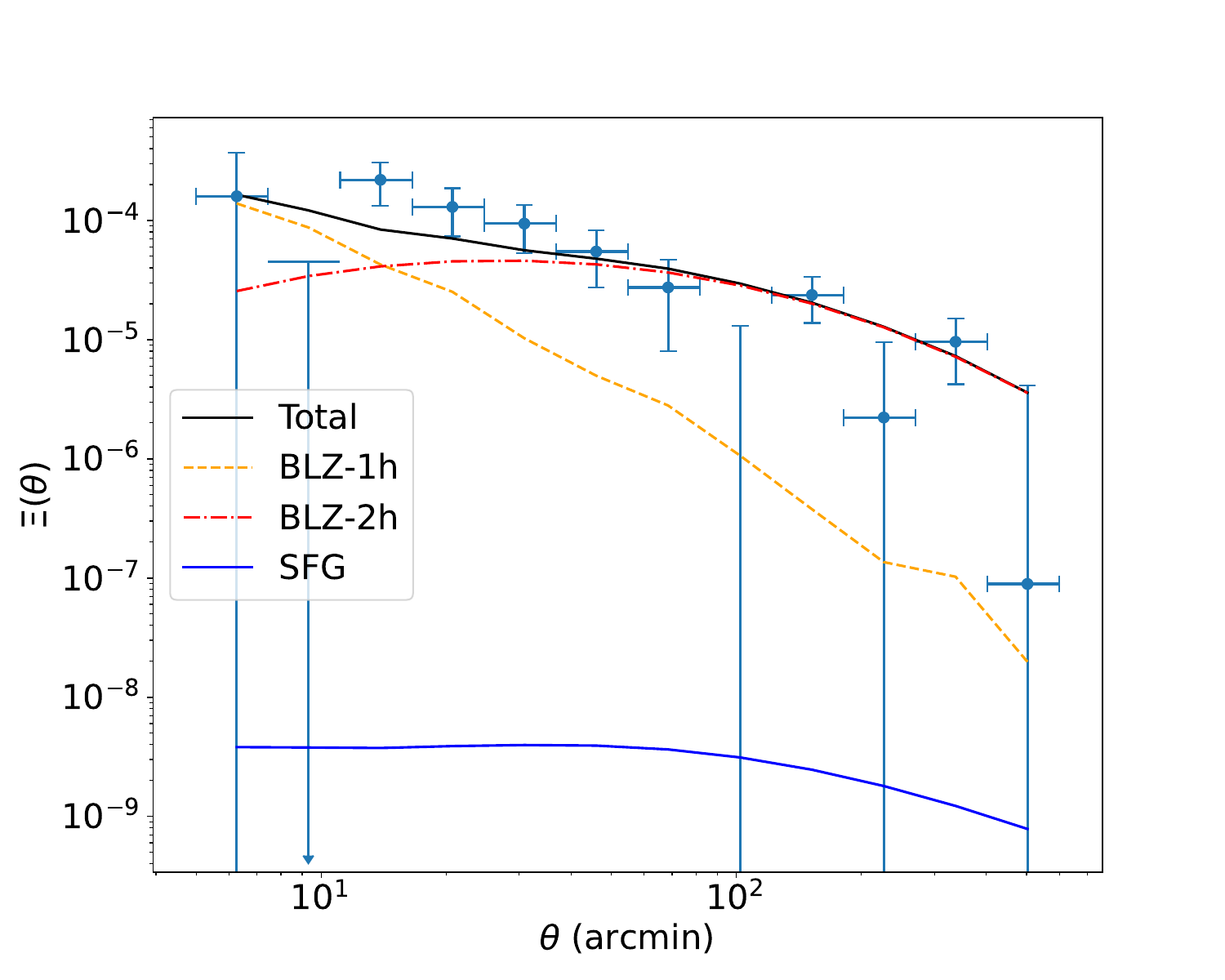}
    \caption{Constraints on the normalization parameters and spectral indices in a model containing blazar and SFGs on left, alongside a description of the angular evolution of the cross-correlations and the best-fit physical models on the right.}
    \label{fig:SFGinclusiveplots}
\end{figure*}
The model of SFG is based on the GLF described in Ref. \cite{DES:2019ucp}. As a first case, we add the contribution to the cross-correlation signal of such reference SFG model with a free normalization. The \phys\ model of Eq.~(\ref{eq:physmdl}) becomes:
\begin{align}
&\Xi_{\rm phys}^{ar}(\theta) \ \langle I_a \rangle =  A_{\rm BLZ}^{\rm 1h}\  \hat\Xi_{\rm BLZ,1h}^{ar} (\theta,\mu_{\rm BLZ},p_1)+  A_{\rm BLZ}^{\rm 2h}\  \hat\Xi_{\rm BLZ,2h}^{ar} (\theta,\mu_{\rm BLZ},p_1) + A_{\rm SFG}\ \hat\Xi_{\rm SFG}^{ar} (\theta), 
\label{eq:physmdlsfg}
\end{align}
where the parameter $A_{\rm SFG}$ sets the amplitude of the SFG correlation. The constraints on the parameters are shown in the left panel of Fig.\ref{fig:SFGinclusiveplots}. 
The SNR does not vary with respect to the BLZ-only case reported in Table~\ref{tab:chi2comp}, i.e., ${\rm SNR}=7.2$, and the contribution from SFGs is negligible, see right panel  of Fig. \ref{fig:SFGinclusiveplots}.
This was expected since the \g-ray spectrum of this reference SFG model is a power-law with spectral index of 2.7, something very different from the properties of the measurement described with the \pheno\ model.

We then allowed more flexibility to the SFG model, including a log-parabolic spectrum and splitting the 1-halo and 2-halo components, so that Eq.~(\ref{eq:physmdlsfg}) can be re-written as:
\begin{align}
&\Xi_{\rm phys}^{ar}(\theta) \ \langle I_a \rangle =  A_{\rm BLZ}^{\rm 1h}\ \hat\Xi_{\rm BLZ,1h}^{ar} (\theta,\mu_{\rm BLZ},p_1)+  A_{\rm BLZ}^{\rm 2h}\  \hat\Xi_{\rm BLZ,2h}^{ar} (\theta,\mu_{\rm BLZ},p_1)\ + \nonumber \\ & A_{\rm SFG}^{\rm 1h}\  \hat\Xi_{\rm SFG,1h}^{ar} (\theta,\mu_{\rm SFG},\mu_E) + A_{\rm SFG}^{\rm 2h}\ \hat\Xi_{\rm SFG,2h}^{ar} (\theta,\mu_{\rm SFG},\mu_E), 
\label{eq:physmdlsfg2}
\end{align}
where $A_{\rm SFG}^{\rm 1h}$ and $A_{\rm SFG}^{\rm 2h}$ are the normalizations on the 1-halo and 2-halo SFG components, while $\mu_{\rm SFG}$ and $\mu_E$ refer to the spectral and curvature indices of the log-parabola in energy. 
Due to cosmic ray activity in dense interstellar matter, SFG can exhibit a log-parabolic behaviour in energy owing to $\pi_0$ decay. This peaks around the GeV (see e.g. Ref. \cite{roth2021diffuse}) and so we set the pivot energy of the log-parabola to 1 GeV, which, together with the prior on $\mu_{\rm SFG}\subset[1.5,3]$, ensures the turnaround of the spectrum to be below a few GeV. 

Performing the MCMC scan with the above model, we obtained similar results as with the simple power-law case introduced at the beginning of the section. The SNR remained unchanged at 7.2, the presence of SFGs comes out to be subdominant, and the SFG parameters are unconstrained. 
To obtain some preference for an SFG contribution, we should move the turnaround energy of the curved spectrum to above 10 GeV, something that lacks physical motivation.
\section{Misaligned AGNs}
\label{sec:mAGN}
\begin{figure*}[t]
    \centering
    \includegraphics[width=0.40\textwidth]{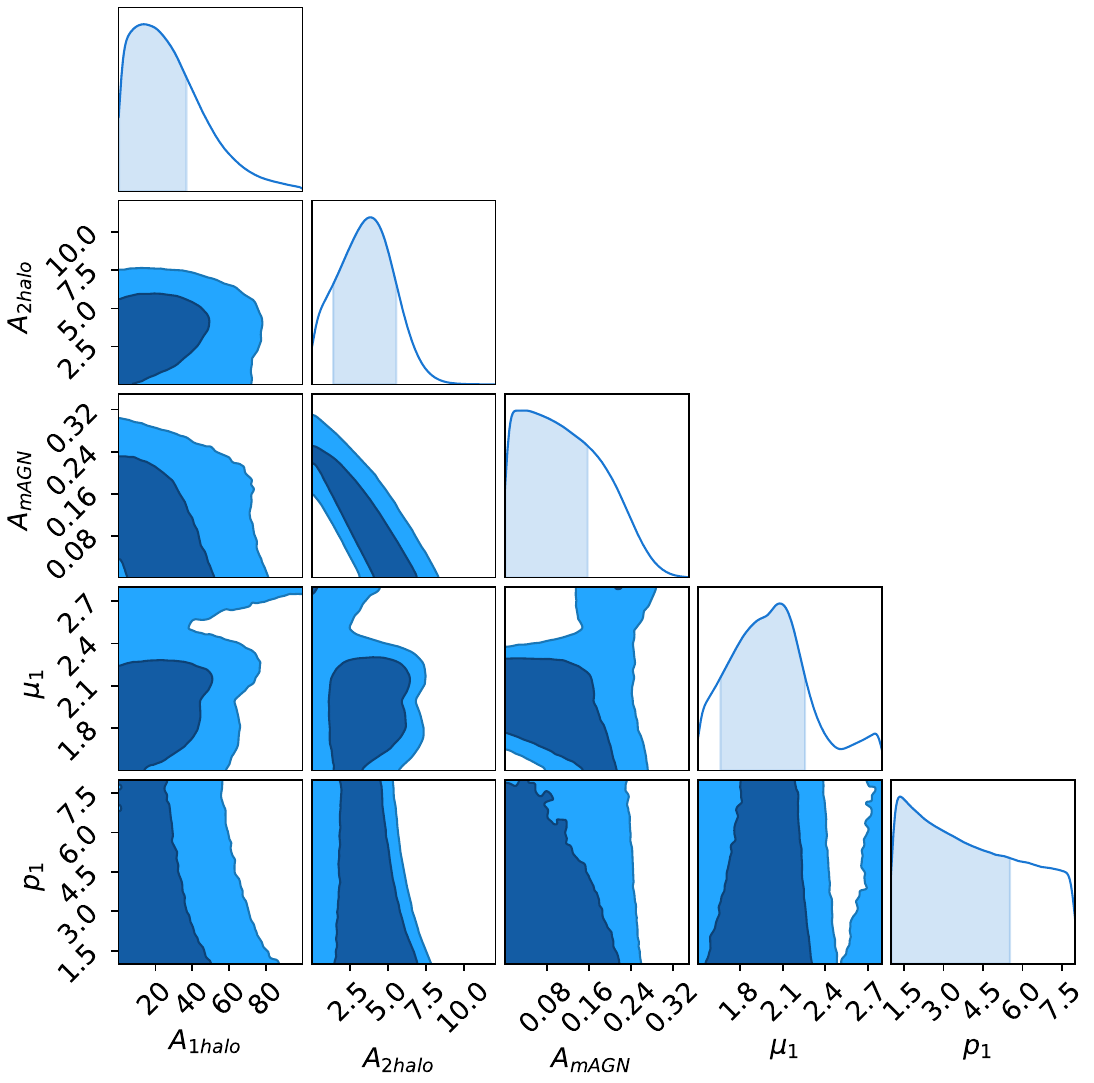}
    \includegraphics[width=0.45\textwidth]{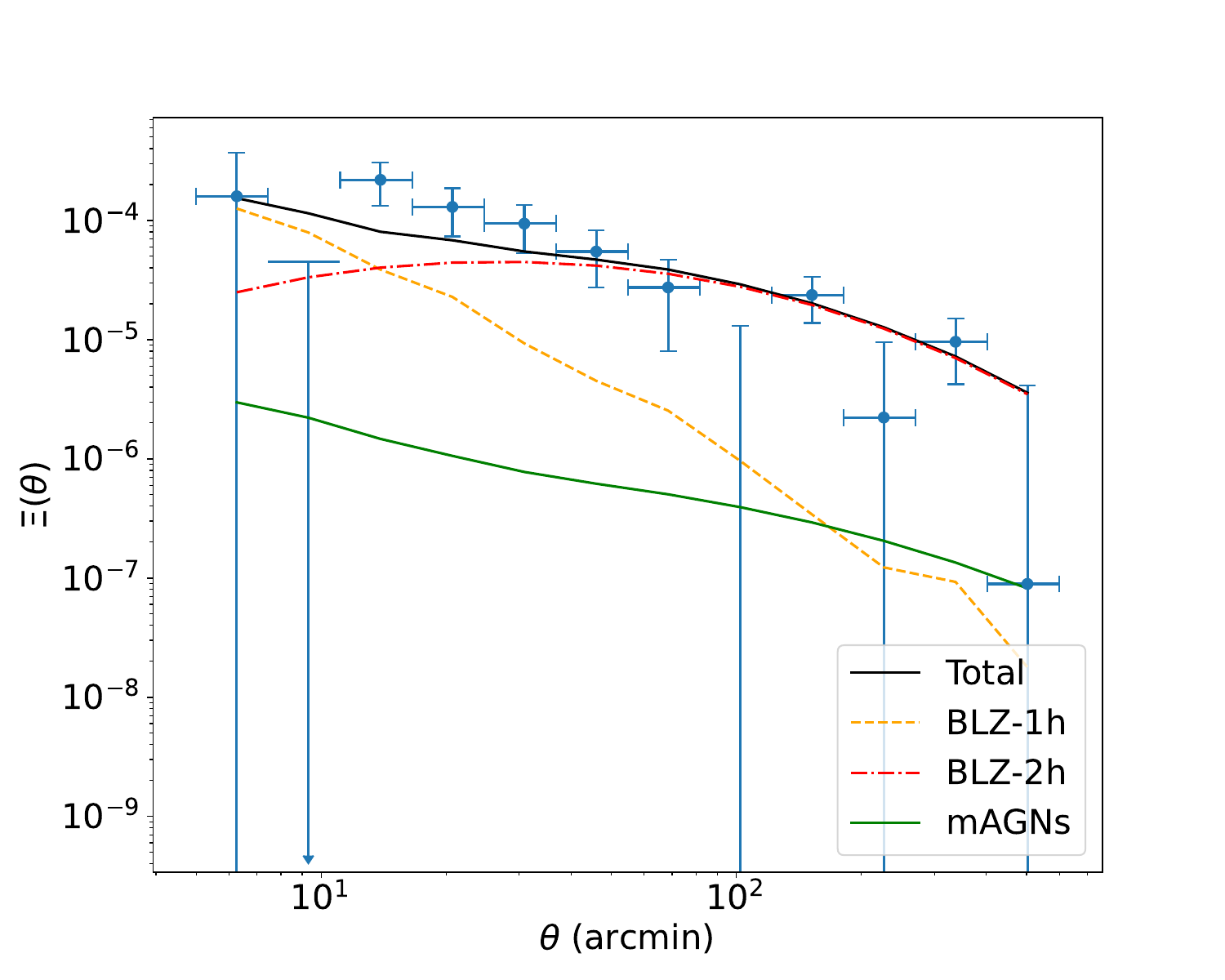}
    \caption{Constraints on the normalization parameters and spectral indices in a model containing blazar and mAGNs on left, alongside a description of the angular evolution of the cross-correlations and the best-fit physical models on the right.}
    \label{fig:mAGNinclusiveplots}
\end{figure*}

For the mAGN model, we follow Ref.~\cite{stecker2019extragalactic}, where the GLF is built from the radio luminosity function described in Ref.~\cite{yuan2018determining}. In a process similar to the one taken for SFGs, we consider the contribution to the cross-correlation signal of this reference mAGN model including a free normalization $A_{\rm mAGN}$, which alters Eq.~\ref{eq:physmdl} as:
\begin{align}
&\Xi_{\rm phys}^{ar}(\theta) \ \langle I_a \rangle =  A_{\rm BLZ}^{\rm 1h}\  \hat\Xi_{\rm BLZ,1h}^{ar} (\theta,\mu_{\rm BLZ},p_1)+  A_{\rm BLZ}^{\rm 2h}\  \hat\Xi_{\rm BLZ,2h}^{ar} (\theta,\mu_{\rm BLZ},p_1) + A_{\rm mAGN}\ \hat\Xi_{\rm mAGN}^{ar} (\theta)\;. 
\label{eq:physmdlmagn}
\end{align}
The constraints on the new set of parameters are reported in the left panel of Fig.~\ref{fig:mAGNinclusiveplots}.
 The inclusion of mAGN does not improve the SNR, which remains stable at 7.2 as in the BLZ-only case (see Table ~\ref{tab:chi2comp}). As discussed for SFGs, this result meets expectations considering that the reference mAGN model has a \g-ray energy spectrum described by a power-law with spectral index of 2.3. In fact, such an index, while providing a relatively harder spectrum with respect to SFGs, describes a significantly softer spectrum compared to the best-fit value we obtain when describing the data with the phenomenological model. We conclude that both SFGs and mAGNs, as modelled under the aforementioned reference cases, can be considered to have negligible contributions to the cross-correlations signal compared to blazars. We note that also the UGRB small scale anisotropy in the LAT band is heavily dominated by blazars, as pointed out in Ref.~\cite{2022ApJ...933..221K}. Future UGRB measurements extending at lower energies may open up new regimes in which SFG and/or mAGN can be expected to contribute more.

\section{WIMP dark matter} \label{sec:DM}
\begin{figure*}[t]
    \centering
    \includegraphics[width=0.40\textwidth]{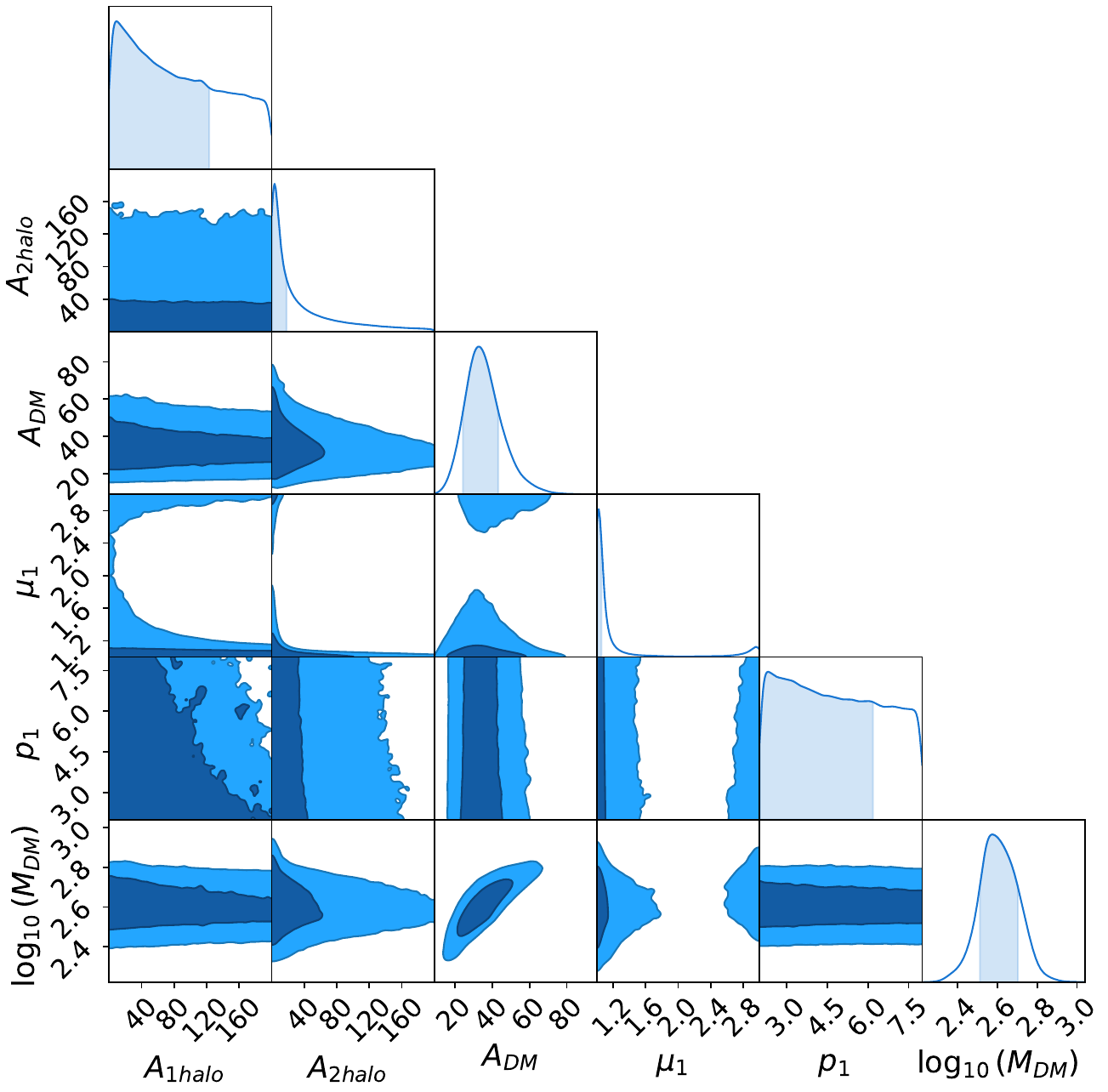}
    \includegraphics[width=0.45\textwidth]{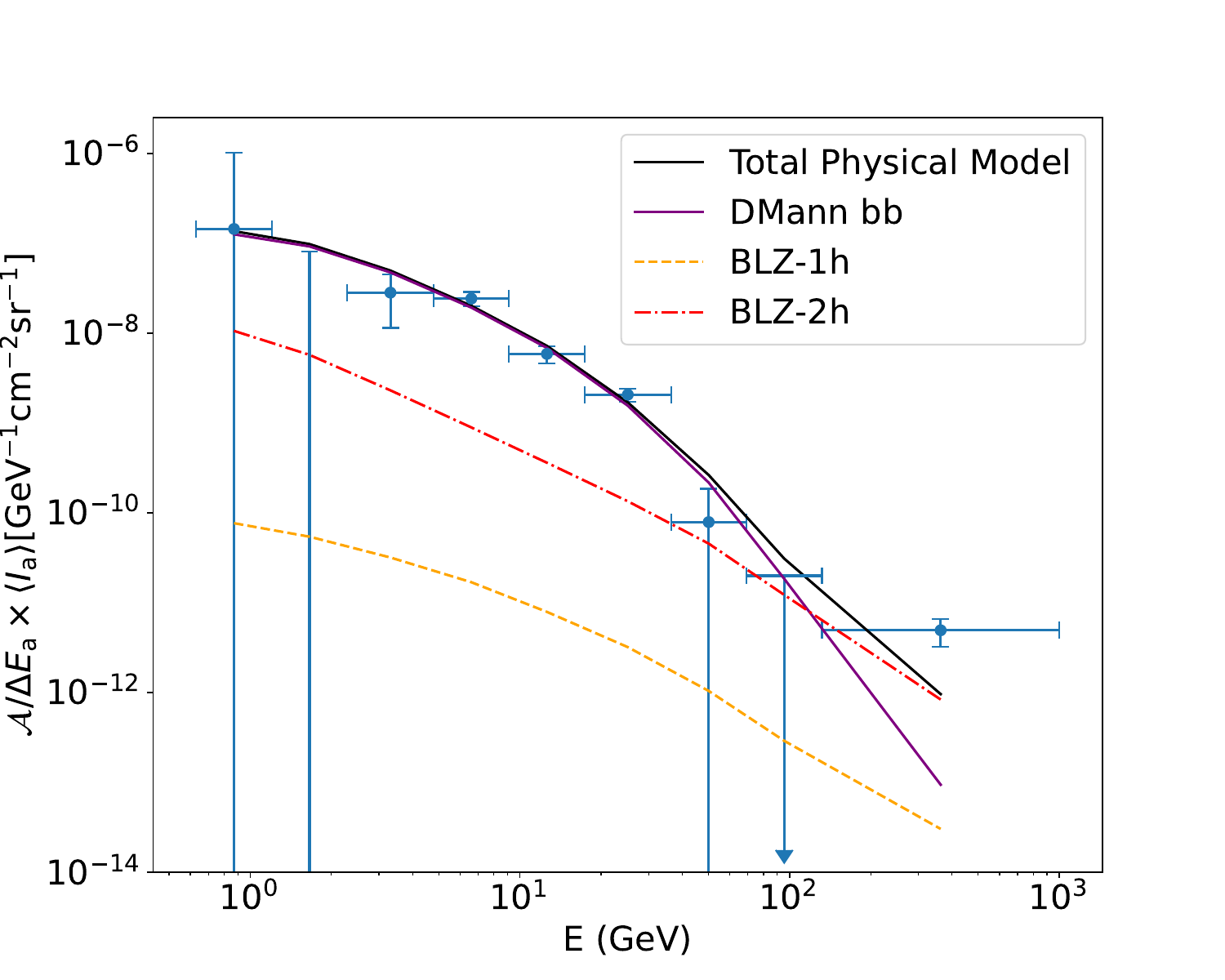}
    \caption{Left: Constraints on the normalization parameters and spectral indices in a model containing blazar and potential contributions from DM annihilation. Right: A description of the energy dependence of each component.}
    \label{fig:DMplusBLZ}
\end{figure*}

Finally, we present a case including the astrophysical contribution from blazars as described in Eq.~(\ref{eq:physmdl}) together with an additional term from particle DM annihilation. Thereby the model reads:
\begin{align}
&\Xi_{\rm phys}^{ar}(\theta) \ \langle I_a \rangle =  A_{\rm BLZ}^{\rm 1h}\ \hat\Xi_{\rm BLZ,1h}^{ar} (\theta,\mu_{\rm BLZ},p_1)+  A_{\rm BLZ}^{\rm 2h}\ \hat\Xi_{\rm BLZ,2h}^{ar} (\theta,\mu_{\rm BLZ},p_1) + A_{\rm DM}\ \hat\Xi_{ {\rm DM}}^{ar}(\theta; \mdm), 
\label{eq:physmdldm}
\end{align}
where the DM contribution depends on the DM mass $\mdm$ along with the velocity-averaged annihilation rate $\langle \sigma_{\rm ann}\upsilon \rangle$ expressed in terms of the "thermal" annihilation cross-section $\langle \sigma_{\rm ann}\upsilon \rangle_{\rm th} = 3\times10^{-26} {\rm cm^{-2} s^{-1}}$, i.e., $A_{\rm DM} = \langle \sigma_{\rm ann}\upsilon \rangle/\langle \sigma_{\rm ann}\upsilon \rangle_{\rm th}$. As a benchmark final state of annihilation, we adopt the $b\Bar{b}$ channel. All the ingredients entering the computation of the DM contribution are detailed in Ref. \cite{DES:2019ucp}.
The triangle plot is shown in Fig. \ref{fig:DMplusBLZ}. We find the best-fit annihilation cross-section and mass to be $\langle\sigma_{\rm ann}\upsilon \rangle  = 32^{+10}_{-8} \langle\sigma_{\rm ann}\upsilon \rangle_{\rm th}$,  and $\mdm = 363^{+138}_{-39.5}$ GeV, which is also in agreement with results for the $b\Bar{b}$ annihilation channel found in Ref.~\cite{DES:2019ucp}. 
We find that the SNR for this \phys\ model is 8.9, which is analogous to the value obtained by the log-parabolic phenomenological model. Quantitatively, the $\Delta\chi^2 = 27$ between the DM-inclusive and a BLZ-only model would point towards a significant preference in favor of the former. 
However, there are two caveats related to such an inference. 
First, as discussed in the main text, the DM component might be effectively fitting a curved \g-ray spectrum which is actually produced by a different astrophysical mechanism. It is also quite unlikely that the astrophysical contribution becomes as subdominant as in Fig.~\ref{fig:DMplusBLZ}, namely, this would require a peculiar redshift behaviour of the UGRB sources to suppress their correlation with lensing.
Second, the fit requires a large $A_{\rm DM}$, and such large values of the annihilation cross-section can be in tension with constraints from other probes, e.g.,  dwarf spheroidal galaxies~\cite{ackermann2015searching} and the Large Magellanic Cloud~\cite{Regis:2021glv}.
On the other hand, the value of $A_{\rm DM}$ has to be taken with a grain of salt, since there is an uncertainty in the overall amplitude of the cosmological DM signal, primarily associated to the poorly known contribution of subhalos (see for example Ref.\cite{fermi2015limits}).

A follow-up analysis, including DES Y3 galaxy clustering and more statistics will help in clarifying the viability of a particle DM contribution. 

\section{Data Verification and Robustness Checks}\label{sec:robustness}

In order to make sure that the cross-correlation described in Eq. \ref{eq:crossshear} is correct, we have tested the estimator using four datasets, three of which were dominated by noise while the remaining dataset contained the cross-correlation signal. The noise dominated datasets were chosen as: i) the cross component of the galaxy shapes (known as B-modes or $\gamma_\times$), which should yield a null detection when cross-correlated with \Fermi\ \g-ray maps; ii) a realization of the shape noise that we obtained by randomly rotating the tangential components of the signal and cross-correlating it with the \g-ray maps; iii) randomly reshuffled pixels of the masked \g-ray maps cross-correlated with the tangential ellipticity. Along with the true signal, the four datasets were blinded and randomly assigned the names W, X, Y, and Z. The results of a full MCMC analysis considering the log-parabola model as a fit for each of the blinded datasets (using the same covariance that was used in the data analysis) are reported in Table \ref{tab:blinding}.  
\begin{table}[!htbp]
\centering
\begin{tabular}
{|l|l| l|l|l|}
\hline
            & \textbf{W} & \textbf{X}    & \textbf{Y}    & \textbf{Z}     \\ \hline
$\Delta\chi^2$$$ &2.07 & 4.78 & 3.36 & 78.64 \\ \hline
SNR         & 1.44 & 2.19 & 1.83 & 8.86  \\ \hline
\end{tabular}
\caption{The SNR and $\Delta\chi^2$ results for the log-parabola \pheno\ model with respect to the null hypothesis for each data vector in the blind analysis. Only Z shows a large preference for the models, with X and Y compatible with null signals.}
\label{tab:blinding}
\end{table}
We find that W, X, and Y yield extremely low SNRs and $\Delta \chi^2$ values, while Z has values that correspond to those obtained in the data analysis in Table \ref{tab:chi2comp}. The results obtained here are compatible with our expectations: there are no spurious detections, and the presence of a signal occurs only where it is truly possible, with clear distinctions between the null and true signals. We note here that the estimator was also tested through an extensive blinding procedure in Ref. \cite{DES:2019ucp}, and has been consequently verified to be a robust method to determine the cross-correlation signal.

\begin{figure}[!ht]
\centering
\subfloat{
\scalebox{0.30}{
    \centering
    \includegraphics{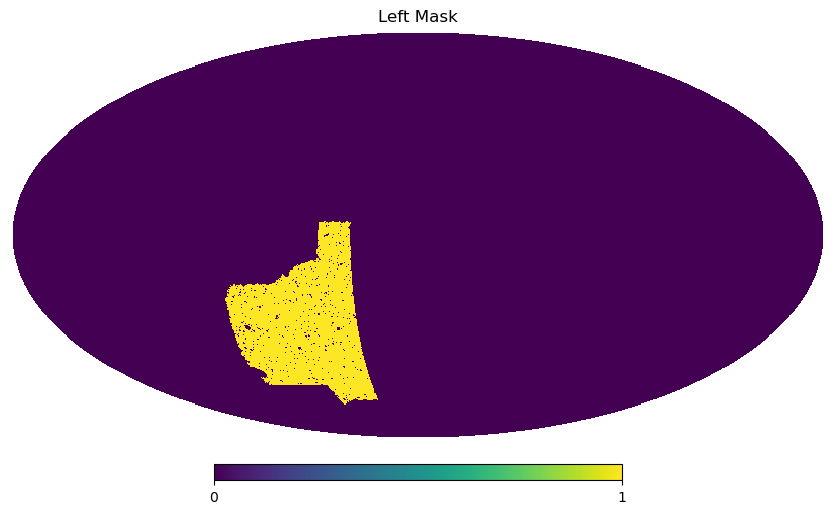}

}
}
\subfloat{
\scalebox{0.30}{
    \centering
    \includegraphics{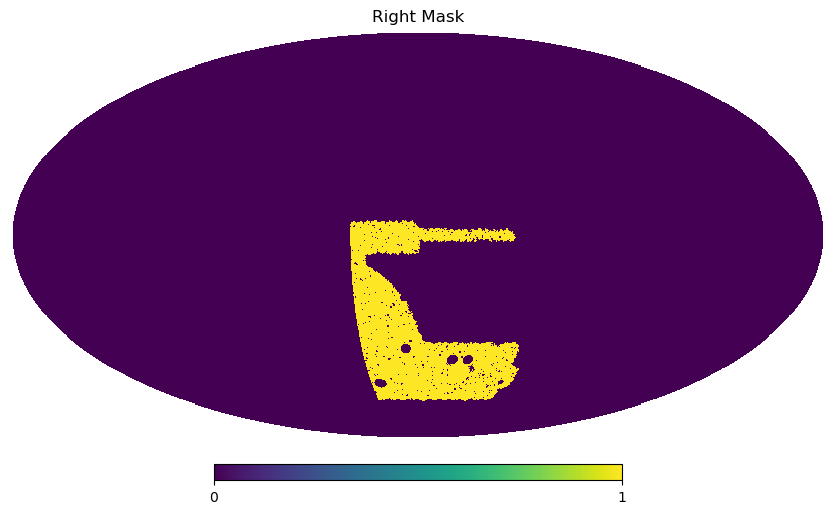}

}
}
\\
\subfloat{
\scalebox{0.30}{
    \centering
    \includegraphics{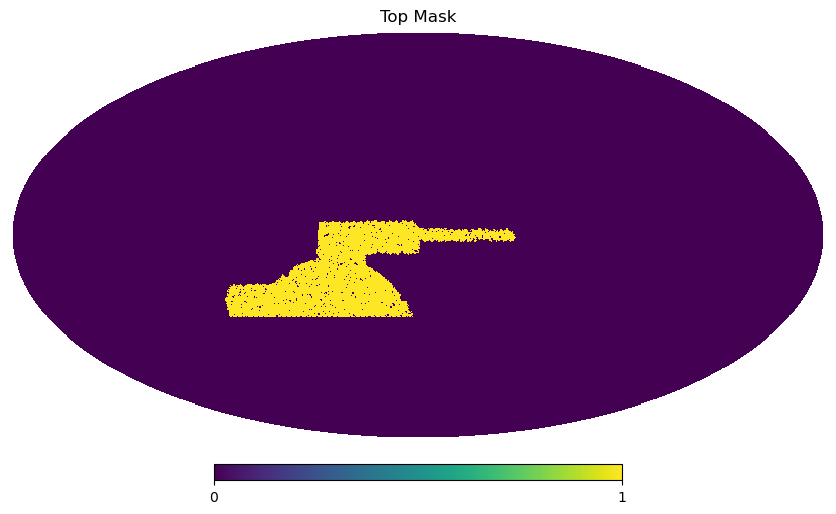}

}
}
\subfloat{
\scalebox{0.30}{
    \centering
    \includegraphics{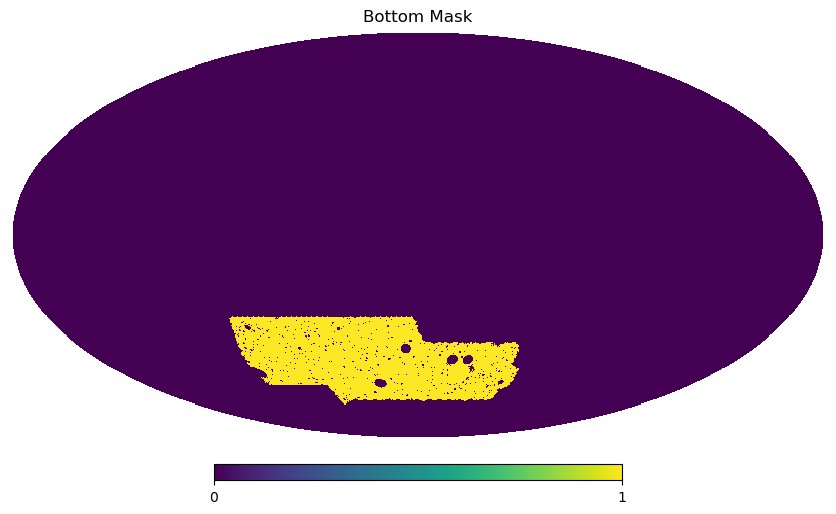}

}
}

\caption{Masks used for the "quadrant" test.}
\label{quadtestmaps}
\end{figure}

\begin{figure}[!ht]
    \centering
    \includegraphics[width=0.5\linewidth]{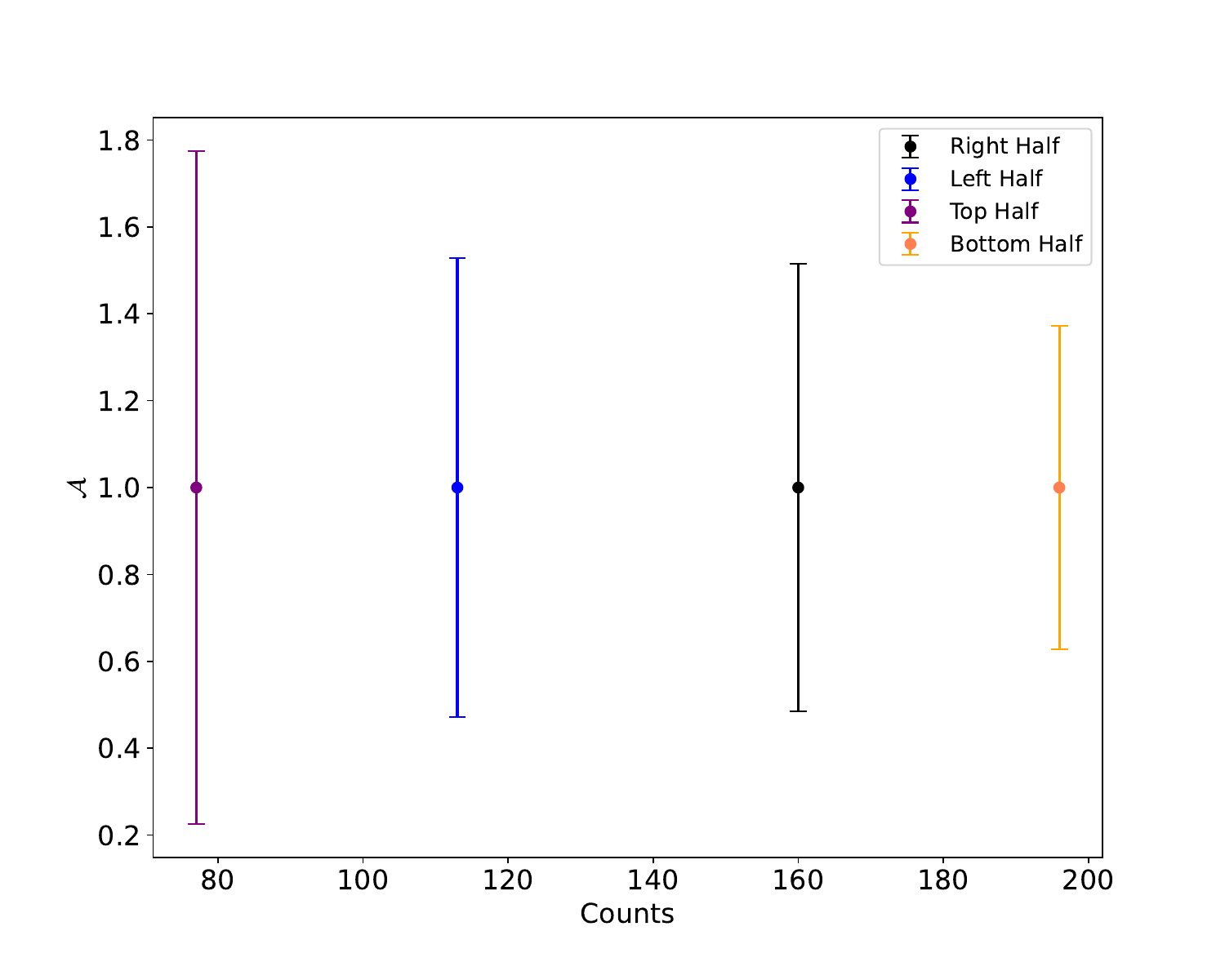}
    \caption{The amplitudes as a function of the \g-ray counts for each of the quadrants depicted in Fig. \ref{quadtestmaps}.}
    \label{fig:quadtest}
\end{figure}

We now turn our attention to the ninth energy bin in the dataset. As shown in Table \ref{tab:enbins}, the \Fermi\ statistics are scarce in the final energy bin, leading to a large shot noise component, which can prevent a reliable signal detection. Therefore, the signal that we have obtained for this energy range could be the consequence of lurking pixels or sets of pixels. In order to test this, we conduct a "quadrant" test that proceeds as follows: we mask the \g-ray map using a combined mask of the DES Y3 footprint as well as the foreground subtracted \Fermi\ map, and further divide the resultant mask into four parts, each of which covers half of the DES footprint, but oriented differently, as shown in Fig. \ref{quadtestmaps}. 
We then measured the amplitudes of the resultant cross-correlation signals as detailed in Section \ref{sec:res}. The covariance used for this test has been obtained using the same method described in Appendix \ref{sec:cov}, i.e, combining shape noise simulations with Gaussian realisations of the large-scale structure. The resulting amplitudes for each of these quadrants, as a function of the photon counts in each quadrant, are shown in Fig. \ref{fig:quadtest}. A mild scaling of the errors with the number of counts is present, with the stongest constraints found for the highest count number. The amplitudes, are very consistent is size, confirming that the signal we have obtained does not appear to be a spurious effect brought in  by particular group of pixels, but appears to be a true cross-correlation signal. 

\begin{figure}[!htbp]
\subfloat{
    \scalebox{1}{\includegraphics[width=0.33\linewidth]{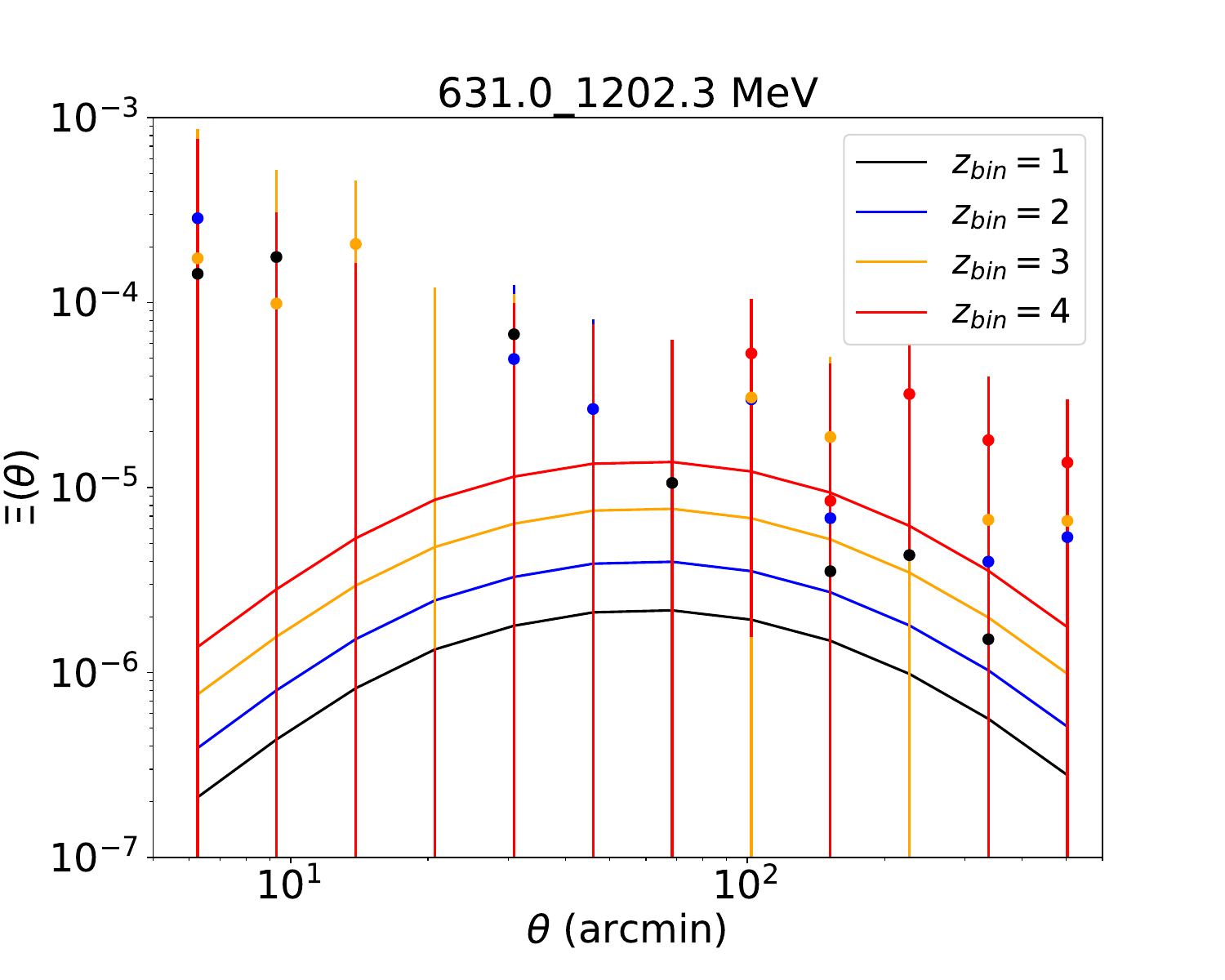}

}

\scalebox{1}{
    \includegraphics[width=0.33\linewidth]{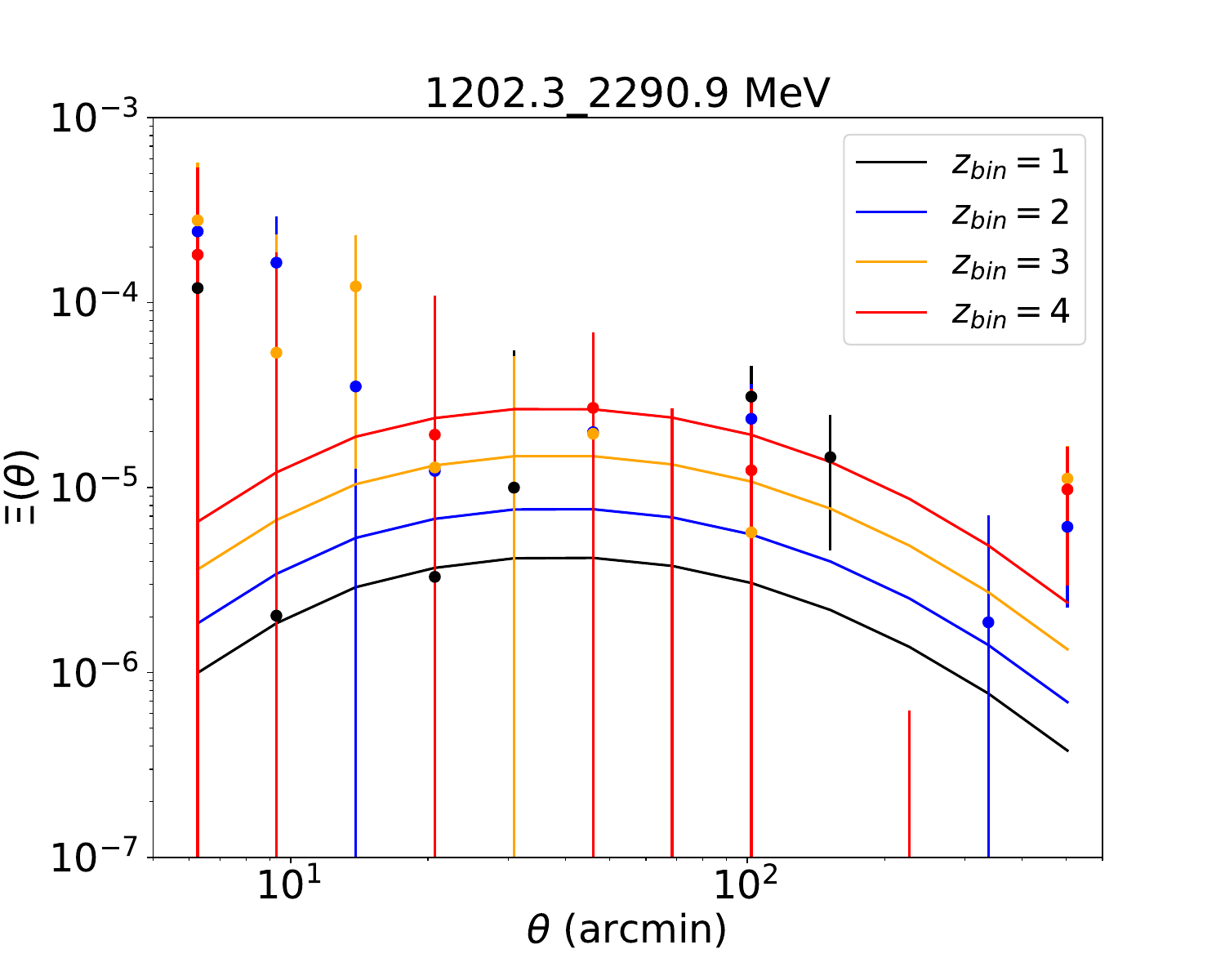}

}

\scalebox{1}{
    \includegraphics[width=0.33\linewidth]{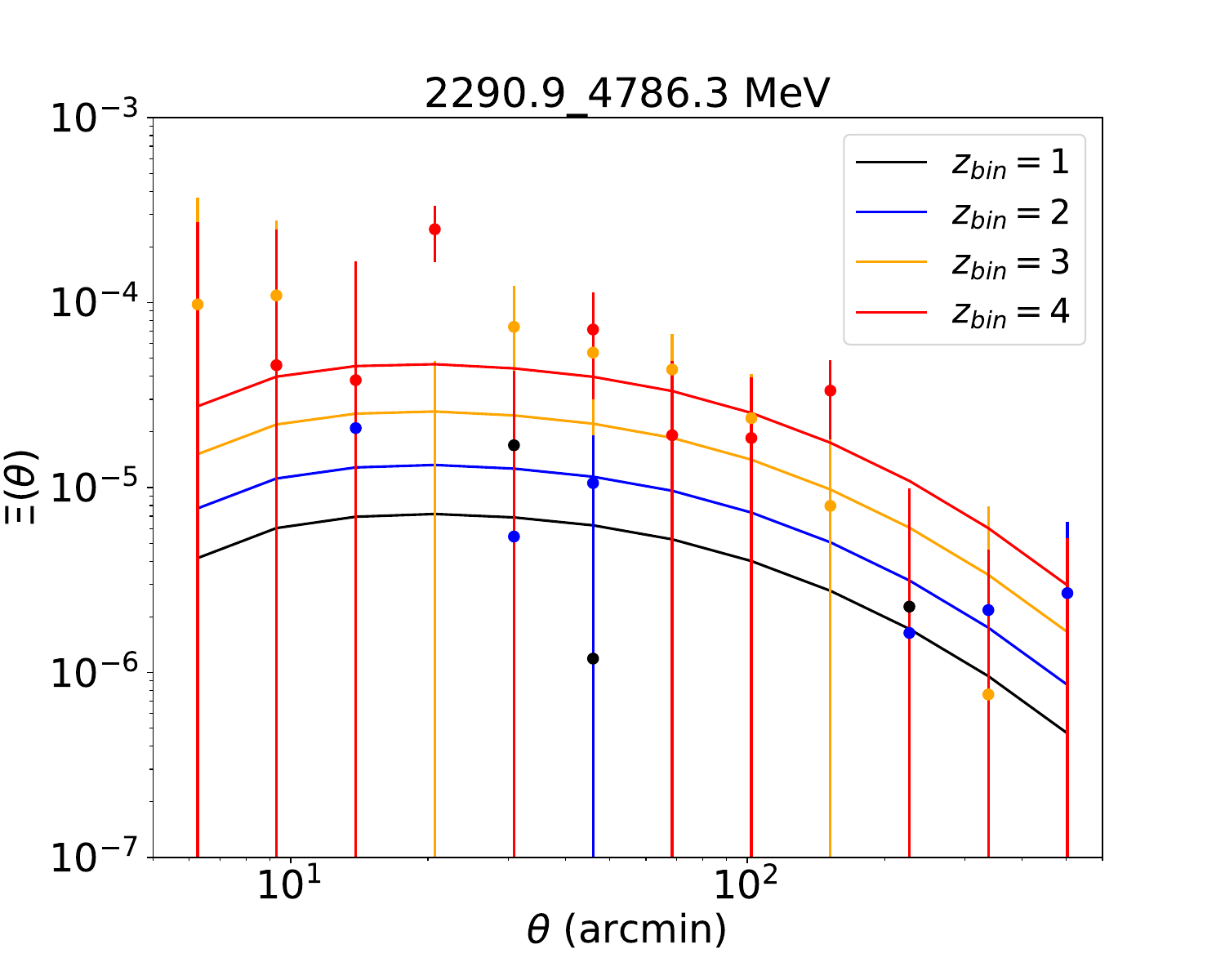}

}
}
\\
\subfloat{
\scalebox{1}{
    \includegraphics[width=0.33\linewidth]{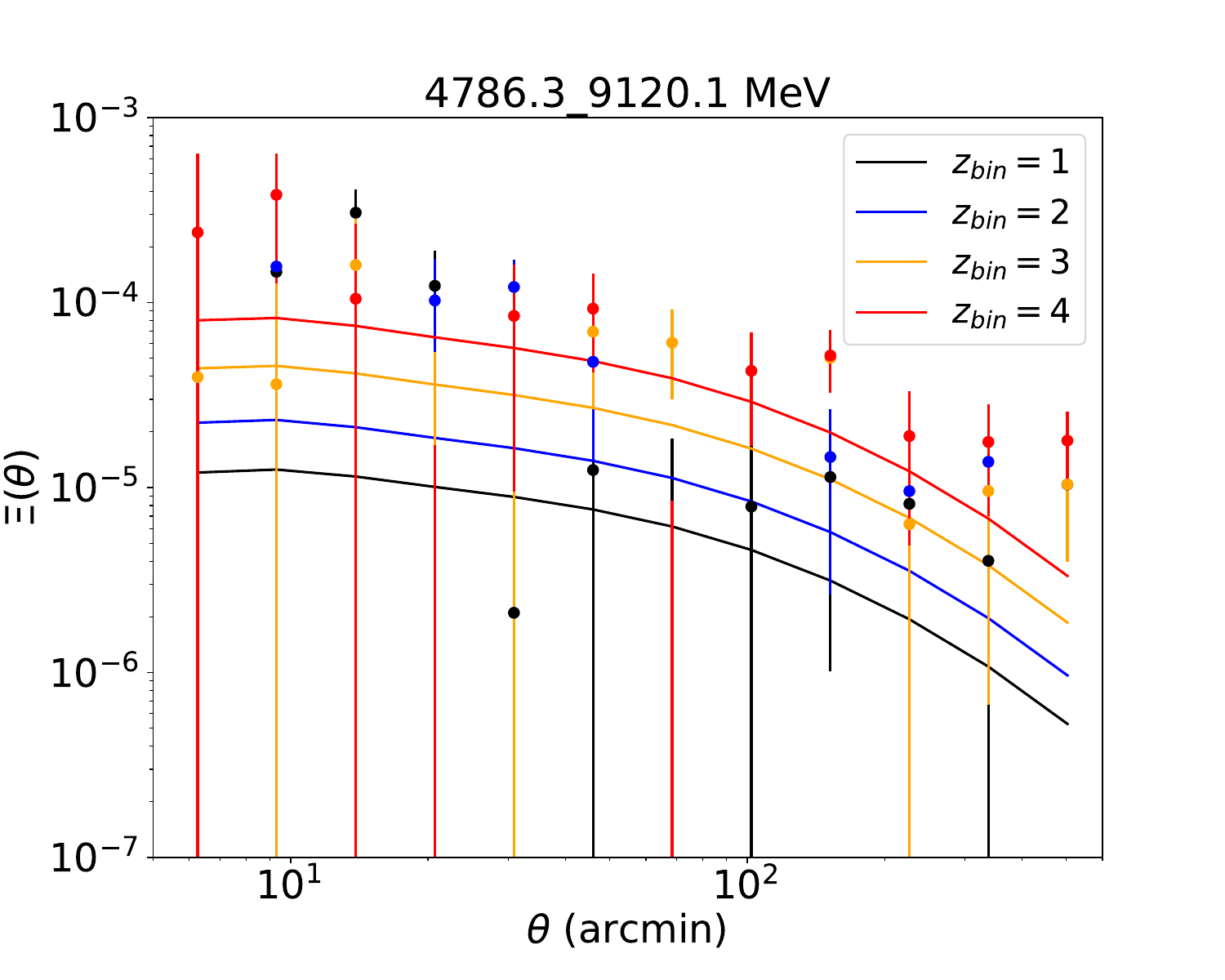}

}

\scalebox{1}{
    \includegraphics[width=0.33\linewidth]{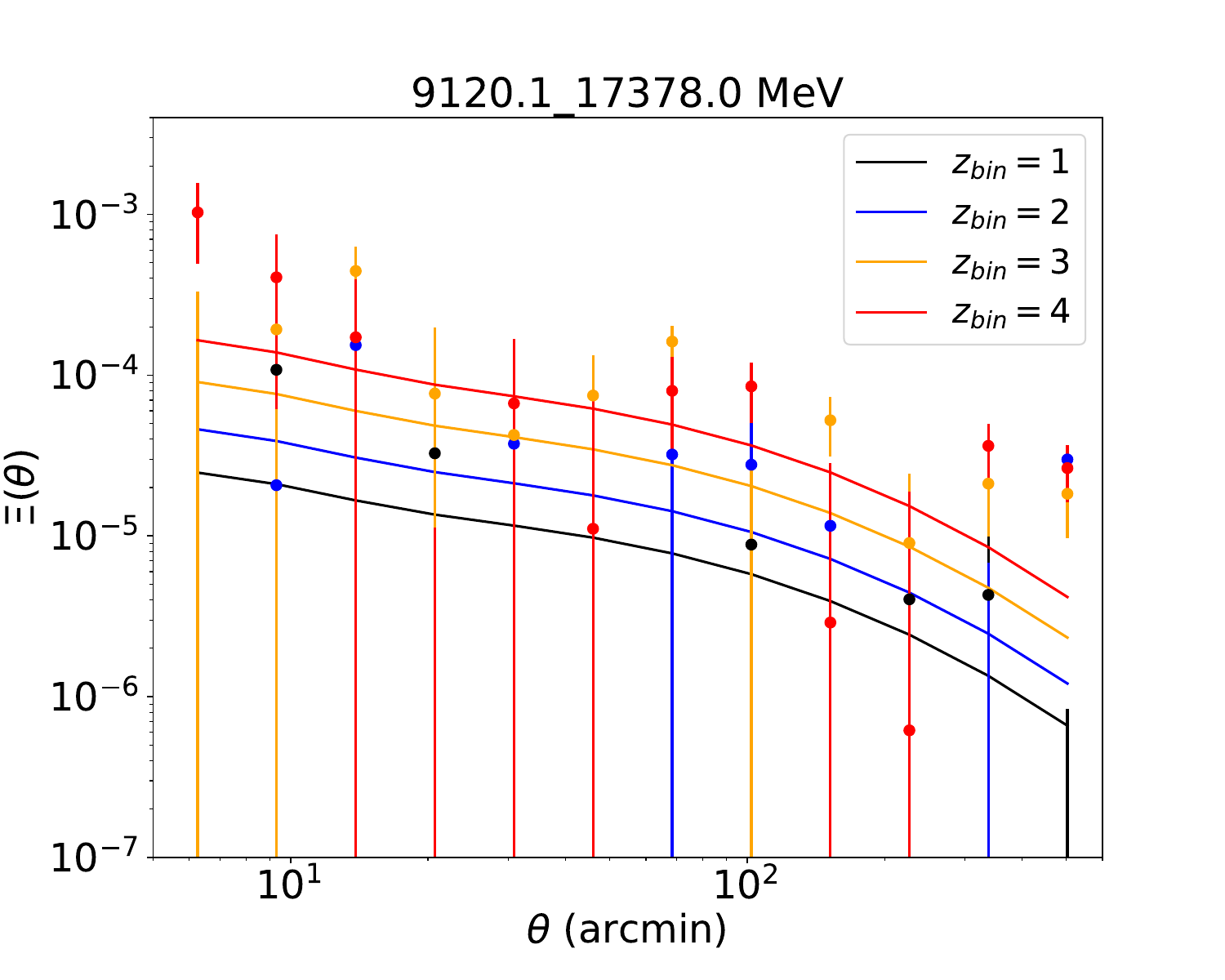}

}

\scalebox{1}{
    \includegraphics[width=0.33\linewidth]{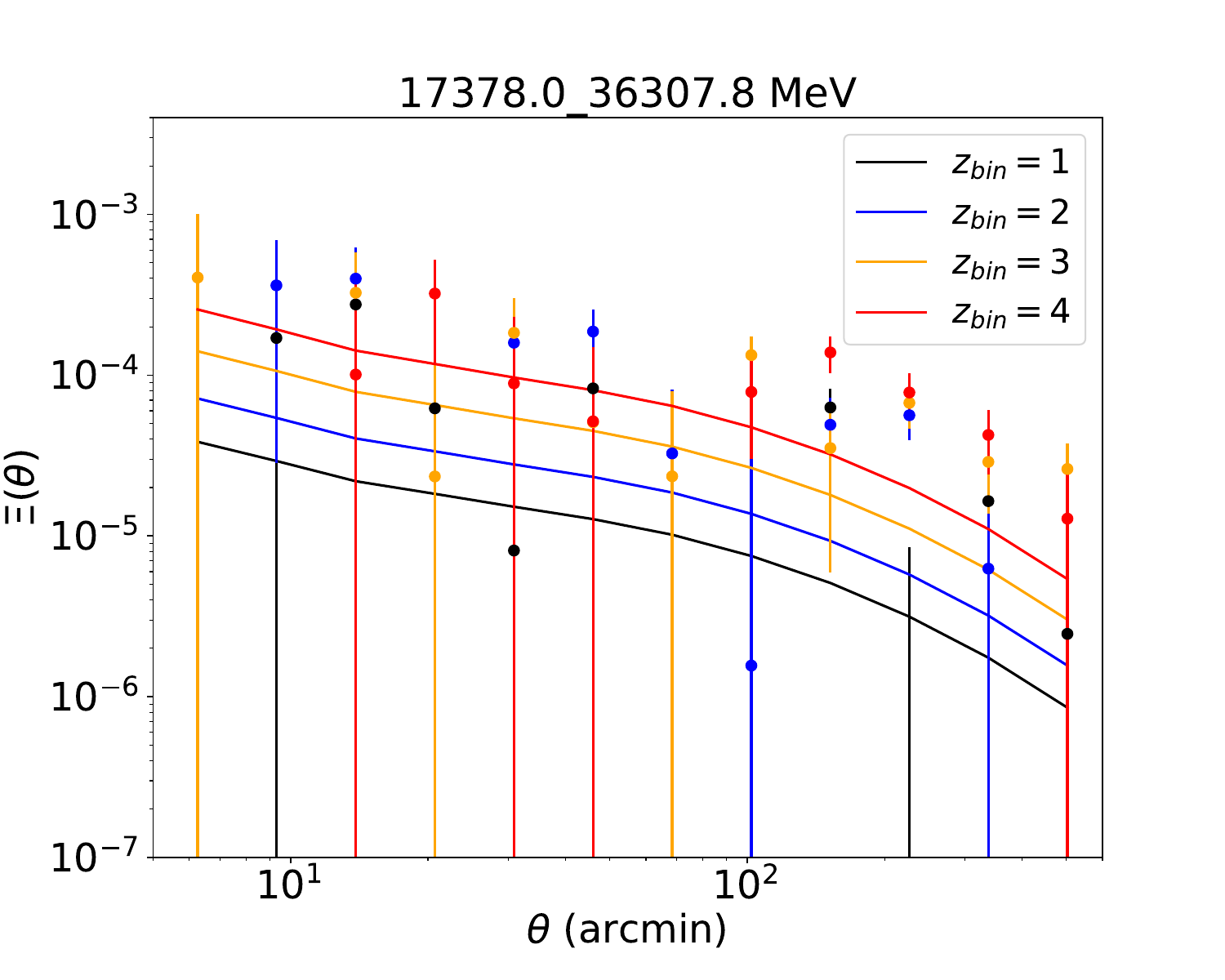}

}
}
\\
\subfloat{
\scalebox{1}{
    \includegraphics[width=0.33\linewidth]{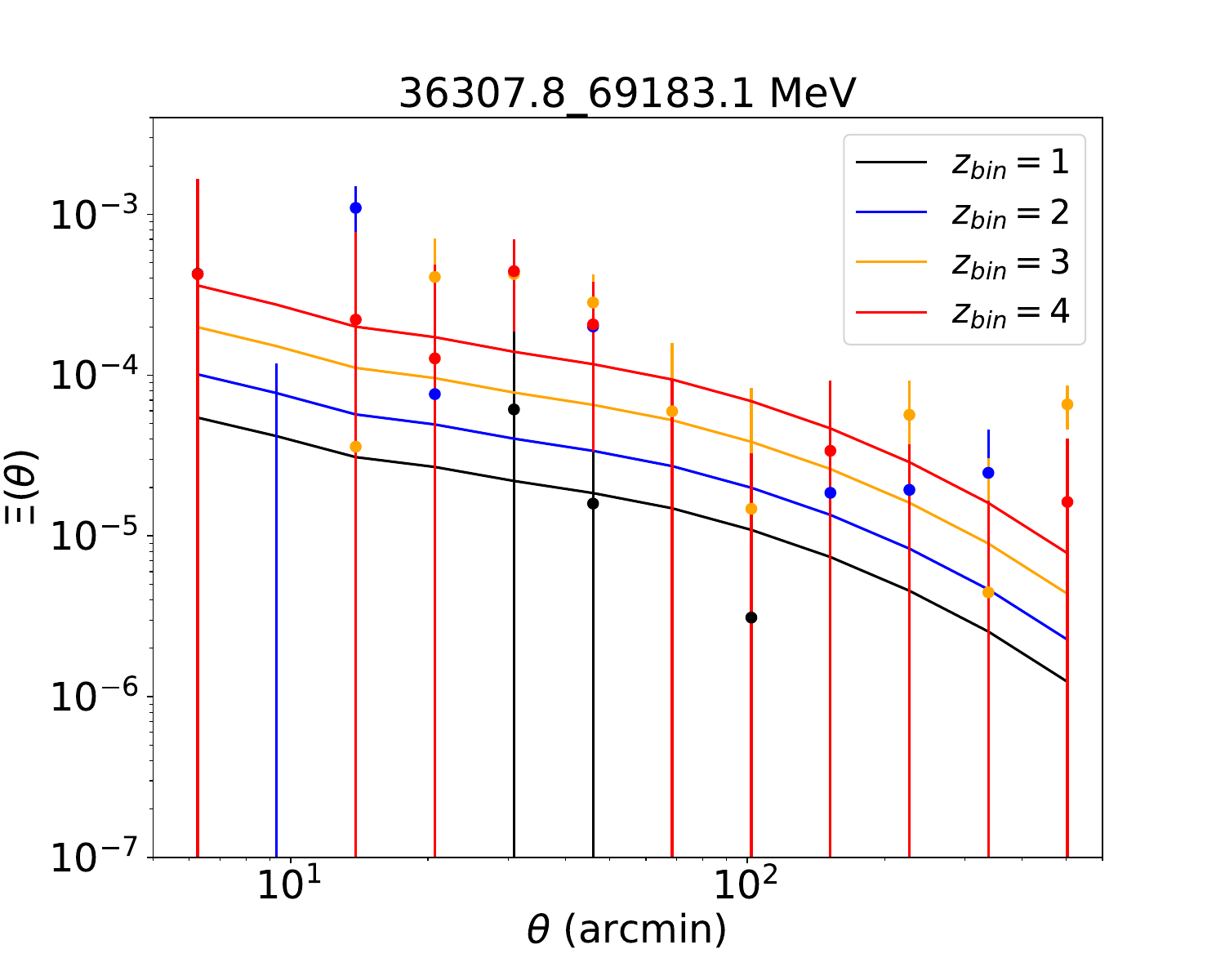}

}

\scalebox{1}{
    \includegraphics[width=0.33\linewidth]{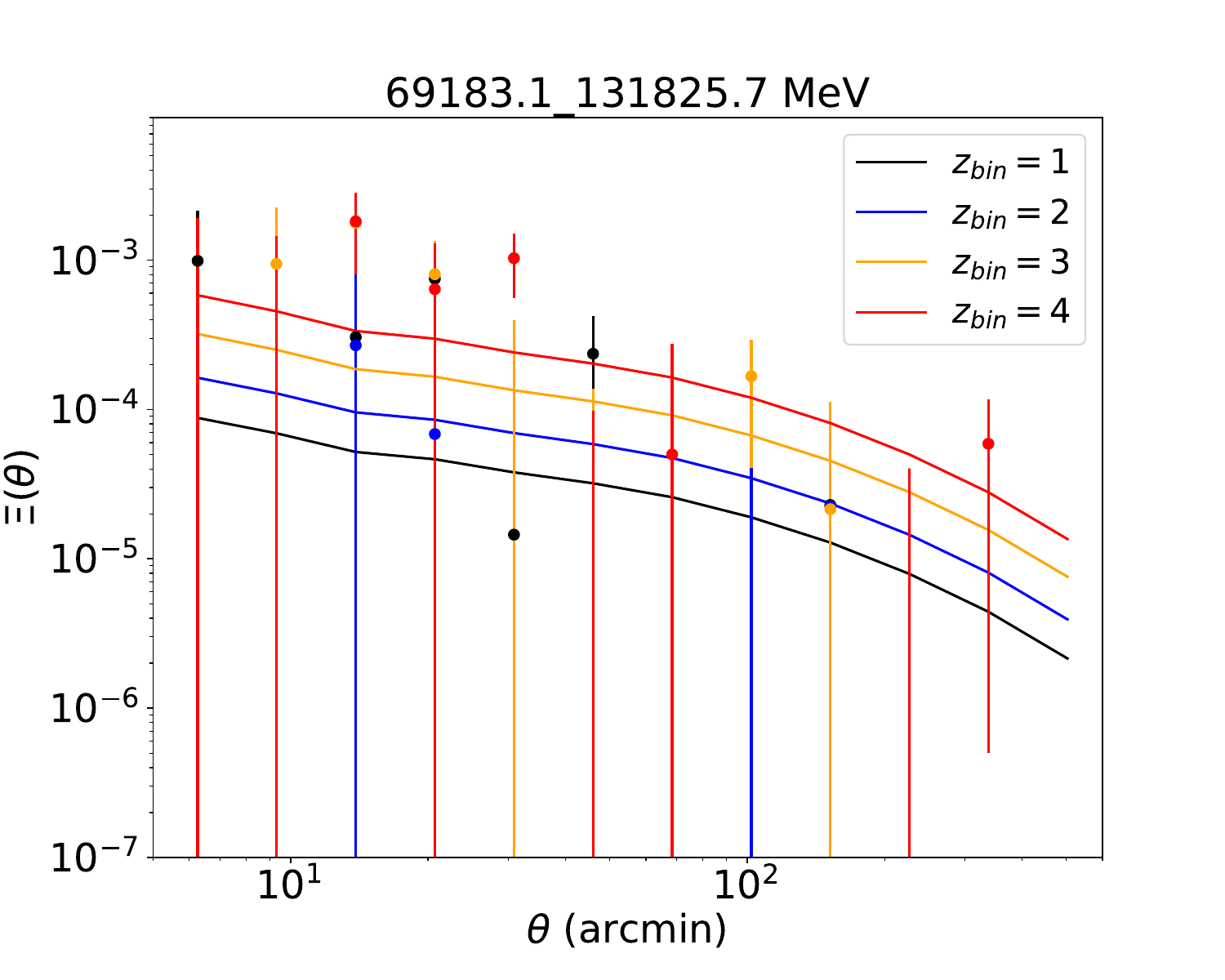}

}

\scalebox{1}{
    \includegraphics[width=0.33\linewidth]{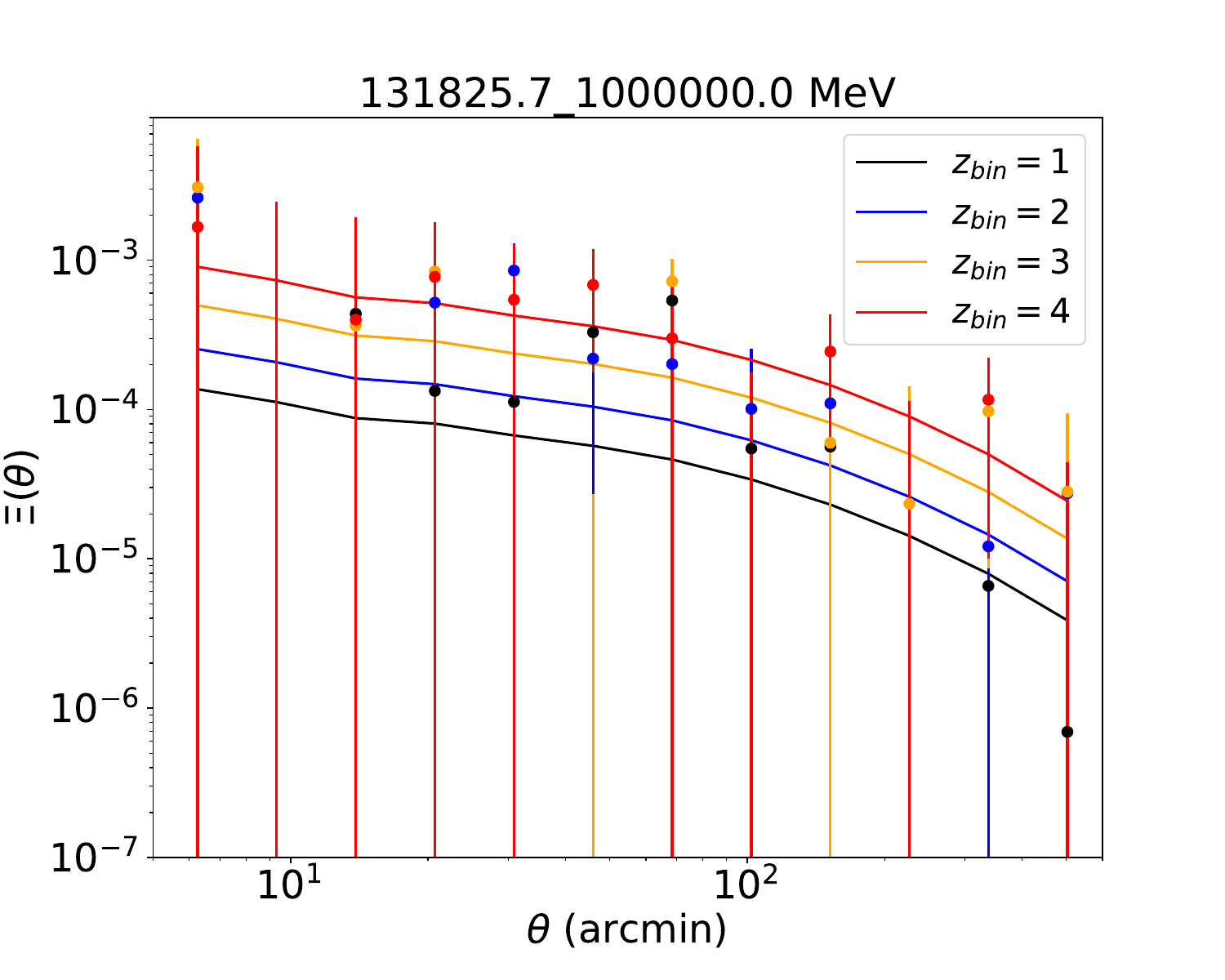}

}
}
\caption{Phenomenological model fits for the cross-correlations between the tangential shear and photon flux intensities, based on the estimator described in Eq.~\ref{eq:crossshear}.}
\label{modelallfits}
\end{figure}

Finally, in Fig. \ref{modelallfits} we report the breakdown, in terms of energy and redshift bins, of the measured cross-correlations signal, together with the best-fit power-law phenomenological model. By looking at the different panels, we notice that the signal appears to become stronger as the energy and redshift increase, with the lowest energy-redshift bin seeming to possess only a visible 2-halo component, showing that the SNR for the total model (i.e the sum of the 1-halo and 2-halo components) is higher at larger angular bins as depicted in Table \ref{tab:chi2comp}. We find here, visually, that as we move to larger energy and redshift bins, the 1-halo terms start to emerge, as the signal on the lower end of the angular scale becomes stronger relative to the lower energies and redshifts.

\end{document}